\documentclass[review,number,sort&compress]{elsarticle}

\usepackage{graphicx}
\usepackage{amssymb}
\usepackage{amsmath}
\usepackage{nicefrac}
\usepackage[hyphens]{url}
\usepackage{xcolor}
\usepackage[normalem]{ulem}

\journal{Nuclear Instruments and Methods }

\begin{document}

\begin{frontmatter}

\title{Self-triggered radio detection and identification of cosmic air showers with the OVRO-LWA}

\author[caltech]{Ryan Monroe}
\author[jpl]{Andres Romero Wolf}
\author[caltech]{Gregg Hallinan}
\author[uci,hu]{Anna Nelles}
\author[caltech]{Michael Eastwood}
\author[caltech]{Marin Anderson}
\author[caltech]{Larry D'Addario}
\author[caltech]{Jonathon Kocz}
\author[caltech]{Yuankun Wang}
\author[caltech]{Devin Cody}
\author[ovro]{David Woody}
\author[nrao,unm]{Frank Schinzel}
\author[unm]{Greg Taylor}
\author[harvard]{Lincoln Greenhill}
\author[swin,ucb]{Daniel Price}

\address[caltech]{California Institute of Technology, 1200 E California Blvd MC 249-17, Pasadena, CA 91125, USA}
\address[nrao]{National Radio Astronomy Observatory, P.O. Box O, Socorro, NM 87801 USA}
\address[unm]{Department of Physics and Astronomy, University of New Mexico, Albuquerque, NM 87131 USA}
\address[swin]{Centre for Astrophysics \& Supercomputing, Swinburne University of Technology, PO Box 218, Hawthorn, VIC 3122, Australia}
\address[ucb]{Department of Astronomy, University of California, Berkeley, 501 Campbell Hall \#3411, Berkeley, CA, 94720, USA}
\address[harvard]{Harvard-Smithsonian Center for Astrophysics, 60 Garden Street, Cambridge MA 02138 USA}
\address[jpl]{Jet Propulsion Laboratory, California Institute of Technology, 4800 Oak Grove Dr, Pasadena, CA 91109, USA}
\address[ovro]{California Institute of Technology, Owens Valley Radio Observatory, Big Pine, CA 93513, USA}
\address[uci]{University of California, Irvine, Irvine, CA 92697, USA}
\address[hu]{Humboldt University of Berlin, Unter den Linden 6, 10099 Berlin, Germany}

\begin{abstract}
A successful ground array Radio Frequency (RF)-only self-trigger on 10 high-energy cosmic ray events is demonstrated with 256 dual-polarization antennas of the Owens Valley Radio Observatory Long Wavelength Array (OVRO-LWA). This RF-only capability is predicated on novel techniques for Radio Frequency Interference (RFI) identification and mitigation with an analysis efficiency of 45\% for
shower-driven events with a Signal-to-noise ratio 
$\gtrsim$
5 against the galactic background noise power of individual antennas. This technique enables more efficient detection of cosmic rays over a wider range of zenith angles than possible via triggers from in-situ particle detectors and can be easily adapted by neutrino experiments relying on RF-only detection. This paper discusses the system design, RFI characterization and mitigation techniques, and initial results from 10 cosmic ray events identified within a 40-hour observing window. A design for a future optimized commensal cosmic-ray detector for the OVRO-LWA is presented, as well as recommendations for developing a similar capability for other experiments -- these designs either reduce data-rate or increase sensitivity by an order of magnitude for many configurations of radio instruments.
\end{abstract}


\end{frontmatter}

\section{Introduction}

The detection of radio signals associated with cosmic ray air showers has become a standard technique for low-frequency radio telescope arrays over the past decade. Dedicated arrays such as
the Auger Engineering Radio Array (AERA) \cite{2016PhRvD..93l2005A}, as part of the Pierre Auger Observatory
\citep{2015NIMPA.798..172P} or Tunka-Rex \citep{2015NIMPA.802...89B}, as well as commensal searches on low-frequency radio-astronomy telescopes such as LOFAR \citep{2013A&A...560A..98S},
regularly measure energy and composition. Radio detection of air showers
combines the precision of optical methods with the almost full up-time of particle detectors~\citep{2016PhR...620....1H}. Radio detection
has been shown to yield very precise $X_{max}$ measurements (a tracer of composition) \citep{2014PhRvD..90h2003B}, as well as excellent energy resolution \citep{2016PhRvL.116x1101A},
which are the key parameters needed to understand the origin of cosmic
rays. 

To date, ground-based radio arrays have relied upon co-located particle detectors to identify air shower events. However, direct radio detection is expected
to be more efficient because radio waves have a much longer range
of propagation compared to the particles in the air shower, which
are stopped after a few kilometers in the atmosphere. This is particularly
a limitation for more inclined showers~\citep{2017EPJWC.13501015K}, as the efficiency of the particle
detectors is limited by the trigger efficiency. Radio telescopes like LOFAR
that are able to measure air shower signals \citep{2013A&A...560A..98S}
have built particle arrays to facilitate the trigger \citep{2014NIMPA.767..339T}.
While the particle detectors also provide additional information for
composition studies \citep{LOFARNature}, these additional hardware
requirements complicate the search for cosmic ray signals in non-dedicated
radio arrays and limit the efficiency of air shower detection. 

Most experiments have chosen to rely on an external trigger due to
the challenges inherent to a trigger on the radio pulse itself, a
\emph{self-trigger}. In most environments, simply triggering on broadband pulses
that are randomly occurring in time and direction by means of a threshold above
the noise floor will yield extremely high trigger rates due to Radio Frequency Interference (RFI) caused by local
human activity. Outside of remote regions in Antarctica, experiments
aiming to detect neutrinos and cosmic rays with an independent self-trigger have been challenging due to this RFI environment\citep{2017APh....90...50B,2010PhRvL.105o1101H,2012JInst...7P1023A,2011APh....34..717A}.

Early experiments to detect air showers have employed a self-trigger,
but all ultimately had to verify their data using a particle array,
partly due to limited knowledge of the nature of the radio emission
pattern, and partly due to the complexity of finding a viable discriminator
\citep{2009APh....31..192A,LOPESselftrigger,2012JInst...7P1023A,2013NIMPA.725..133K}.
All working self-triggers had to severely limit the acceptance of
signals due to local noise sources, such as transmitters, or airplanes
\citep{2016JInst..11P1018T}. This can be overcome by using an external
trigger, which was then deemed more reliable. 

Separately, it has been proposed to build radio arrays to detect the radio signal
following the interaction of a $\tau$-neutrino of an energy above
$10^{15}$\,eV \citep{2017EPJWC.13502001M,2017ICRC...35..234L}.
In order to reach the necessary sensitivity to detect these cosmogenic
neutrinos \citep{1966PhRvL..16..748G,1966JETPL...4...78Z},
these experiments will need to employ a radio self-trigger \citep{2008PhRvL.100u1101A,2016PhRvL.117x1101A}.
The trigger has to be all the more effective at rejecting
RFI due to the expected much lower rate of $\tau$-neutrino events, a process which has to happen at least semi-locally in the case of distributed systems, as data rates will be limited. Also, the computational power
of autonomous stations will be restricted, which will limit the complexity of
procedures to identify the air shower signals. It will, therefore, be necessary
to know the requirements for a successful self-trigger when designing
the detector system.

Independently of cosmic ray science, there has been an explosive growth
in low-frequency arrays for astronomy. These arrays, including
LOFAR \citep{2013A&A...556A...2V}, MWA \citep{mwa_instr_paper},
PAPER \citep{paper_paper}, the LWA \citep{5109716,fig:another_lwa} and OVRO-LWA, offer excellent collecting
area and computing resources, conducive to commensal air shower detection, but in many cases are limited by both
their impulsive RFI\footnote{Radio Frequency Interference} environment
\citep{astron_data_quality, 2013A&A...549A..11O}, as well as the fact that their array
configurations were not optimized for the task of cosmic ray detection.
As demonstrated by LOFAR, accessing these arrays can provide the opportunity to
perform powerful commensal air shower searches at low cost. These arrays would obviously benefit from the ability to self-trigger.

This paper presents the demonstration of a successful design of a cosmic ray self-trigger
system using only the radio signal implemented on the OVRO-LWA. First, we will summarize the radio instrument (Section~\ref{ovro-desc}). We then discuss an overview of the techniques used in Section~\ref{sec:Summary-of-technique}, as well as discriminating characteristics of cosmic rays in Section~\ref{sec:Detection-of-cosmic}. Details on these methods are provided in Sections \ref{sec:Firmware} and \ref{enu:Software}.
A brief characterization of the types of RFI seen, as well as successful
cosmic ray detections, are covered in Section~\ref{sec:Characterization-of-events}.
Recommendations for future instrument design and ``Lessons Learned'' are discussed in the Appendix.

%%%%%%%%%%%%%%

\section{The OVRO-LWA\label{ovro-desc}}

This experiment was performed by re-purposing an existing instrument:
the Owens Valley Radio Observatory Long Wavelength Array (OVRO-LWA) to perform cosmic
ray detections without modifying any hardware. The system was originally
designed as a fully cross-correlating interferometer, for which it
continues to be used for astronomical purposes, transient
science \citep{2018ApJ...864...22A}, and cosmology \citep{2018AJ....156...32E}. As such, the array is at a site which has no particle
detectors or fluorescence telescopes (the standard methods of detecting
cosmic ray events). 

\begin{figure}[hbt!]
\centering
\includegraphics[width=0.8\columnwidth]{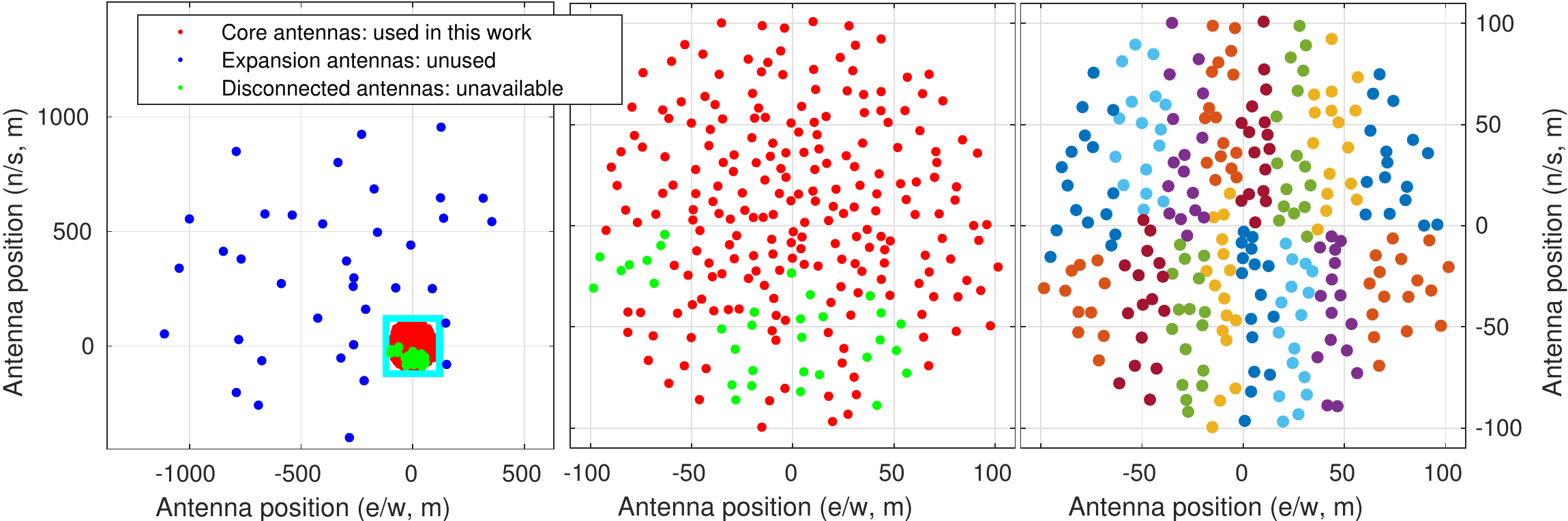}
\caption{\textbf{Left:} a summary of the 288 OVRO-LWA antennas currently installed. Red antennas were used in this work. Blue antennas could be used in this work, but were excluded in favor of the dense core. Green antennas are physically present and available in an alternative array configuration, but are not used in this work. \textbf{Center:} the same contents, zoomed on the cyan box.  The 251 "core" antennas, randomly distributed across a diameter of 200\,m, are shown. \textbf{Right:} The same antennas as displayed in the center figure. Each adjacent cluster of antennas which share the same color are assigned to the same FPGA, and are therefore grouped together for the purposes of initial triggering, which limits the performance of on-chip RFI mitigation when using the technique described in Section \ref{subsec:On-chip-RFI-suppression}.}
\label{fig:array_layout}
\end{figure}

The OVRO-LWA is located at the Owens Valley Radio Observatory, near
Bishop, CA, USA, where it has substantial shielding from
major population centers, courtesy of the Sierra and Inyo mountains.
As a consequence, the environment is very good for shielding against intentionally transmitted
RFI. However, power lines in the valley are responsible for impulsive,
band-limited RFI which makes the detection of cosmic rays challenging.
Emphasizing cosmology and transient science, the array consists of
288 dual-polarization antennas (of which only 256 are in active use)
distributed across a 1.53\,km-diameter region. A 200\,m diameter core contains 251 of these antennas
and was
the focus of this pilot effort\footnote{Once completed, the full array will consist of 352 antennas spread across a 2.6\,km region.}. The distribution of these antennas is shown in Figure~\ref{fig:array_layout}. OVRO-LWA inherits its antenna and analog electronics from the New Mexico Long Wavelength Array stations (LWA, \citep{5109716}), which are similar (but with a smaller, 100 m diameter layout,
and an emphasis on high time-resolution beam forming as well as all-sky imaging). Each antenna sees nearly the entire
sky, with a 3\,dB point of approximately $40^\circ$ elevation angle. RF signals are baseband-sampled at $F_{s}\sim196.6$\,MSPS, for a Nyquist frequency of $\sim98.3$\,MHz.  This is performed with 8-bit ADC16x250-8 Analog to Digital Converter (ADC) boards \citep{adc16x250}.

%At OVRO-LWA, each antenna is baseband
%sampled at $F_{s}\sim196.6$\,MSPS, for a Nyquist frequency of %$\sim98.3$\,MHz.
%This is performed with 8-bit ADC16x250-8 Analog to Digital Converter
%(ADC) boards \citep{adc16x250}, each an on-board HMCAD1511 \citep{hmcad1511}
%ADC. In the band of interest ($25-75$\,MHz), both the power observed by each antenna and the system %temperature are dominated by the galactic background (Figure~\ref{fig:Typical_spectrum})\textendash{}although the upper and lower portions of the band are %severely contaminated, the center is galactic background dominated at most frequencies. Signals are then processed by the Large-Aperture Experiment to Detect the Dark Ages
%(LEDA) back-end, consisting of 16 CASPER ROACH2 Field Programmable Gate Arrays (FPGAs); each contains a Virtex 6 XC6VSX475T FPGA (32 inputs or 16 antennas
%per FPGA). During normal correlator operations, these FPGAs perform
%the ``F-engine'' \citep{2017arXiv171100466E,2017arXiv171106665A,2014JAI.....350002K}
%portion of the correlator, which applies a Polyphase Filterbank to form 4096 channels. Subsequently, a 10\,GbE / 40\,GbE Ethernet
%switch is used to transfer data to a set of GPU computers for further
%processing, which is not relevant for this work.

The OVRO-LWA (full technical details available at \citep{2015JAI.....450003K}) provides an ideal testing ground for both the development
of a self-trigger, and verifying the utility of existing arrays for detecting cosmic rays. With few computational restrictions, different
avenues can be explored and it can be determined whether air shower detection
without the confirmation from particle detectors is at all feasible, let
alone the more challenging task of radio detection of $\tau$ neutrinos.
Having all signals brought back to a central location (as opposed
to beam-forming in the field, for instance), as well as possessing a versatile digital back-end \citep{2015JAI.....450003K}, enable a
wide variety of techniques to be quickly prototyped and validated.
 As well as being a powerful and commensal cosmic ray experiment,
running a sensitive and efficient self-trigger at the OVRO-LWA can
be a first step in proving the feasibility of proposed neutrino arrays
and is an important step towards an independent method of air shower
detection. 

%%%%%%%%%%%%%%%%%

\section{\label{sec:Summary-of-technique}Summary of Technique}

\begin{figure}[hbt!]
\centering
\includegraphics[width=0.8\columnwidth]{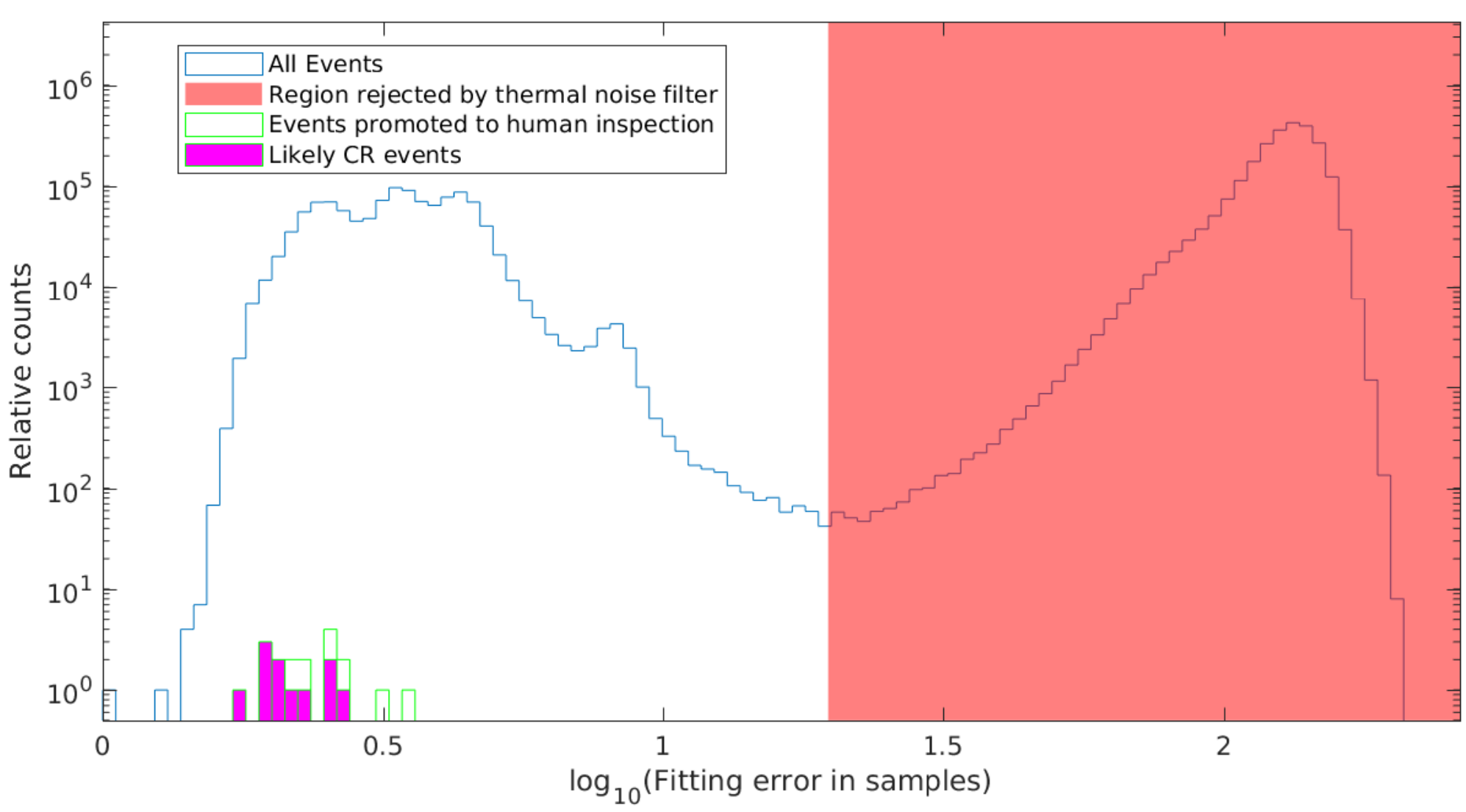}
\caption{Histogram of time-of-arrival residuals from the direction of arrival fitting. The median residual was time-variable, but typically $3\sim4$ samples. High fitting residuals (chosen to be $>12$ times the median residual) usually indicate a spurious trigger caused by random thermal noise\textendash{}alongside other basic instrumental checks, this is the first cut performed outside the FPGA.}
\label{fig:fit_errors}
\end{figure}

For this demonstration, we re-use the existing hardware in its current
configuration for the task of cosmic ray detection. The standard F-engine
firmware is replaced with firmware which is designed to detect impulsive
events while identifying and filtering known RFI sources. Preliminary
detections are made at an individual FPGA level: initially, each FPGA
processes data from the 16 antennas it services (which are spatially
localized to rectangular regions roughly 25\,m by 100\,m in extent),
and makes a preliminary decision using only this data. After detecting
a promising impulsive event, the FPGA transmits an appropriate time-stamp
to all 15 other FPGAs, which respond by dumping their data for that
time window over 10\,GbE to a host server. That server stores the candidate
to disk for offline processing. After on-chip RFI filtering, the trigger
is sensitive enough that 60\% (typical conditions) to 80\% (ideal
conditions) of on-chip triggered events are caused by random thermal
noise \textendash{} evidence that the system is close to thermally limited (Figure \ref{fig:fit_errors}). %This is also illustrated in Figure~\ref{fig:fit_errors}

Candidate events are then processed in MATLAB to produce the higher level
statistics necessary for identification of true cosmic ray events.
Primary flagging metrics include observed power at each antenna, direction,
and range of source arrival, quality of a spherical wavefront model
in source fitting, and time-domain clustering of events (see Section~\ref{enu:Software}). This flagging
reduces the number of impulsive candidates by a factor of $\sim$220,000.
In the 40-hour dataset presented in this paper, 16 events are sufficiently promising to warrant human inspection.  Of these, 10 events (62\%) are classified as cosmic rays, a further 2 events (13\%) are unable to be confirmed or rejected as air shower events and 4 events (25\%) are most likely caused by RFI.  This is equivalent to 5-6 cosmic-ray driven events per day under good RFI conditions.

The largest challenge in this project is filtering RFI events from
cosmic ray events: under good RFI conditions, the system captures approximately
400K impulsive candidates for every cosmic ray ($\sim$500~Hz, versus $\sim$hourly).
This issue is worsened by the fact that this system was not designed
for cosmic ray detections. There is insufficient communication bandwidth
between the FPGAs and the Ethernet switch with which they are connected,
making transmission of the entire raw ADC data-stream impossible. For this reason and others,
triggering must be performed on an individual FPGA-basis with minimal communication,
severely limiting the systems ability to reject RFI. The use of hierarchical beamforming \textendash{} conceived after the bulk of this work was finished \textendash{} might relax this constraint (see Section~\ref{sec:Future-experimental-design}).

\section{\label{sec:Detection-of-cosmic}Identification of cosmic rays}

\begin{figure}[hbt!]
\centering
\includegraphics[width=0.8\columnwidth]{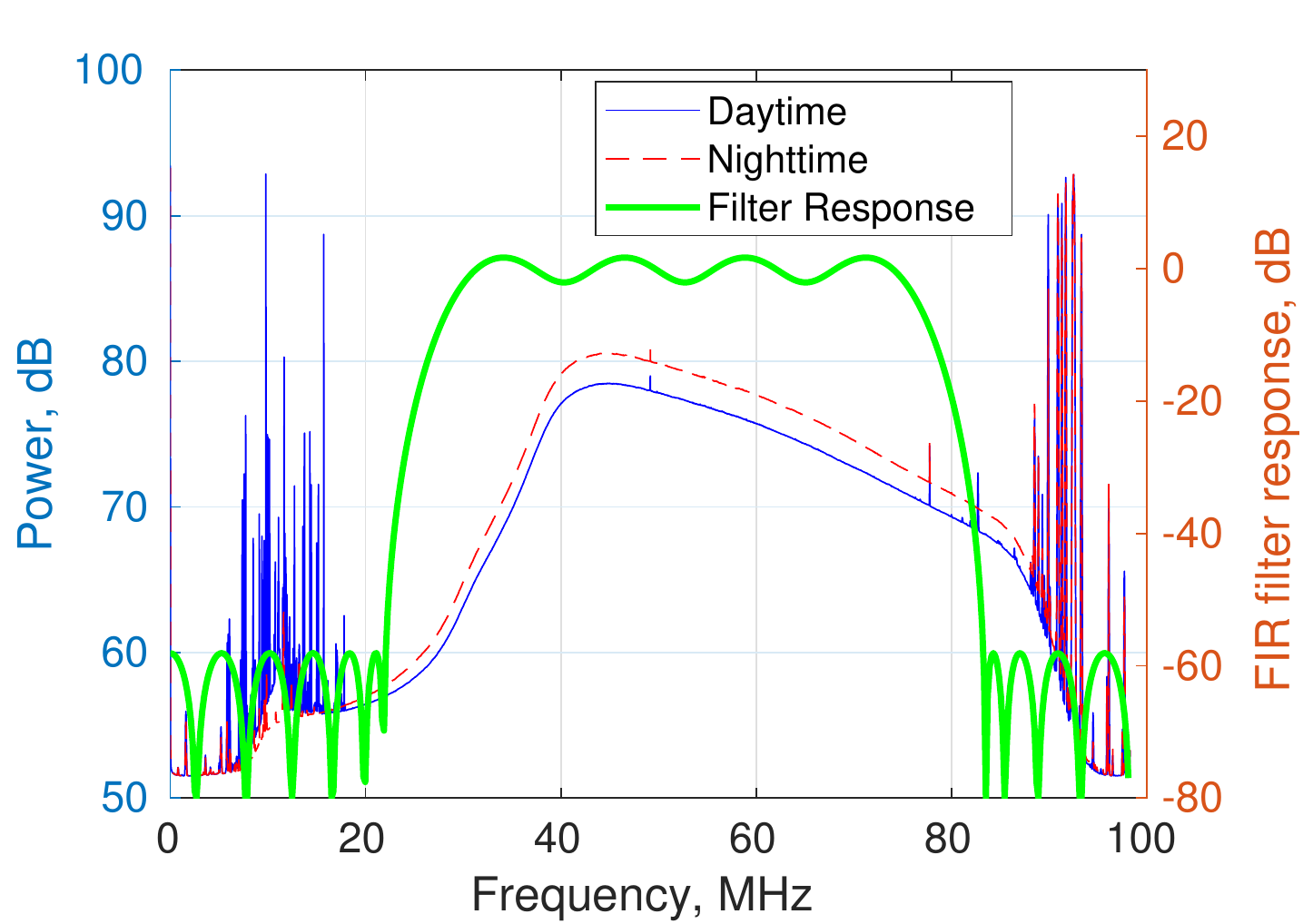}
\caption{Typical spectra as seen from
the OVRO-LWA. The on-chip integrators used to produce this spectrum saturate
at 93 dB, meaning that the true power of the RFI at high frequencies
may be higher than what is shown in this figure. Also shown is the
FIR filter response used in Sections \ref{sec:Firmware} and \ref{enu:Software}.}
\label{fig:Typical_spectrum}
\end{figure}

Although the RFI conditions of the OVRO-LWA site do not interfere with the standard imaging mode, the impulsive RFI environment is very challenging for the detection of cosmic ray events.
Figure~\ref{fig:Typical_spectrum} shows a typical spectrum, which is galactic noise dominated in the 20-85 MHz band except for a few narrow band RFI sources. The low frequencies are convenient for air shower detection, as radio emission in the atmosphere peaks in this band and the radio emission beam pattern is wider compared to higher frequencies. Approximately 500 impulsive RFI events
(with more power than the galactic background noise) occur per second (see Section~\ref{subsec:On-chip-RFI-suppression}), compared to the cosmic ray incidence rate of $\sim$hourly
\citep{cr_rates1,cr_rates2}. For this reason, aggressive use of descriptive
statistics must be used to filter out promising cosmic ray candidates
from spurious events. Motivated by radio measurements \citep{2013A&A...560A..98S}, these properties can broadly be described
as:
\begin{itemize}
\item Input signal is roughly a band-limited impulse \citep{2013AIPC.1535..105N}, convolved with the instrumental impulse response.
\item Time and direction of arrival is usually not coincident with a known
or detected RFI source (mostly stationary sources and airplanes),
which can be clustered in either direction or time+direction.
\item Power distribution in the axis of ray travel is similar to that described
in \citep{2015JCAP...05..018N} \textendash{} for this work, we searched
for a localized region of power with a FWHM\footnote{Full-Width-Half-Max} of 100-200\,m on the
ground. Events arriving at low elevation angles could have much broader extent, but this initial work focused on those events which were most easily discriminated by the OVRO-LWA core.
\item Polarization properties roughly match what is expected given an event's
direction of arrival and the local geomagnetic field \citep{2014JCAP...10..014S} \textendash{} this was used for final validation of events.
\item Finally, many events are discarded if it is likely that event power is insufficient for quality parameter estimation, or instrumental errors (such as dropped data due to network bandwidth saturation) are likely to result in ill-conditioned fits.
\end{itemize}
The system described below was designed to use these properties with
the goal of both detecting cosmic rays while rejecting as many RFI
events as possible, and as early as feasible in the signal chain so as to manage data-rates and make more sensitive detections.

%%%%%%%%%%%%%%%

\section{\label{sec:Firmware}Trigger}

Here we discuss the algorithms used in the FPGA firmware to detect
cosmic rays and perform preliminary RFI filtering. For definitions
of symbols used in this section, refer to Table \ref{tab:Firmware-symbol-definitions.}. 

%%%%%%%%%%%%

%%%%%%%%%%%%%%%%%%%%%%%

\begin{table}[hbt!]
\centering
\begin{tabular}{ll}
\hline
Symbol & Meaning\\
\hline
$k$ & Input index ($0\sim31$ or $0\sim511$)\\
$v_{k}\left[n\right]$ & Sampled voltage for k'th input \\
$v_{filt,k}\left[n\right]$ & Filtered voltage for k'th input \\
$h\left[m\right]$ & Coefficients for 40-tap symmetric FIR filter \\
$p_{k}\left[n\right]$ & Filtered power for k'th input \\
$p_{smooth,k}\left[n\right]$ & Filtered power for k'th input (smoothed) \\
$T_{k}$ & Triggering threshold for k'th input \\
$E_{I,k}\left[n\right]$ & Input-wise event for k'th input \\
$D_{k}$ & Corrective delay for k'th input (in samples) \\
$E_{D,k}\left[n\right]$ & Event for k'th input (delayed) \\
$N_{bl}$ & Number of samples to block for consecutive pulses \\
$E_{B,k}\left[n\right]$ & Event for k'th input (after repeated event blocking) \\
$N_{sus}$ & Event sustaining window (in samples) \\
$E_{sus,k}\left[n\right]$ & Event for k'th input (additionally sustained) \\
$N_{trig}\left[n\right]$ & Number of input triggers required for FPGA trigger \\
$C\left[n\right]$ & FPGA-wise event detection \\
\hline
\end{tabular}
\caption{Firmware symbol definitions.}
\label{tab:Firmware-symbol-definitions.}
\end{table}

%%%%

\begin{figure}
\includegraphics[width=0.8\columnwidth]{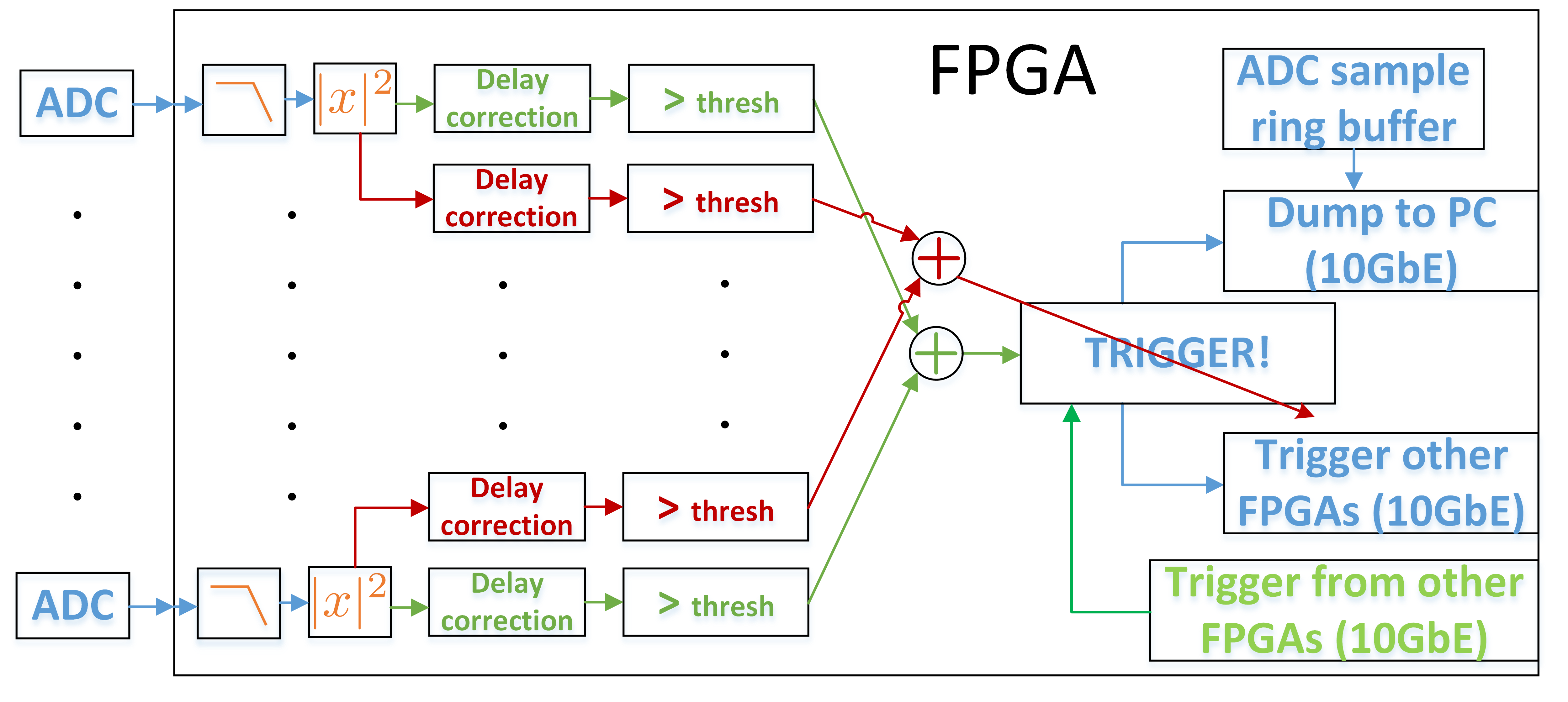}
\caption{Simplified block diagram of the FPGA
system. Full details described in Section~\ref{sec:Firmware}.}
\label{fig:FPGA_block_diagram}
\end{figure}

As mentioned in Section \ref{sec:Summary-of-technique}, each ROACH2
processes 16 antennas, for 32 inputs per FPGA. A circular buffer stores
raw ADC samples for triggered dumping, while the signals are processed
in parallel to detect impulsive events received by the system. A simplified
block diagram of the firmware is shown in Figure~\ref{fig:FPGA_block_diagram}. All math
used in this work is real-valued.

%%%%%%%%

\subsection{Filtering and power smoothing\label{filterSmooth}}

First, a 40-tap symmetric-coefficient bandpass FIR filter $h\left[m\right]$  (eq. \ref{eq:filter})
is used to filter the ADC voltage samples ($v_k[n]$ for each antenna indexed by $k$) to the range {[}30,75{]}\,MHz
(Figure~\ref{fig:Typical_spectrum}), which was chosen to avoid narrow-band,
high-duty-cycle RFI sources. Xilinx FPGAs have special features which
half the filter resource requirements for symmetric FIR filters, driving
that decision. Afterward, filtered voltages are squared to compute
power $p_k[n]$ (eq. \ref{eq:power}), and subsequently smoothed with a 4-tap running
average filter (eq. \ref{eq:pSmooth}), producing $p_{smooth,k}[n]$; the value of 4~samples (20\,ns)
over the band-limited impulse time of 7\,samples (35\,ns) in this
system was chosen to prioritize accurate time-of-arrival estimation
over maximum sensitivity, justified because downstream processing
could not handle such low SNR events. This smoothing and triggering
process does not explicitly filter for impulsive events, but the short integration length
prefers them over events with similar power levels
but longer duration (and therefore, lower instantaneous power).

%%%%%%%%%%%%%%%%%
\subsection{Thresholding and RFI mitigation\label{thresholdRfi}}

Each of these signals which exceeds a threshold (eq.~\ref{eq:pThreshold}) is registered as an ADC-wise event (at this level, each ADC input is treated completely independently \textendash{} it is not until Eq.~\ref{eq:fpgaTrig} as well as Section~\ref{enu:Software} that multiple inputs for a single temporal epoch are treated jointly; this limits the sensitivity of this technique). Cable
length-delays are then removed from each signal using pre-computed values 
(eq.~\ref{eq:delay}). In order to facilitate a primitive time-of-arrival
estimate on-chip, all ADC-wise events which occur within $N_{bl}$
of a preceding event are suppressed: otherwise, a powerful event could
trigger the RFI blocking system (Section~\ref{subsec:On-chip-RFI-suppression})
multiple times, resulting in powerful events being blocked regardless
of incident direction (eq.~\ref{eq:blocklen}); $N_{bl}$ is only non-zero
in the directional RFI blocker, as discussed in Section \ref{subsec:On-chip-RFI-suppression}.
Since events will not typically be detected by all ADCs simultaneously
even after cable delay\footnote{Due to geometric effects},
event detections are sustained for an appropriate number of samples (eq.~\ref{eq:sustain}) $N_{sus}$.  This value was set to 140, but could have been set as small as $100*\nicefrac{F_{s}}{c}\approxeq66$~samples,
where $c=3e8\,\nicefrac{\text{m}}{\text{s}}$ is the speed of light,
and 100\,m is the maximum extent of the core OVRO-LWA antennas
that a single FPGA services. If after thresholding, suppression and sustaining, more
than $N_{trig}$ ADCs have sustained events at any given time, that
FPGA is said to have made an impulsive event detection, denoted $C\left[n\right]$ (eq.~\ref{eq:fpgaTrig}). 

\subsection{Trigger and dump via Ethernet}

The time-stamp, which is synchronized by a global 1PPS\footnote{1 Pulse Per Second }
signal, is then transmitted to all other FPGAs via a
10 GbE interface, which synchronously halt writing to their circular
buffers and transmit an event record (meta-data as well as 4096 raw
ADC samples \textendash{} equivalent to $20\,\mu s$ near the trigger
time) to the host PC, which stores the events using the ``tcpdump''
tool. A logical ``Rising Edge''
operation is applied to the signals after thresholding individual
ADC detections to facilitate the RFI blocking algorithm: discussed below, in Section~\ref{subsec:On-chip-RFI-suppression}. There is 2.6\,ms of dead time after each trigger: likely limited by burst network capacity. Better system design could easily remove this limitation.

\subsection{FPGA signal processing: summary}

The operations described above are approximately summarized here. Equations (\ref{eq:filter}$\sim$\ref{eq:pSmooth}) are covered in Section~\ref{filterSmooth}, while Section~\ref{thresholdRfi} covers equations~(\ref{eq:pThreshold}$\sim$\ref{eq:fpgaTrig}).  Equations \eqref{eq:re1} and \eqref{eq:blocklen} are handled out-of-order to accommodate RFI mitigation needs.

\begin{align}
\mbox{(bandpass filter)\,\,}v_{filt,k}\left[n\right] & = & v_{k}\left[n\right]*h\left[m\right]\label{eq:filter}\\
\mbox{(compute power)\,\,}p_{k}\left[n\right] & = & v_{filt,k}\left[n\right]^{2}\label{eq:power}\\
\mbox{(smooth power)\,\,}p_{smooth,k}\left[n\right] & = & \sum_{m=0}^{3}p_{k}\left[n-m\right]\label{eq:pSmooth}\\
\mbox{(threshold)\,\,}E_{I,k}\left[n\right] & = & p_{smooth,k}\left[n\right]>T_{k}\label{eq:pThreshold}\\
\mbox{(rising edge)\,\,}E_{I,k}^{\prime}\left[n\right] & = & E_{I,k}\left[n\right]\&\neg E_{I,k}\left[n-1\right]\label{eq:re1}\\
\mbox{(delay)\,\,}E_{D,k}\left[n\right] & = & E_{I,k}\left[n-D_{k}\right]\label{eq:delay}\\
\mbox{(block repeat events)\,\,}E_{B,k}\left[n\right] & = & E_{D,k}\left[n\right]\&\neg\max_{s\in\left\{ 1\dots N_{bl}\right\} }\left\{ E_{D,k}\left[n-s\right]\right\} \label{eq:blocklen}\\
\mbox{(sustain detections)\,\,}E_{sus,k}\left[n\right] & = & \max_{s\in\left\{ 0\dots N_{sus}+1\right\} }\left(E_{B,k}\left[n-s\right]\right)\label{eq:sustain}\\
\mbox{(count events)\,\,}C\left[n\right] & = & \left(\sum_{k=0}^{31}E_{I-dl,k}\left[n\right]\right)>N_{trig}\label{eq:fpgaTrig}
\end{align}

\subsection{Practical considerations}

There are some run-time practicalities to making this system work.
The thresholds $T_{k}$ must be scaled to account for differences
in system gain across antennas and polarizations. This is done once at startup by selecting
a random selection of 4096~samples from each of the 512~array
inputs, and enforcing that each $T_{k}$ is inversely proportional
to the power in that input: $T_{k}=G\left(\sum_{n=0}^{4095}v_{k}\left[n\right]^{2}\right)^{-1}$.
As background RFI event rate varies, G is adjusted slowly (specifically,
scaled by a factor of 2\%$\sim$50\% every six seconds, depending
on how far away from the target range the current dump rate falls)
such that the typical number of detected events falls within a user-defined
region (12-50 events/sec in this case: 80\% of those are typically
random thermal noise, whereas almost all of the remainder are RFI).
The upper bound was set by the computational power available for post-processing,
while the lower bound was set as a quarter of the upper bound to avoid
rapid adjustments in threshold values. In terms of trigger threshold, this is equivalent to a typical threshold variation of about 30\%. Brief bursts of RFI occasionally increase this threshold by up to another 40\%, which the system quickly recovers from.

\subsection{\label{subsec:On-chip-RFI-suppression}On-chip RFI suppression}

As the system design was iterated, it was found that the large number
of RFI events was outstripping the processing ability of the software
system. In order to mitigate this problem, a parallel (on-FPGA) algorithm
was designed to block incident RFI coming from particular directions
\textendash{} this strategy was feasible because almost all of the
RFI came from only three distinct, stationary sources (see Section
\ref{sec:Characterization-of-events}). This was implemented by replicating
equations \eqref{eq:delay} through \eqref{eq:fpgaTrig} three times, gating
the main detection on a lack of activity on these ``RFI blockers''.
Each of these directional blocking modules is configured with delays
describing the timing of signals with regard to their arrival at the
FPGA (equivalent to the sum of aforementioned cable length delay and
the geometric delay appropriate for sources in the given direction).
A detection with one of these modules blocks an FPGA trigger for the
past and future $\sim2.5\,\mu s$. A few minor departures from the
original event trigger are also necessary for the RFI blocker: first,
the original sustaining period of $N_{sus}=140$~samples was designed
to capture any impulsive events: for the RFI blocking module, this is adjusted to $N_{sus}=6$~samples,
a number which (per captured data) allows the RFI system to detect
the majority of events from the chosen source, while blocking events
from only a trivial fraction of the sky. Additionally, equations \eqref{eq:re1}
and \eqref{eq:blocklen} (which were not originally necessary) were added, which temporarily suppress
subsequent detections \textendash{} transforming the detectors from
simple power thresholds into a measure of time-of-arrival (otherwise,
powerful events would trigger the RFI detectors multiple times, causing
these powerful events to be blocked regardless of direction). 

The blocking technique has some limitations that can be improved in the future: random thermal noise would occasionally
trigger the RFI detection of individual inputs shortly before the arrival
of a true RFI event. Per (eq.~\ref{eq:blocklen}), this would preclude
that input triggering at the appropriate time for the true event,
allowing sufficiently weak RFI events (such that it was seen by a
marginal number of RFI detection inputs) to slip this filter. The
RFI blocker suppresses roughly 99.7\% of events which would otherwise
trigger the system, allowing the detection threshold to be reduced
by a factor of roughly 2.8, for the same candidate event rate (Figure~\ref{fig:eventRates}). Although many RFI events are not blocked by this filtering mechanism, the remaining events can be handled in subsequent software post-processing.
A more effective RFI filtering technique is described and prototyped
in Section~\ref{sec:Future-experimental-design}.

%%%%%%%%%%%%%%%%%%

\section{\label{enu:Software}Analysis}

Here we discuss the software algorithms used to detect cosmic rays.
For definitions of symbols in this section, refer to Table \ref{tab:Software-symbol-definitions.}.

\begin{table}[hbt!]
\centering
\begin{tabular}{ll}
\hline
Symbol & Meaning\\
\hline
$F_{s}=196.608\thinspace\text{MHz}$ & ADC sampling rate\\
$t_{center}$ & Estimate of event arrival time, in samples\\
$k\in\left\{ 0..511\right\} $ & Input index (note: across entire array)  \\
$p_{smooth,k}\left[n\right]$ & Filtered power for k'th input (smoothed)  \\
$N_{k}$ & Noise estimate for k'th input \\
$P_{k}$  & Event power estimate for k'th input \\
$t_{k}$ & time of arrival estimate for k'th input \\
$d_{k}$ & Distance from the $k$'th inputs antenna to event \\
$\mbox{flag}_{k}$ & Flags discarding the k'th input \\
$S=\left[S_{x},S_{y},S_{z}\right]^{\top}$ & Position of event source \\
$A_{k}=\left[A_{x,k},A_{y,k},A_{z,k}\right]^{\top}$ & Position of antenna serving $k$'th input \\
$G_k$ & Estimate of relative sample delay implied by fit \\
$R\left[n\right]$ & Estimated pulse power profile \\

\hline
\end{tabular}
\caption{Software symbol definitions. Note that $G_k$ has a new definition in this section.}
\label{tab:Software-symbol-definitions.}
\end{table}

\subsection{Filtering and time-of-arrival estimation}

\begin{figure}[hbt!]
\centering
\includegraphics[width=0.8\columnwidth]{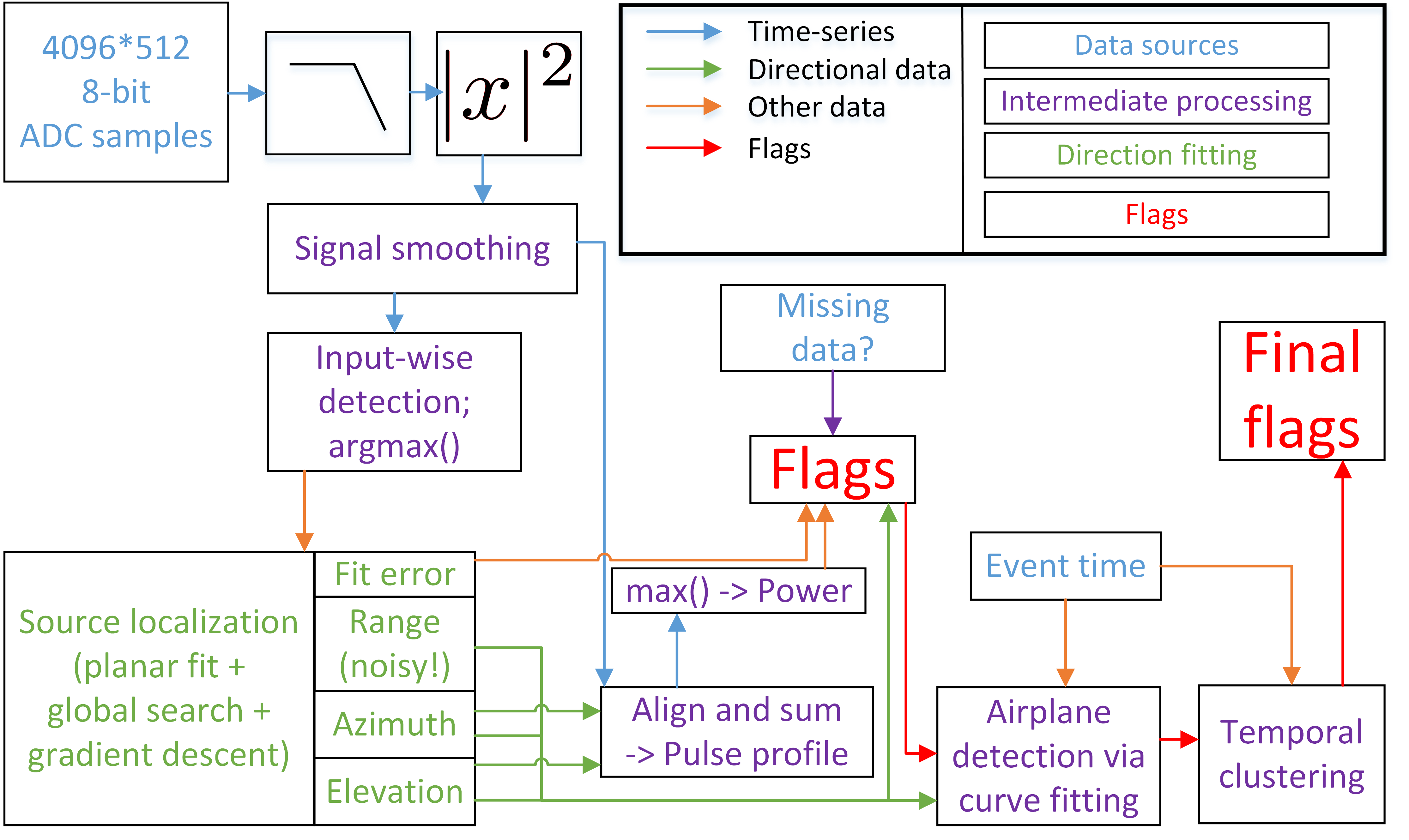}
\caption{Simplified block diagram of the
software processing pipeline.}
\label{fig:software_block_diagram-1}
\end{figure}

After receiving event data from the FPGAs, candidate events are then
processed to produce higher-level statistics. A simplified summary
is shown in Figure~\ref{fig:software_block_diagram-1}. For each record,
ADC samples have the same time-domain filter and power operations applied as in the FPGA system, reproducing $p_{smooth,k}\left[n\right]$.
The time of peak power $t_{center}$ of the event is estimated as in (eq~\ref{eq:toaEstCent}):
all subsequent analysis is constrained to 250~samples ($1.2\,\mu s$)
centered around $t_{center}$ to minimize the effects of noise and non-impulsive
events in detections. Additionally, an estimate of the background
noise in each input $N_k$ is made by computing (eq~\ref{eq:noiseEst}): the
mean of several samples of $p_{smooth,k}$ far\footnote{Because it was before the triggering time, events only rarely
fall near these samples} from $t_{center}$. The event power $P_k$ on each input is estimated with
\eqref{eq:powEst}. Any of the 512~inputs with an estimated Signal
to Noise Ratio (SNR, estimated as the ratio between input event power
and background noise power) less than 5 is flagged (eq.~\ref{eq:snrFlag})
and ignored for further processing \textendash{} any event with fewer
than 50 un-flagged inputs is rejected as too weak (or otherwise unfit) for analysis or (most rejected events are actually spurious triggers from random thermal noise; this was typically 80\% of events, but in poor RFI conditions could be as little as 25\%). Event time of arrival for a given input is
estimated as (eq.~\ref{eq:toaEstOne}). Flagging is done at this stage
because the sample-resolution peak-search function will produce a ``time-of-arrival''
regardless of the quality of the input signal: subsequent stages would
otherwise fail due to a large number of outliers contaminating the
true signal.

\subsection{\label{subsec:Direction-of-arrival}Direction of arrival fitting}

A spherical wave model is used to fit for the direction of event arrival as well as distance. With the positions of each antenna as well as the time of arrival at those
antennas, each input is treated as a measurement of the distance from
the antenna to the source. Estimates of cable length are subtracted
from the data and the resultant times are converted into units of
meters traveled at the speed of light by multiplying by $\nicefrac{c}{F_{s}}$. As a crude outlier rejection, a robust plane fit is performed on the tuples  $\left(A_{x,k},A_{y,k},d_{k}\right)$\footnote{The altitude of the antennas is ignored for this step: the array
is very nearly planar}, using the MATLAB function ``pcfitplane'' \citep{pcfitplane-desc:,pcfitplane-impl}.
Inputs which fell sufficiently
far from the robust plane fit were also flagged \textendash{} the
threshold was set such that events which were farther than the edge
of the array were not impacted. The 3D position (distance r and direction S) of the source is then
estimated as (eq~\ref{eq:optimFun}), using the planar fit as a starting
point. The search was performed in $S,\log_{10}\left(d_k\right)$ to reduce the dynamic range of the distance estimate. Sequential 1D global searches were
done in azimuth, elevation and $\log_{10}\left(d_k\right)$
(three passes), followed by a call to the MATLAB tool ``fminunc'' \citep{fminunc-desc:}. Additionally,
by aligning all un-flagged power signals using the geometric delays $G_k$
implied by the source position (eq.~\ref{eq:geometric_delay_calculation})
and summing across non-flagged inputs, a pulse profile $R[k]$ for the event
is constructed (eq.~\ref{eq:refSignal}) for visual inspection. The results of the fitting procedure are the key metrics by which candidate cosmic ray events
are automatically filtered: azimuth, elevation, range, fitting residual,
$N_{inputs-detected}$, peak pulse power.

\subsection{Software processing: summary}

Most of the process described above can be summarized as
\begin{align}
\mbox{(global time estimate)\,}t_{center} & = & \underset{n}{\arg\max}\sum_{k=0}^{511}p_{smooth,k}\left[n\right]^{2}\label{eq:toaEstCent}\\
\mbox{(input noise estimate)\,}N_{k} & = & \frac{1}{301}\sum_{n=100}^{400}p_{smooth,k}\left[n\right]\label{eq:noiseEst}\\
\mbox{(input power estimate)\,}P_{k} & = & \underset{n}{\max\,}\left(p_{smooth,k}\left[n\right]\right)\label{eq:powEst}\\
\mbox{(input TOA estimate)\,}t_{k} & = & \underset{n}{\arg\max\,}\left(p_{smooth,k}\left[n\right]\right)\label{eq:toaEstOne}\\
\mbox{(input flag by SNR)\,}\mbox{flag}_{k} & = & \nicefrac{P_{k}}{N_{k}}<5\label{eq:snrFlag}  ~\\
\mbox{(fit plane)\,}[\sim,\mbox{inliers}] & = & \mbox{pcfitplane}\left(t_{k},A_{k}\forall k|\neg\text{flag}\right)\label{eq:planeFit}\\
\mbox{(flag on fitting residual)\,}\mbox{flag}_{k} & = & \mbox{flag}_{k}|\neg\mbox{inliers}\label{planeFlag}\\
\mbox{(fit location)\,}\left\{ S,r\right\}  & = & \underset{S,r}{\arg\min}\sum_{k|\neg\text{flag}}\left((S-A)-\nicefrac{t_{k}\cdot c}{F_{s}}-r\right)^{2}\label{eq:optimFun}\\
\mbox{(estimate delays)\,}G_{k} & = & (S-A)\frac{F_{s}}{c}\label{eq:geometric_delay_calculation}\\
\mbox{(estimate pulse profile)\,}R\left[n\right] & = & \frac{\sum_{k|\neg\text{flag}}p_{k}\left[n-G_{k}\right]}{\sum_{k}\neg\mbox{flag}}\label{eq:refSignal}
\end{align}

\subsection{RFI detection and flagging}

Remembering that there are hundreds of thousands of RFI events for
every cosmic ray, the descriptive statistics from Section \ref{subsec:Direction-of-arrival}
must be used aggressively to remove non-air-shower events. The median
high-SNR event has an RMS time-of-arrival residual across inputs of
$\sim3$~samples (15~ns, about half of which is due to signal processing
losses). The distribution is bimodal, however, with a number of events
having SNR sufficient for detection, but not localization with this
algorithm (Figure~\ref{fig:fit_errors}). For this reason, events with over 12 times the median
time-of-arrival/direction fitting residual are discarded. Additionally, events which were only
seen on a small number of inputs (for Ethernet bandwidth: roughly a uniformly-distributed 22\% of data, or signal-strength: due to random thermal noise
reasons) are cut.
Of those that remain, most are clustered in a few azimuthal directions:
these are likely RFI. Additionally, most RFI events are found to come from a low elevation
angle, and some events can be localized to a short distance from the
array. For this reason, events coming from a few tightly-constrained
directions in azimuth ($97^{\circ}$ out of a total of $360^{\circ}$
in azimuth, or $\sim27$\% of all directions), or low elevations
($<2^{\circ}$) and ranges ($<$1\,km), are filtered, removing all but
one event per 10,000.

\subsection{Airplanes: a disruptive source of RFI}

\begin{figure}[hbt!]
\centering
\includegraphics[width=0.8\columnwidth]{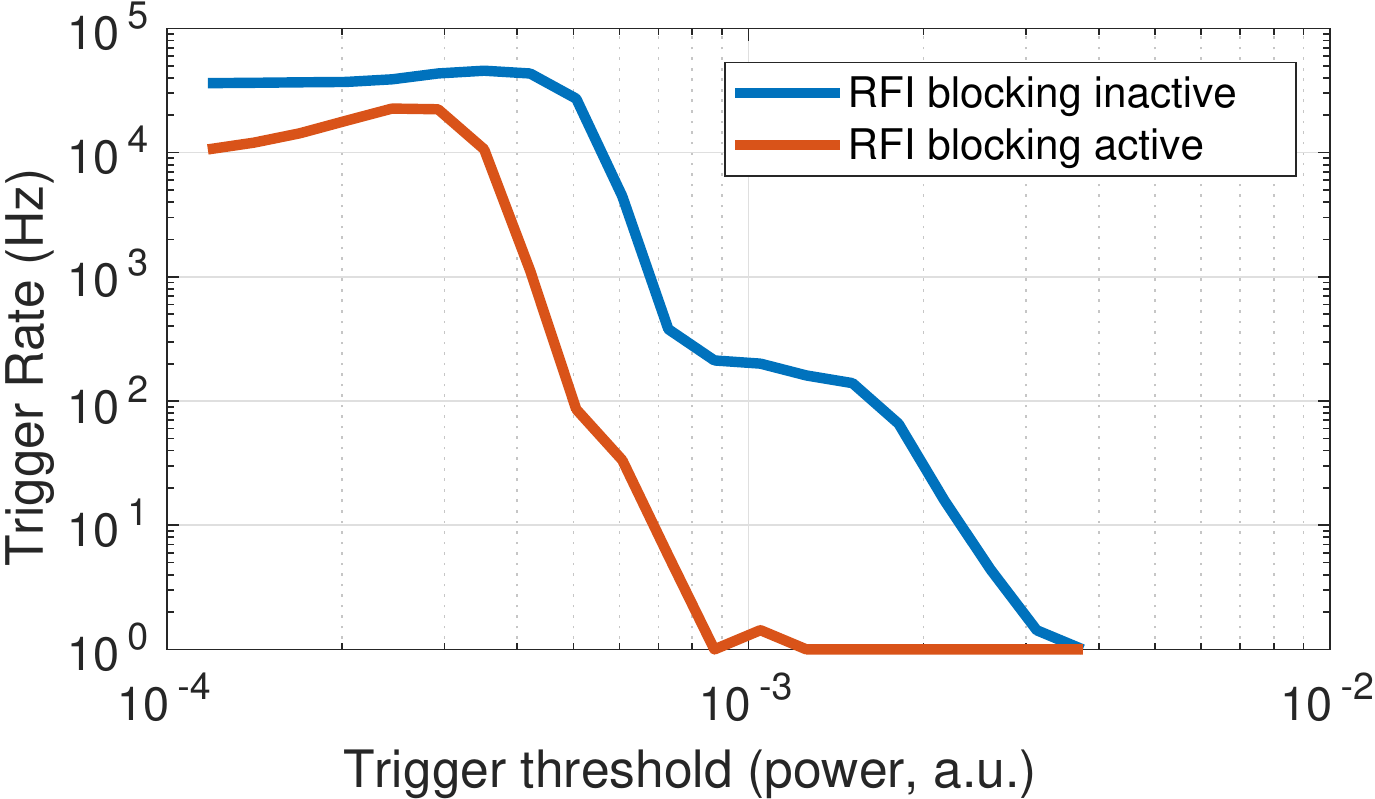}
%\caption{\textbf{Top:} event rates both with and without on-chip filtering applied. Each data point was collected at a distinct time in a time-variable RFI environment, which limits interpretability of the data. \textbf{Bottom:} approximate fraction of events which remained after on-chip RFI mitigation. Because each data point is taken at a distinct time, it is not possible to differentiate between time variable RFI and random thermal noise using the data from this plot.}
\caption{\textbf{Top:} event rates both with and without on-chip filtering applied. Each data point was collected at a distinct time in a time-variable RFI environment, which limits interpretability of the data. Because each data point is taken at a distinct time, it is not possible to differentiate between time variable RFI and random thermal noise using the data from this plot.}

\label{fig:eventRates}
\end{figure}

\begin{figure}[hbt!]
\centering
\includegraphics[width=0.8\columnwidth]{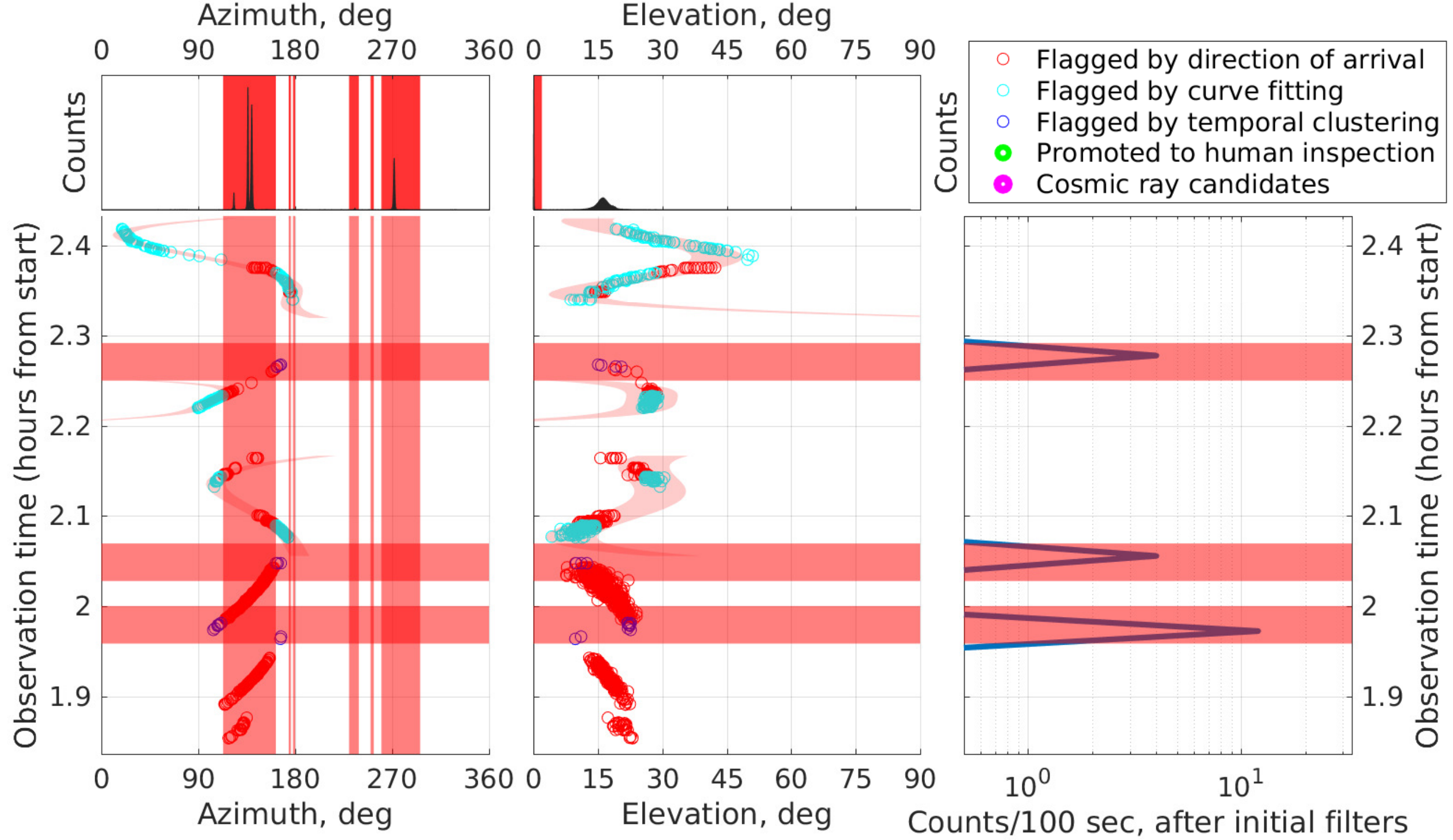}
\caption{\textbf{Upper~left, upper~center:} histogram of events
as a function of Azimuth (degrees North of East) and Elevation (degrees),
respectively. \textbf{Lower~left, lower~center:} scatter-plots of the same
against time. The many smooth traces are likely RFI originating from
(or reflected by) planes. This dataset consist of 30 minutes of observation taken during midday when airplanes
were especially prominent: they were mostly filtered using the time-binning
technique described in Section \ref{enu:Software}. The vertical red
bars are sections of azimuth and elevation which are automatically
flagged. Green circles indicate events which pass preliminary automated filters
and are promoted to final cuts and human inspection, whereas magenta circles indicate events identified as originating from air showers (neither of these classes appear in this plot, but are present later, in Figure~\ref{fig:scatterplot-ss3}). Despite the filtering, almost
all remaining events come from airplanes (flagged in the curved red regions). \textbf{Right:} number of events remaining in a given 100\,s window, after all flags (through airplane detection) are applied. Time-domain clustering of these overhead events
results in the flagging of a substantial portion of the dataset, but application of the airplane mitigation algorithm greatly reduces this fraction of data flagged. Being
taken during daytime, this is one of the worst datasets (from an RFI
standpoint) collected.}
\label{fig:scatterplot-zoom}
\end{figure}

An overwhelming majority ($>99\%$, see Figure~\ref{fig:eventRates}) of the events which remain after the aforementioned
filtering steps trace smooth curves in the \{azimuth, time\} and \{elevation,
time\} space (see Figure~\ref{fig:scatterplot-zoom}. These are suspected to be airplanes, which can also
be seen in OVRO-LWA imaging data. A precise flagging of airplanes was
achieved by performing local and robust curve fits to the \{azimuth,
time\} and \{elevation, time\} spaces. This algorithm was imperfect, especially for airplanes which produced a small number of RFI events or interacted substantially with azimuths or elevations which were already rejected by fixed direction filters. Most remaining airplane-driven events were mitigated by filtering temporally clustered impulsive events. If only temporal clustering were used for airplane mitigation, $5-37\%$ of temporal epochs would be flagged due to airplanes\textendash{}considerable observing time is retained due to specialized flagging: in this dataset, less than 1\% of data was flagged due to bursting RFI. In general,
these airplane-driven events do not interfere with the FPGA trigger
beyond increasing the output data rate slightly.

\begin{figure}[hbt!]
\centering
\includegraphics[width=0.8\columnwidth]{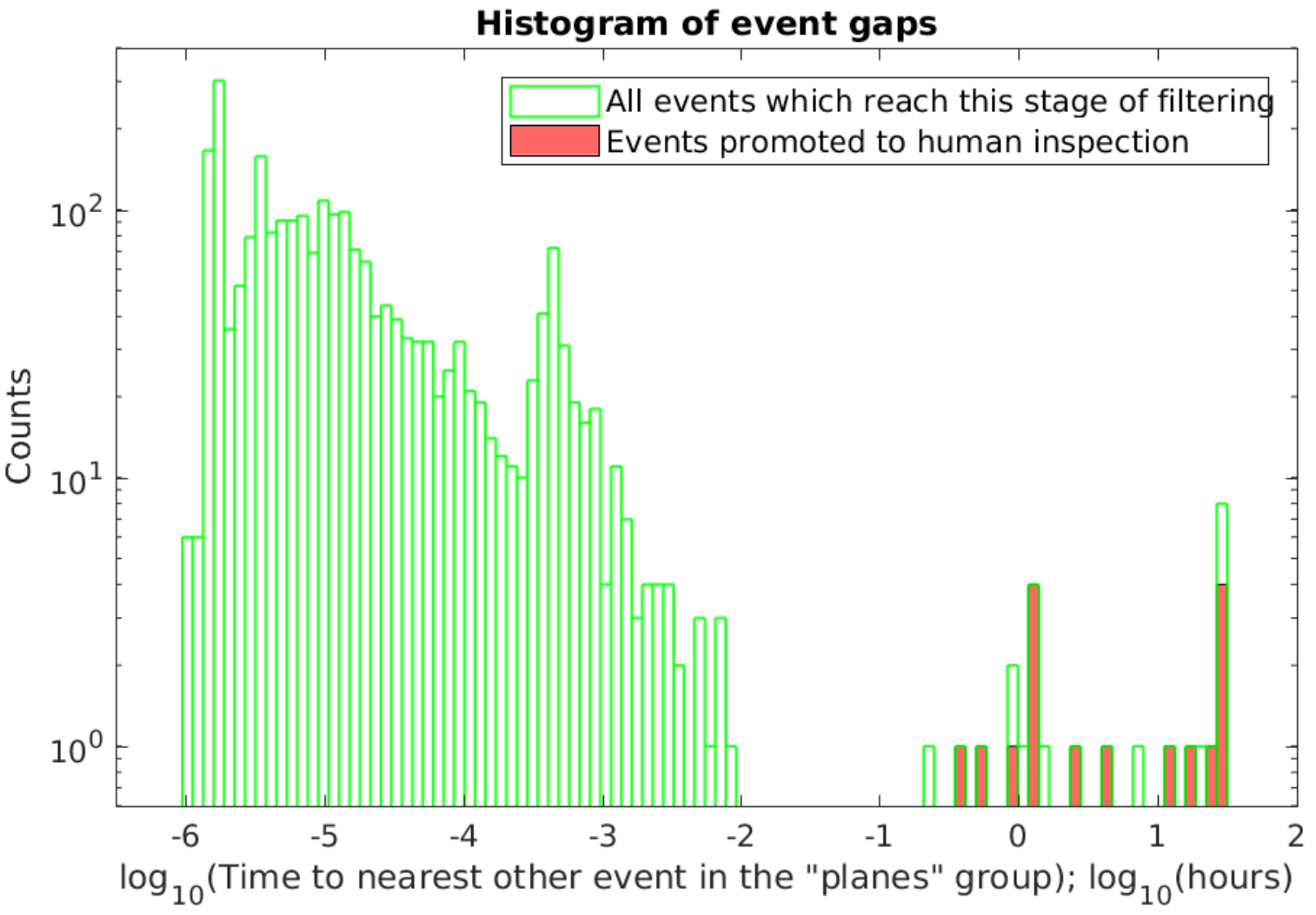}
\caption{A histogram of temporal proximity between flagged airplane events using the method in Figure~\ref{fig:scatterplot-zoom} on the entire 40-hour dataset. The cut illustrated here is performed between the Upper Right and Lower Left plots of Figure~\ref{fig:polarPlots}. The airplane mitigation algorithm does a good job separating RFI events from air shower events. This algorithm cannot be reduced to a single monotonic decision statistic, so this histogram imperfectly visualizes those cuts and true performance is better than what is illustrated here.}
\label{fig:airplaneHist}
\end{figure}

Figure~\ref{fig:scatterplot-zoom} shows the azimuth, elevation, and trigger rate as a function of time for a 30 minute window with high levels of RFI. Airplanes trace out smooth curves in azimuth and elevation vs time. These are identified and removed from the data. Time periods with high trigger rates are flagged and removed from the data. Figure~\ref{fig:airplaneHist} shows a histogram of the time to the nearest event for airplane tracks. The airplane events are clearly clustered in time showing clear separation from other classes of events including our cosmic ray candidates. 

\begin{figure}[hbt!]
\centering
\includegraphics[width=0.8\columnwidth]{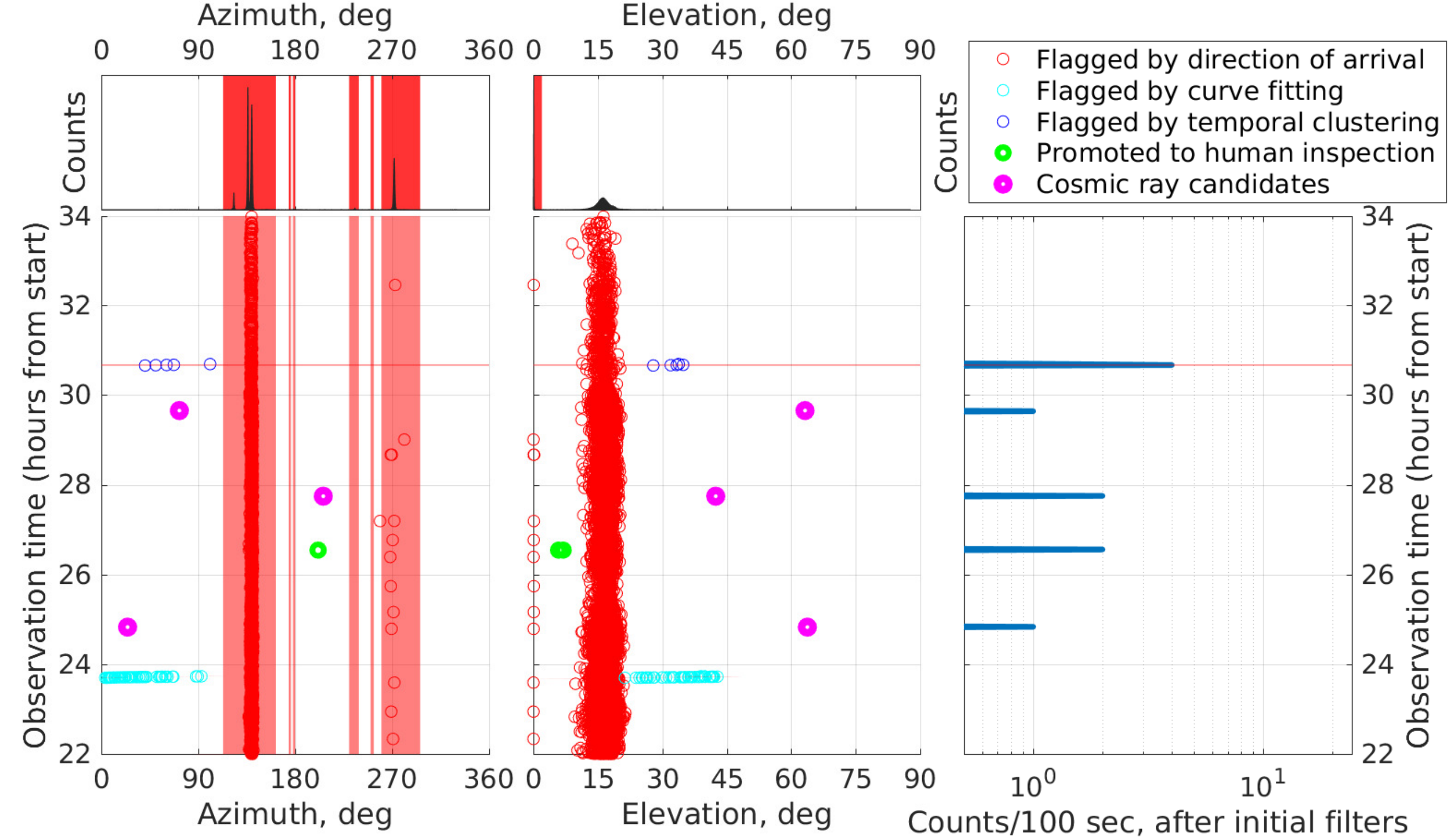}
\caption{This dataset was one of the cleaner recorded. All events in magenta are cosmic ray candidates.}
\label{fig:scatterplot-ss3}
\end{figure}

\begin{figure}[hbt!]
\centering
\includegraphics[width=0.8\columnwidth]{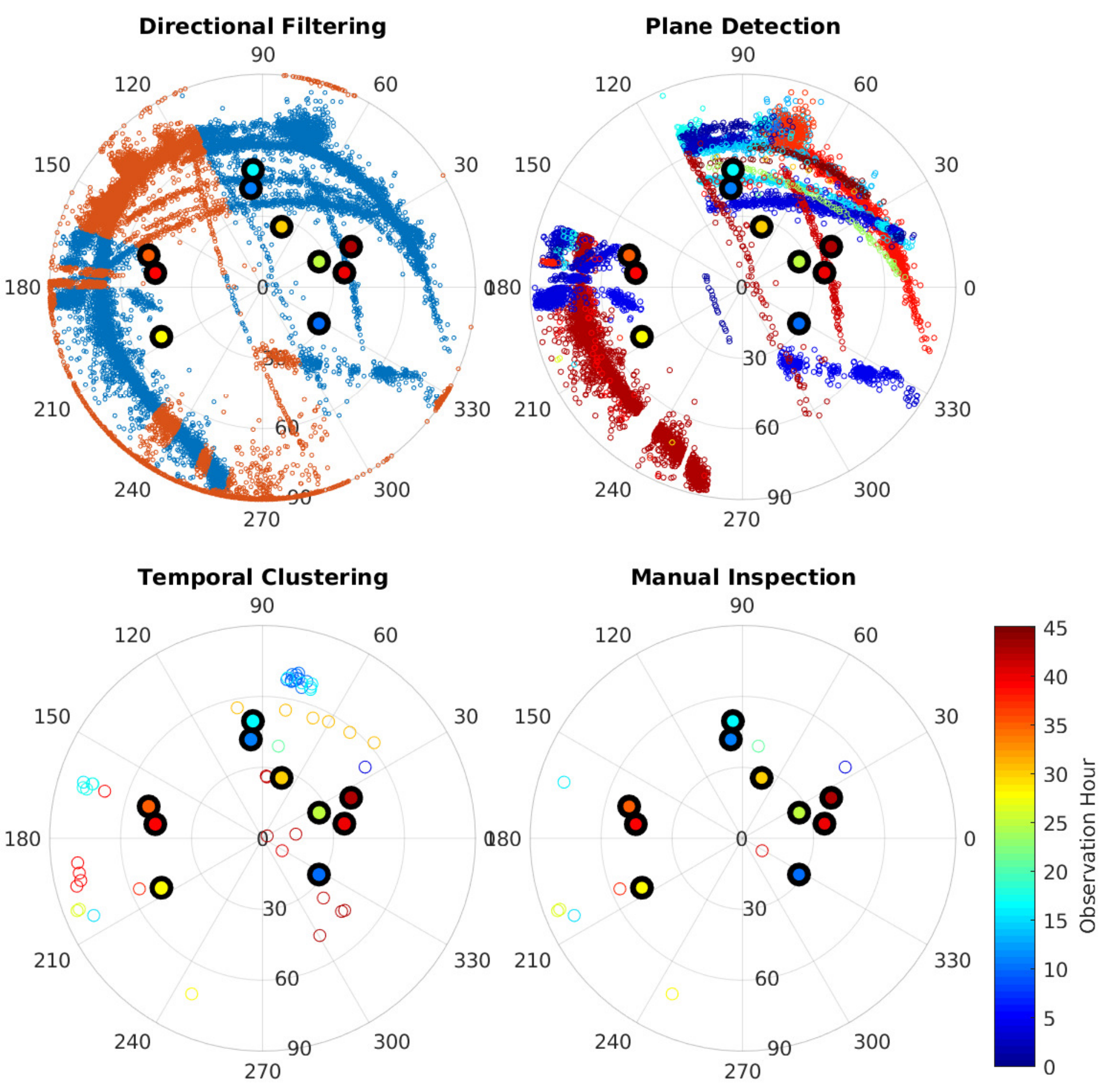}
\caption{A spatio-temporal representation of the sequence of cuts used in the event identification process.  Each point is an event, placed at the appropriate location for its azimuth and zenith angle (North is up).  Each panel of this figure represents a different stage of flagging.  Cosmic ray candidate events are given a black emphasis.  \textbf{Upper~left:} All 473K events in the dataset which pass instrumental and event power checks.  Events marked red are flagged based on azimuth and elevation cuts.  \textbf{Upper~right:} Point color indicates observation hour of event arrival.  Curves of matching color correspond to likely correlated airplane events.  \textbf{Lower~left:} Event colors same as before-- remaining temporally clustered events are flagged regardless of direction of arrival (they still appear to have time-direction structure, but were able to defeat the airplane-fitting routine).  \textbf{Lower~right:} Events which are promoted to final cuts and human inspection. In all figures, the bolded events are the 10 cosmic rays. At the location of observation, earth's local geomagnetic field is oriented $77.4^{\circ}$ N of E, and dips $61.4^{\circ}$ below the horizontal.}
\label{fig:polarPlots}

\end{figure}

The elevation, azimuth, and trigger rates as a function of time for a 12 hour portion of the 40 hour data taking period are shown in Figure~\ref{fig:scatterplot-ss3}. There are clear static sources of RFI in azimuth and elevation. Although occasional bursts of RFI appear, the data is relatively quiet. Events surviving these cuts, including our cosmic ray candidates, are clearly separated from the source of RFI. Figure~\ref{fig:polarPlots} shows a scatter plot of the events in azimuth and elevation to illustrate the sequence of cuts applied. Known sources of RFI are first filtered by their azimuthal clustering. This leaves clear airplane tracks that are identified and cut by curve fitting. The surviving events are flagged for temporal clustering with the remaining events promoted to inspection.

%%%%%%%%%%%%%%%%%%%%%%%%

\subsection{Final cuts \label{manual_inspection}}

Many un-flagged events were associated with airplanes by
visual inspection of \{azimuth, elevation, time\} scatter-plots, but
were missed by flagging algorithms. This was largely due to the flaws in the airplane detection routine's interaction with fixed directional blockers, as well as trend to promote marginal events (trusting human pattern matching to detect moving objects more reliably than a computer, in fear of accidentally discarding an air shower event)\textendash{}
these were flagged by hand. 

\begin{figure}[hbt!]
\centering

\includegraphics[width=0.8\columnwidth]{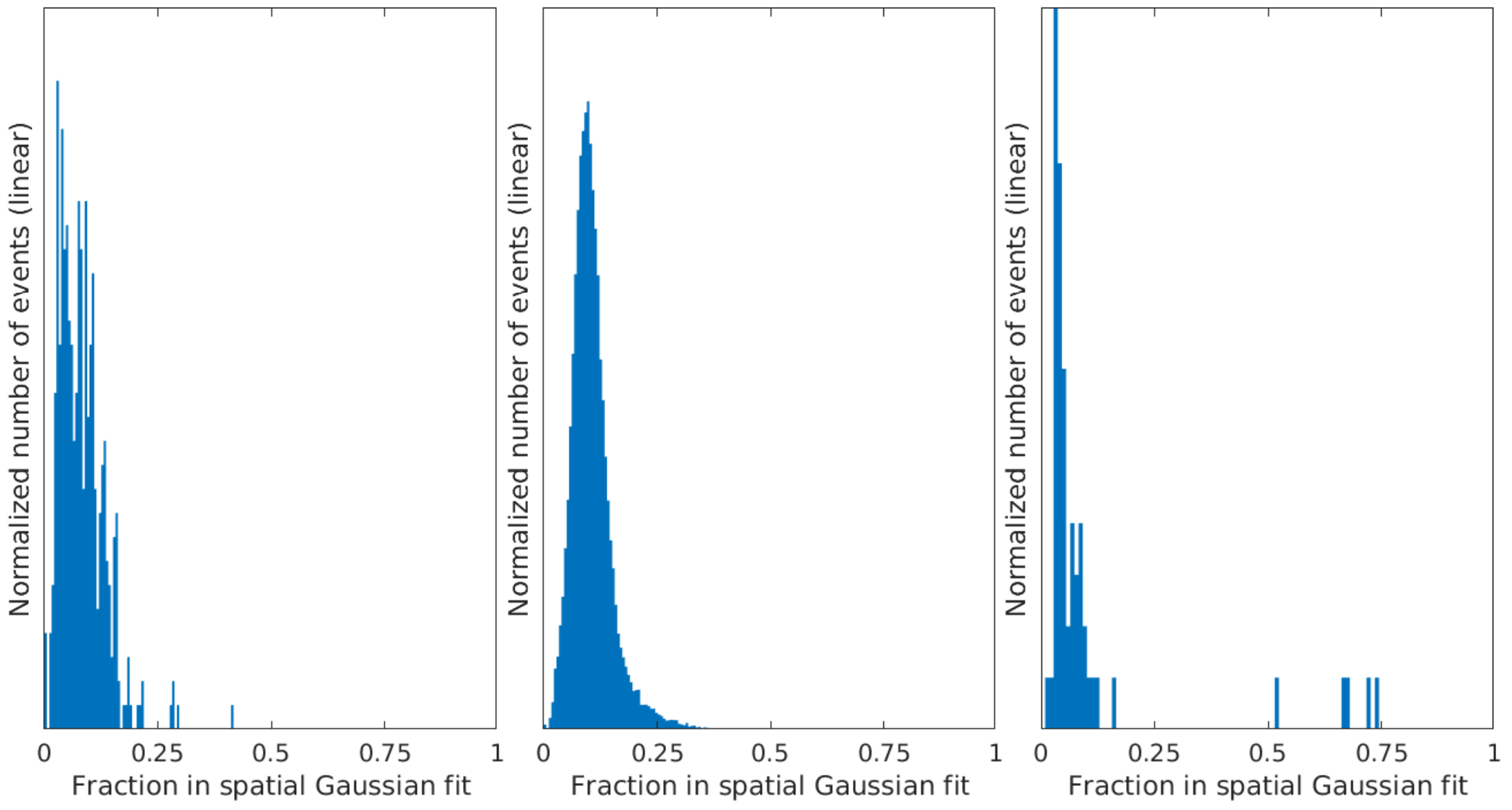}

\caption{A histogram of fractional power captured by a spatial Gaussian fit. The left two figures represent events originating from different locations within the nearby city of Bishop, CA, while the rightmost plot represents a source coming from somewhere South of the array: some from the nearby signal processing shelter, as well as (likely) other more distant sources. Unlike Figure~\ref{fig:rfiPolSpread}, this Figure only contains data from roughly the last four hours of observation for computational reasons. Most RFI produces a roughly uniform illumination on the array, making this a generally good statistic. A notable exception to this is displayed in the right figure: all events with over 50\% of received power captured in the fit are associated with near-field RFI produced by the local signal processing shelter. These events are easily cut using measures of impulsivity and source distance. Power distribution on the array appears to be a good statistic for the detection of air shower events.}
\label{fig:rfiGaussSpread}
\end{figure}

\begin{figure}[hbt!]
\centering

\includegraphics[width=0.8\columnwidth]{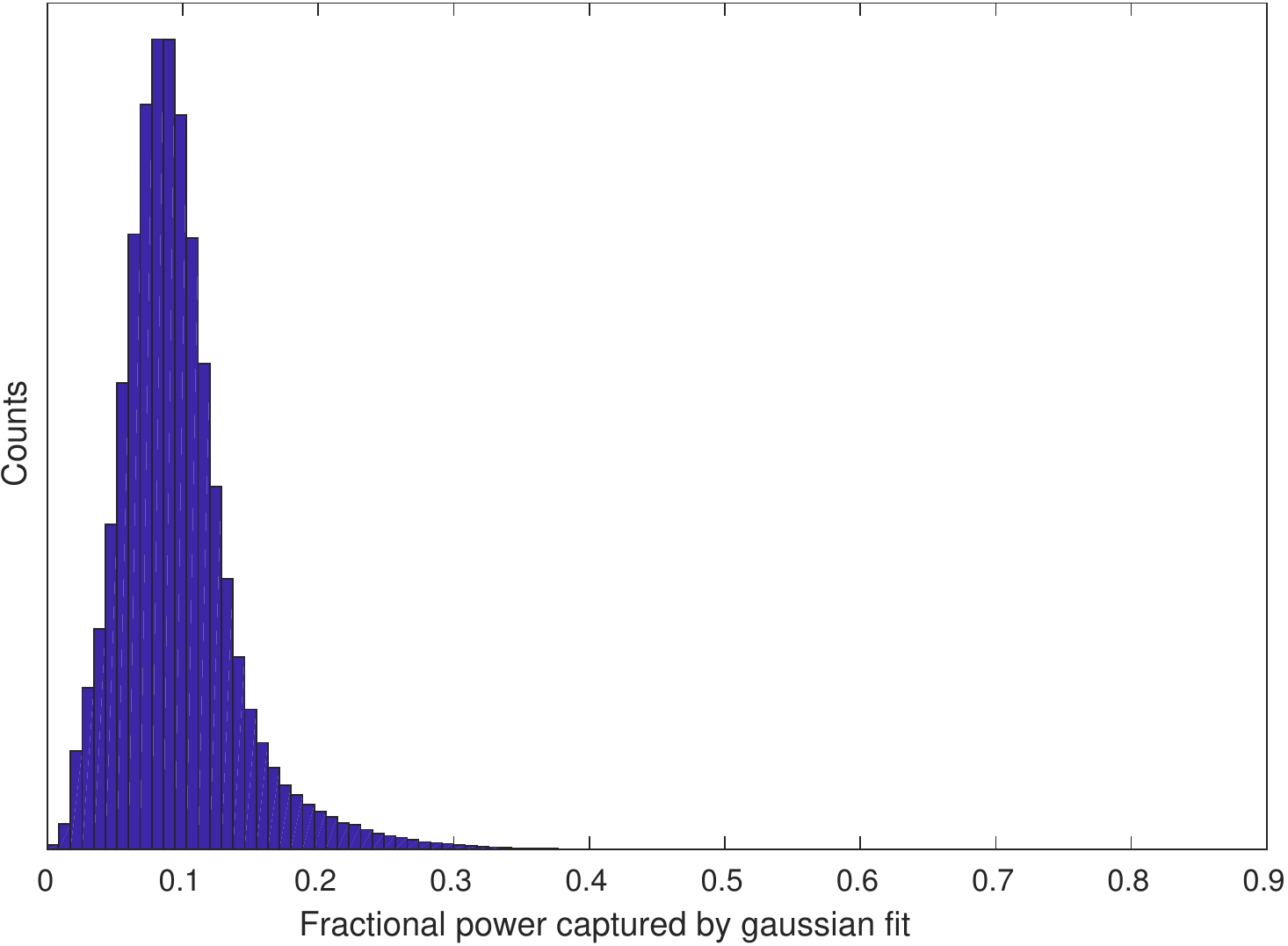}

\caption{A histogram of fractional power captured by a spatial Gaussian fit as calculated in Figure~\ref{fig:polGaussScatter}. This plot covers the roughly 300K events captured in the last four hours of the run.  Only events which passed impulsivity and systematics tests were included in the dataset, including cosmic ray candidates. Since most of those events come from the far-field, they are likely to represent events which have a uniform illumination across the array.  The vast majority of RFI events appear to represent fractional power distributions of 0.2 or less.}
\label{fig:gaussFitSpread}
\end{figure}

The remaining 16 events are classified using two statistics: polarization residual between predicted and observed, and fractional power captured by a Gaussian fit of each event -- a method motivated by recent radio measurements \citep{2013A&A...560A..98S}. The peak power seen by each antenna was taken as a measure of the total power of the event at that location on the ground. A constrained Gaussian + uniform background power model was fit to this spatial power profile $P\left(x,y\right)$, with $x$ and $y$ being the coordinates of the given antenna on the ground (this is not as efficient as going to coordinates aligned with the shower axis, but was sufficient for the purpose of RFI discrimination).  The fraction of total power which was captured by the Gaussian is taken to be a measure of how much this event matches the distribution expected for a cosmic ray. The performance of this statistic is characterized in Figure~\ref{fig:rfiGaussSpread}: it appears to perform well, with the exception of a few well-defined RFI sources which can be cut in other ways. Figure~\ref{fig:gaussFitSpread} furthermore displays a histogram of this statistic for all events captured in the last four hours of the run (including cosmic ray candidates). Very few RFI events achieve a high value for this statistic.

\begin{figure}[hbt!]
\centering

\includegraphics[width=0.8\columnwidth]{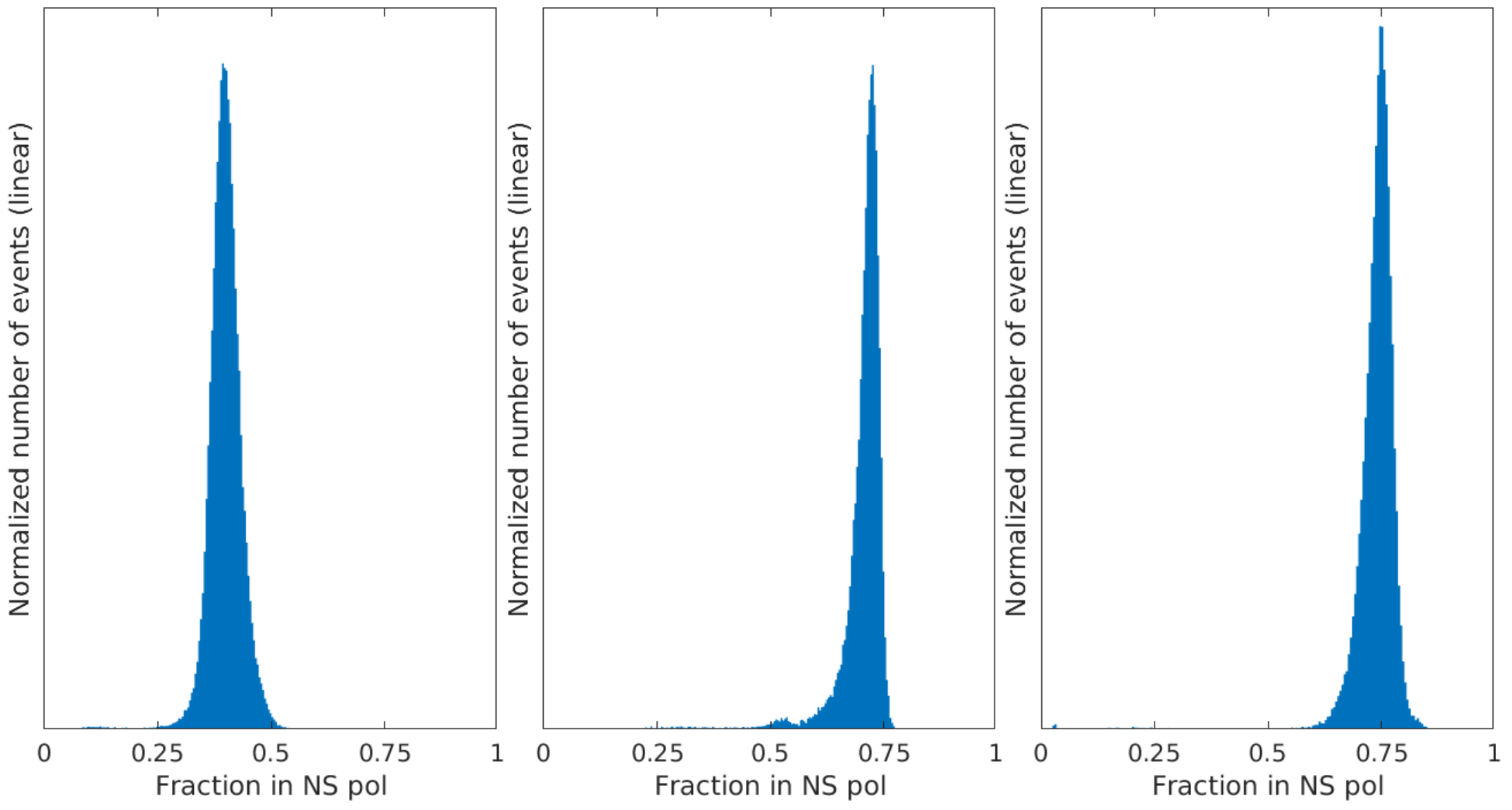}

\caption{A histogram of fractional power received in the N/S-oriented dipoles (as calculated in Figure~\ref{fig:polGaussScatter} for three distinct categories of RFI event, chosen for high event rates in those azimuth angles. The three panels present the same sources as shown in Figure~\ref{fig:rfiGaussSpread}. Alongside instrumental uncertainties, the parameter fitting of events shown here have a FWHM of 3$\sim$6\% of total power, an error contributes to the fitting residuals shown in Figure~\ref{fig:polGaussScatter}}
\label{fig:rfiPolSpread}
\end{figure}

    This was coupled with a measure of the polarization agreement with expectation from  a cosmic ray air shower.  Each OVRO-LWA antenna contains one dipole each oriented in the N-S and E-W direction.  Using the Gaussian fits described above, the ratio between the power captured in the N-S dipole as a ratio of the summed power captured in both Gaussians, is used as a measure of polarization.  This measure is compared to the expected fractional power received in the N-S dipole, modeled by assuming that all power comes from the geomagnetic $v\times{}B$ emission.  A primitive beam model of the OVRO-LWA antennas is used to predict the response in each dipole due to geomagnetic emission from the direction in question. Comparing prediction to reality allows for a measure of the similarity of the polarized signal to expectation. This statistic is limited because it does not model Askaryan emission, and is additionally subject to errors in power estimation, direction of arrival and antenna beam properties. In Figure~\ref{fig:rfiPolSpread}, we show the the polarization fraction in the NS direction measured for sources coming from three distinct azimuth angles which show exceptionally large amounts of RFI.  Each of these plots is expected to mostly contain events coming from a unique physical source: variation in the distribution of polarization measures can be used as an indicator of the quality of this measure in practice, which is to say precise to about 2-3 percent. The remainder is likely due to imperfect modeling or systematics, especially for events originating from directions near parallel to the local geomagnetic field lines, or events with low received power.

\begin{figure}[hbt!]
\centering

\includegraphics[width=0.8\columnwidth]{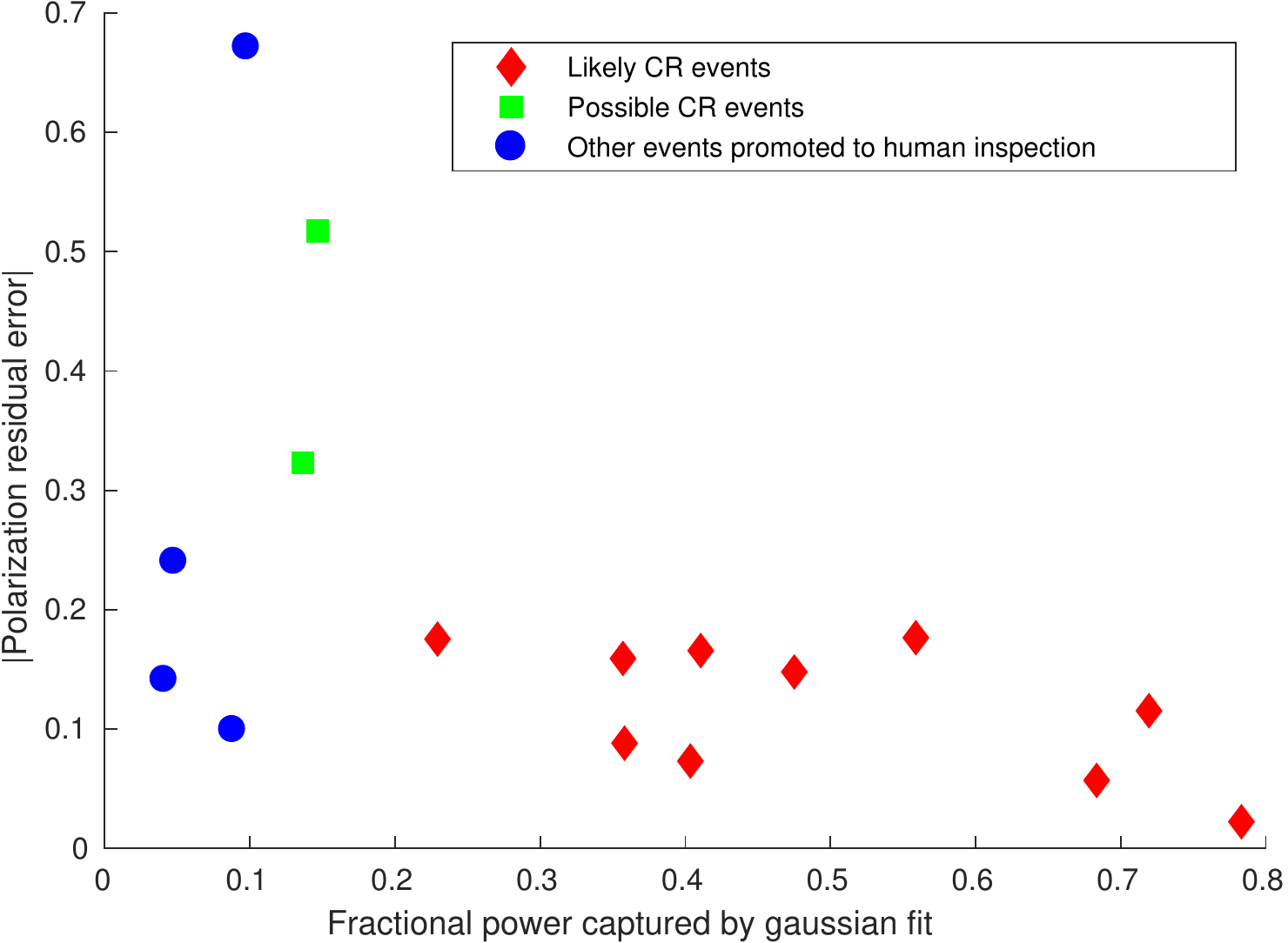}

\caption{Final cuts on the expected cosmic ray polarization and radiation footprint for events which pass all subsequent criteria, as described in Section~\ref{manual_inspection}. Automated final cuts tend to agree with human intuition.}
\label{fig:polGaussScatter}
\end{figure}

     Figure~\ref{fig:polGaussScatter} shows the synthesis of these two statistics.  Intuitively, a cosmic ray event would receive a low polarization deviation and a high fractional power captured by the Gaussian fit \textendash{} events in the bottom-right are likely to be cosmic rays. Unfortunately, both statistics are somewhat flawed: the array polarization is not perfectly known, and the spatial Gaussian is a simplification of more complicated structure, and lacks a compelling gain calibration.  In both cases, a representation of the errors which is both simply described and accurate would be challenging to produce. In order to characterize these performance of this Figure, each event is carefully inspected by hand. All 16 events are band-limited impulses and lack spatio-temporal correlation with other events. Additionally, the spatial distribution of antenna time-of-arrival is inspected to confirm that the event was in the far-field (always true). Each event was inspected for spatial structure which is commensurate with a cosmic ray: a roughly Gaussian spatial distribution is expected. Deviation from the alignment of this asymmetry is permitted if the location of strike on the array results in an inconclusive result. Finally, special attention is given to events for which the power distribution and direction of arrival are roughly compatible with a near-field source: in these two cases, the spatial antenna time-of-arrivals are again carefully inspected for signatures of a near-field source (not found in either case). The polarization signature is ignored during this manual inspection process

     Manual inspection of the 16 events presents 10 events which appear compatible with air showers and four events which are most likely caused by RFI. By human inspection, two events are inconclusive: there is a weak Gaussian signature on each event, but insufficient spatial extent to see the entire spatial structure. Each event comes from a low elevation angle, and the event from azimuth $202.3^\circ$ exhibits a Gaussian shape which is less prominent than that of events with much lower total power. The conflicting and weak evidence for both events makes discrimination challenging in this case. The use of an array with larger spatial extent such as the remainder of the LWA array, having a spatial extent of 1.53\,km, but unused in this work, or the use of hierarchical beamforming as described in \ref{sec:Future-experimental-design}, would have made classification of these events much easier. Both goals are beyond the scope of this demonstration, but will be included in future studies.
     
     As shown in  Figure~\ref{fig:polGaussScatter}, the statistics used for final cuts appear to capture human intuition very well.  Despite the limitations of said statistics, the population of air-shower events is distinct from that of RFI events, typically by substantially more than the typical uncertainty estimated on these decision statistics. All 16 events, grouped by classification per human inspection, are listed in Table~\ref{tab:Cosmic-ray-candidates,}, as well as the Appendix.

\section{\label{sec:Characterization-of-events}Characterization of events}

\begin{figure}[hbt!]
\centering

\includegraphics[width=0.8\columnwidth]{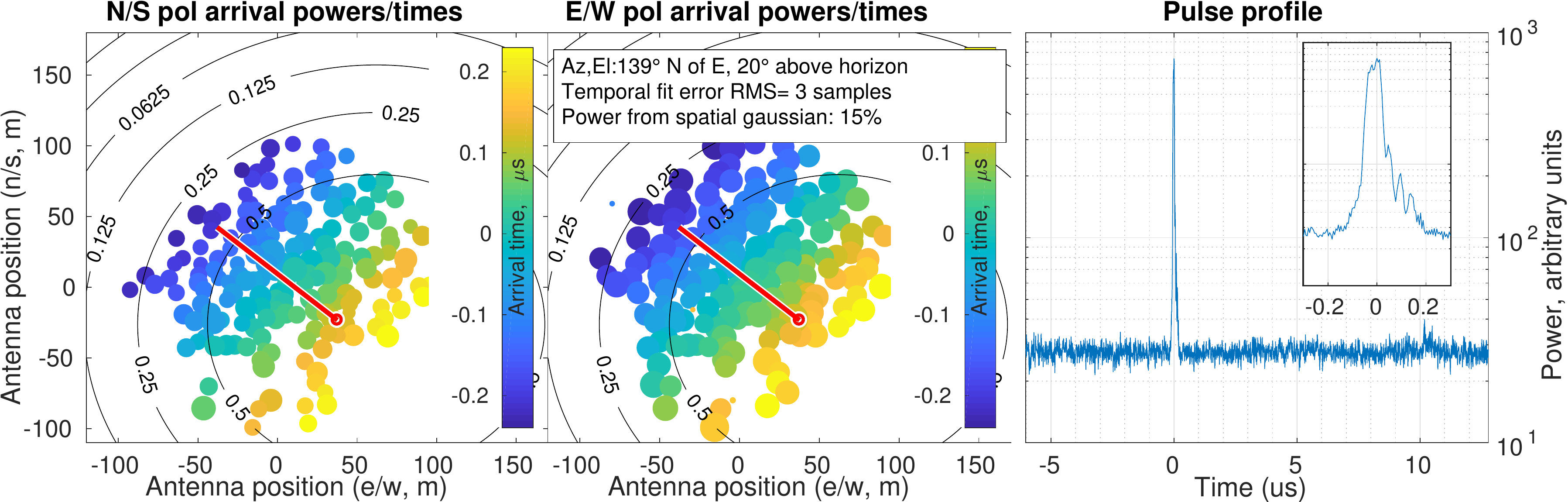}
\caption{Event originating from the nearby city of
Bishop, CA. \textbf{Left,~center:} Calibrated relative power for each antenna/polarization as size of the scatter-plot circle, time of arrival
(in microseconds) in color. One circle per antenna. The drawn arrow indicates azimuthal
direction of travel, but not necessarily location of the strike on
the array. \textbf{Right:} Estimated pulse profile constructed through a power sum across all inputs, using the geometric fit extracted from input time of arrivals (pulse frequency content varies somewhat with distance to shower axis, which may partially describe measured pulse profile shape variation).}
\label{fig:Bishop_event}
\end{figure}

\begin{figure}[hbt!]
\centering
\includegraphics[width=0.8\columnwidth]{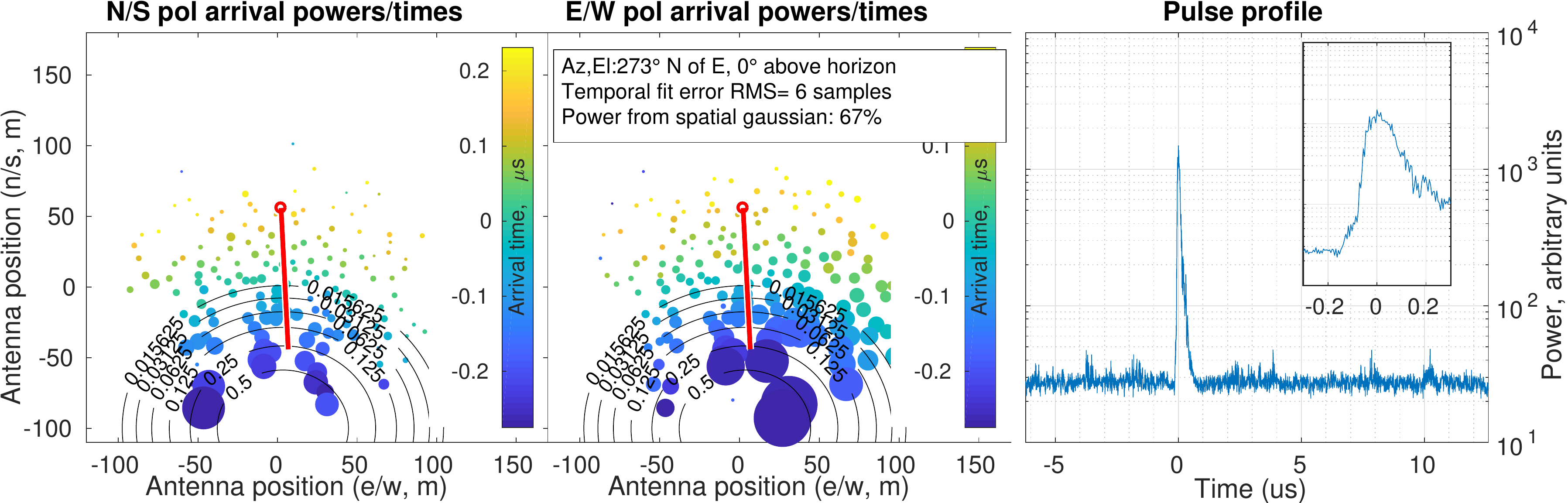}
\caption{``Shelter Event''. Wavefront curvature
is clearly visible in the left two subplots: events originating from this source appear to be the
only ones which are close enough to produce a useful range fit. Note
that the pulse profile has extended structure uncharacteristic of a band-limited
impulse. Full description as per Figure~\ref{fig:Bishop_event}.}
\label{fig:shelter_event}
\end{figure}

\begin{figure}[hbt!]
\centering

\includegraphics[width=0.8\columnwidth]{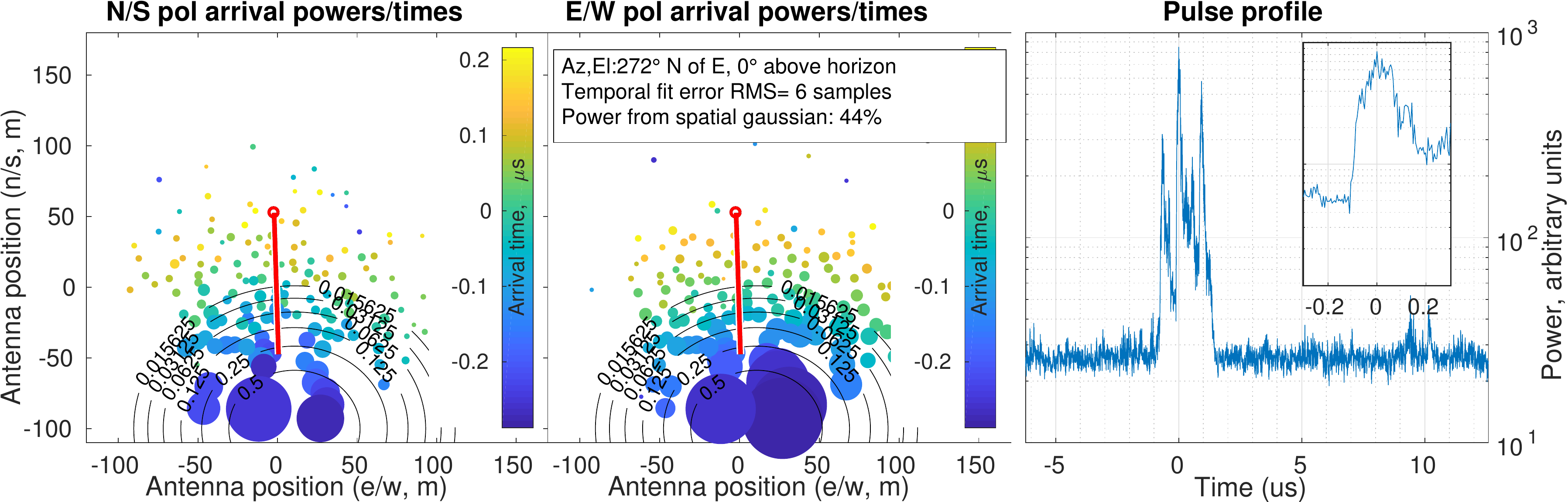}
\caption{Another event originating from A/C unit
on OVRO-LWA signal processing shelter. Note the multiple distinct
components. It is believed that this is caused by a relay toggling
as the A/C unit is activated, judging by past experience as well
as a temporal pattern of events which is synchronous with A/C activity. Full
description as per Figure~\ref{fig:Bishop_event}.}
\label{fig:shelter_event-2}
\end{figure}

\begin{figure}[hbt!]
\centering

\includegraphics[width=0.8\columnwidth]{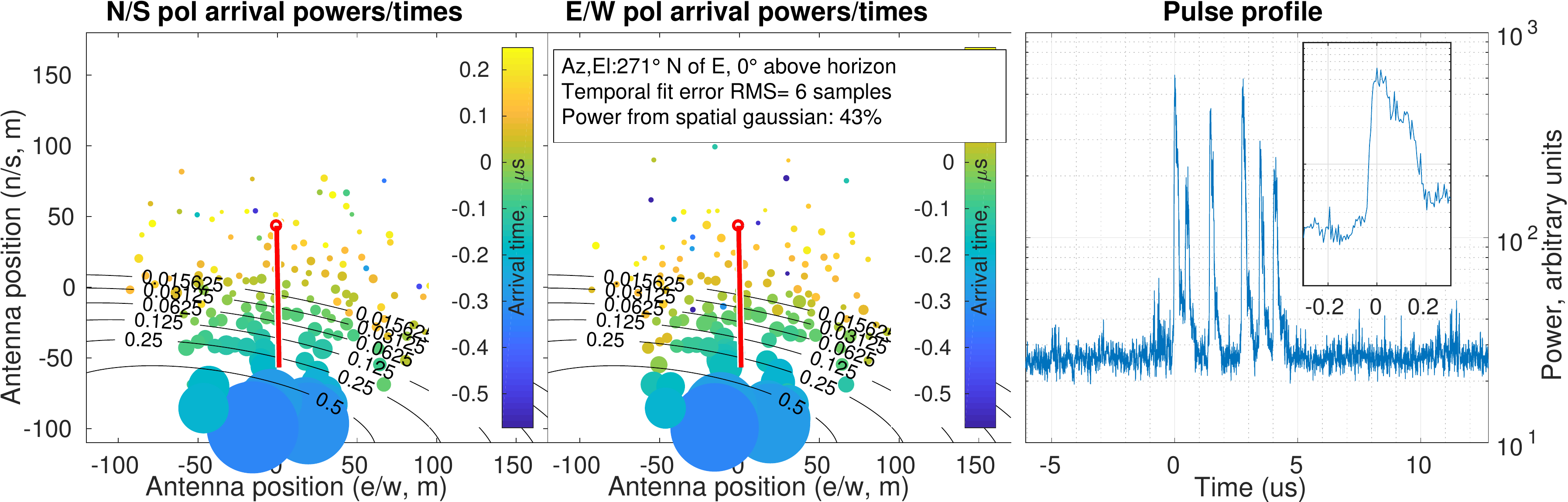}
\caption{Third event likely originating from A/C unit
on OVRO-LWA signal processing shelter. Despite the multiple peaks
in the pulse profile which confused an early stage of the algorithm,
robust plane fitting of the time-of-arrivals allowed successful location
fitting. Full description as per Figure~\ref{fig:Bishop_event}.}
\label{fig:shelter_event-3}
\end{figure}

\begin{figure}[hbt!]
\centering

\includegraphics[width=0.8\columnwidth]{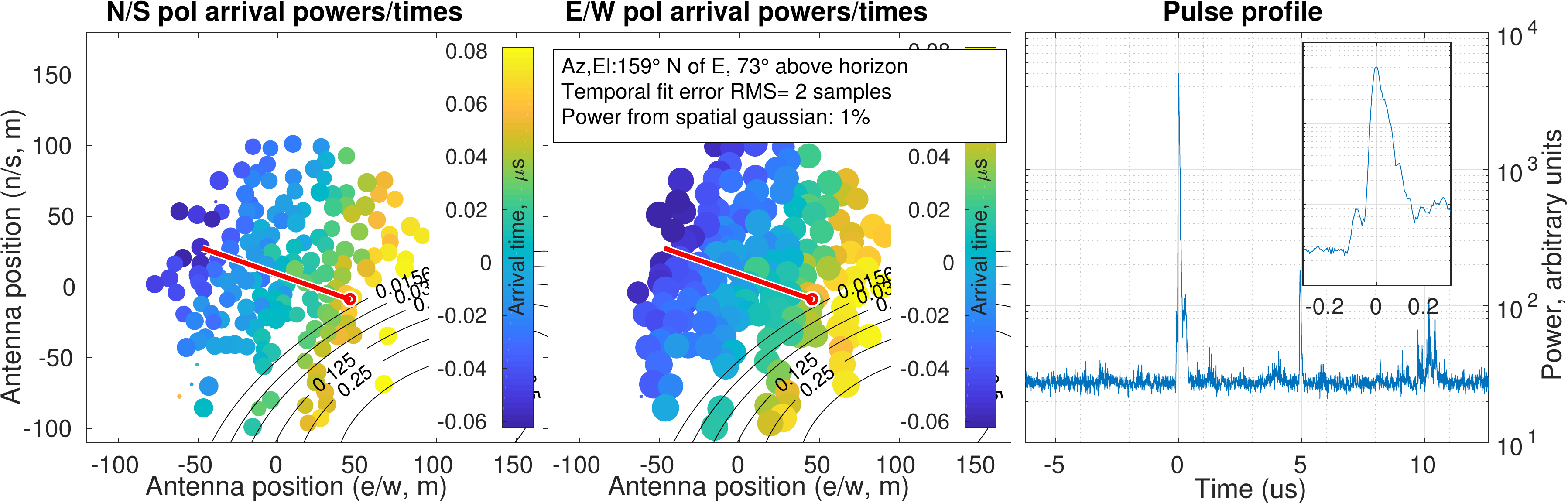}

\caption{Non-cosmic ray event originating from above
the array - presumably an airplane, considering other spatio-temporally correlated events.
Most detected events originating from airplanes had a single component. The event is unlikely
to be caused by reflections from nearby RFI sources because the line-of-sight event would have been brighter and triggered the system first. Full description
as per Figure~\ref{fig:Bishop_event}.}
\label{fig:plane_event-1}
\end{figure}

\begin{figure}[hbt!]
\centering

\includegraphics[width=0.8\columnwidth]{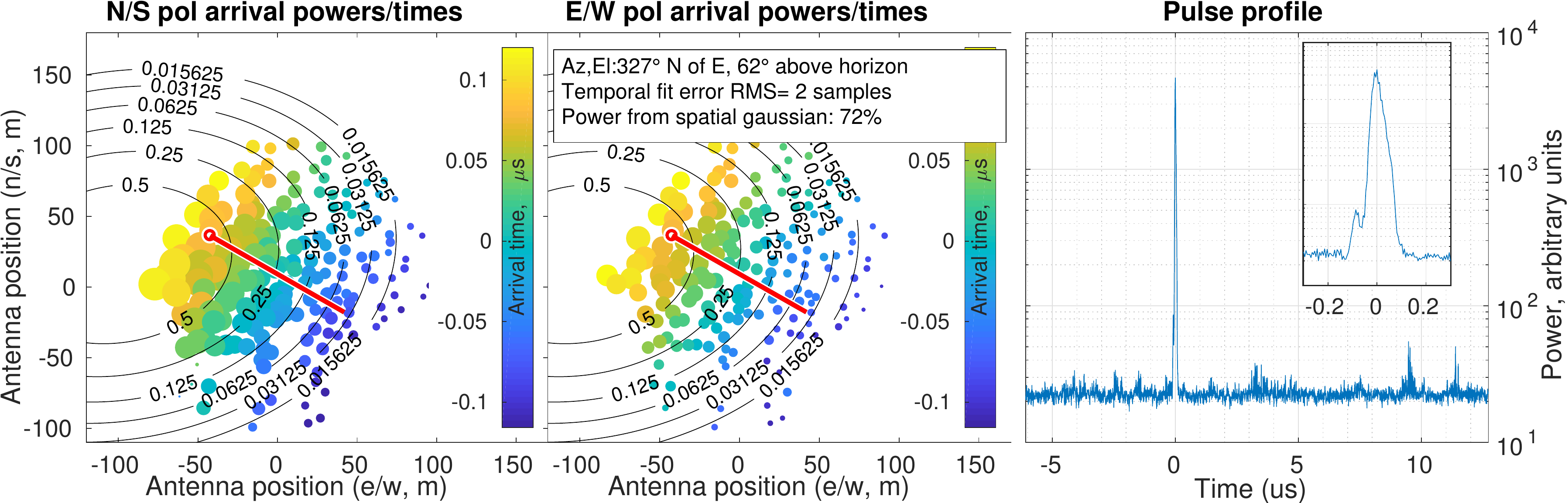}

\caption{Cosmic ray candidate event. The event has maximum power to the northeast, but arrives from the southwest\textendash{}extremely strong evidence that this comes from a beamed source (left and center panels). In addition, this event is a band-limited impulse (right panel) and is not spatio-temporally correlated with other events (source information in text inset, but other event information not displayed here).  Likewise, the polarization signature is compatible with expected models (also not shown).}
\label{fig:CR_event-1}
\end{figure}

\begin{figure}[hbt!]
\centering
\includegraphics[width=0.8\columnwidth]{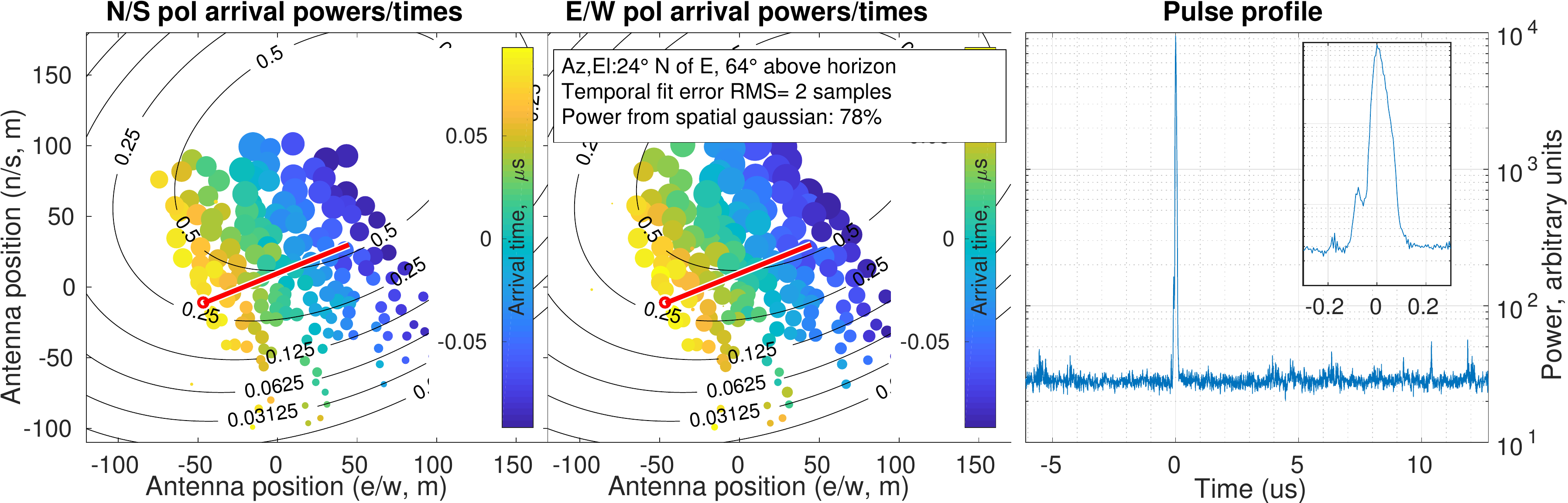}

\caption{Cosmic ray candidate event.}
\label{fig:CR_event-2}
\end{figure}

A multitude of impulsive events was detected by this system. The more
common classes of events are listed here, alongside a prototypical
event overview for at least one instance of each type
\begin{enumerate}
\item Band-limited impulsive events, coming from the nearby cities of Big
Pine and Bishop, or one of another $\sim10$ nearby sites of human
civilization. These are the overwhelming majority of events detected
by the array, and can be easily identified by spatial clustering, but are included here for completeness. These events are highly impulsive but illuminate the array uniformly; Figure~\ref{fig:Bishop_event}.
\item Events originating from A/C unit on the OVRO-LWA signal processing
shelter.  A combination of position fitting and pulse profile FWHM rejects these events from the ``final cuts" dataset, but they are included for completeness. These events can also be highly impulsive. However, coming from a nearby source, they clearly show a power distribution that is high for the closer antennas and low for the more distant ones; Figures \ref{fig:shelter_event}, \ref{fig:shelter_event-2},
\ref{fig:shelter_event-3}.
\item Overhead airplanes, many of which traveling to the nearby Eastern Sierra Regional Airport (16\,km away with line-of-sight) or Mammoth Yosemite
Airport (60\,km away). These are characterized by smooth curves in
azimuth and elevation against time. It is likely that most sites will
experience less RFI from airplanes, provided that they are farther from
a very active airport.  The airplane mitigation algorithm rejects most of these, but some manual spatio-temporal flagging is required to fully mitigate. These events can also be highly impulsive, but, unlike our cosmic ray candidates, they tend to uniformly illuminate the array; Figure~\ref{fig:plane_event-1}.
\item A family of events exhibited all of the properties of cosmic rays,
as discussed above; Figure~\ref{fig:CR_event-1}, \ref{fig:CR_event-2}.
These events are summarized in Table \ref{tab:Cosmic-ray-candidates,}.  There are ten such events for which the polarization is consistent with expectation and which have a footprint consistent with a cosmic ray air signature. Another two are more questionable and are listed as such in favor of data purity, but deserve their own category as "ambiguous events".
\item Ambiguous events:  Most events which are promoted to manual inspection are not associated with any other spatial and/or temporal cluster. However, the challenge remains to discriminate between an isolated RFI event coming from an airplane and one originating from a cosmic ray.  This must be validated using the spatial power distribution of the event as seen by the array\textendash{}cosmic ray events should have a vaguely Gaussian spatial footprint with an orientation and width compatible with the direction and parameters of the cosmic ray.  Strong cosmic ray events at high elevation angles are easily confirmed, but events coming at low elevation angles can have a very broad power distribution -- appearing uniform over the small 200m footprint of the OVRO-LWA core, whereas low power events can have a very weak signature.  In both cases, the discrimination can be challenging. The polarization measure is of limited value, given the estimated $\sim20\%$ uncertainty on true air shower events, and $\sim5\%$ uncertainty for all sources\textendash{}worse for low SNR events and especially low SNR air shower events. The low elevation angle case could be mitigated by using the 1.53\,km wide OVRO-LWA expansion array. Being a demonstration that the OVRO-LWA is capable of detecting cosmic ray events, the entire issue is sidestepped in this work by discarding all events considered to be ``questionable".  Future projects will use the full array, and additionally could constrain power distribution and polarization based on detailed air shower and detector simulations, power and direction of arrival, as well as careful analysis of airplane flight patterns and other backgrounds, allowing for more sensitive discrimination.  However, this is beyond the scope of showing the plausibility of OVRO-LWA as a standalone cosmic ray detector and will be reserved for a future project.
%%%%%%
\end{enumerate}

\begin{table}[hbt!]
\centering
\begin{tabular}{rrrrrrrr}
\hline
Azimuth & Elevation & Power & \% seen$\rightarrow$N/S  & \% expec$\rightarrow$N/S & Residual & \% $\rightarrow$Gaussian  & Class\\
\hline

$327.3^{\circ}$ & $61.7^{\circ}$ 	& 177.8 	& 37 		& 22 	& -15 	& 72 		& CR\\
$96.7^{\circ}$    & $47.9^{\circ}$ 		& 17.7 		& 100 		& 99 	& -1 	& 48 		&CR\\
$94.7^{\circ}$    & $40.3^{\circ}$ 		& 55.1 		& 100 		& 99 	& -1 	& 40 		& CR\\
$24.4^{\circ}$    & $63.8^{\circ}$ 		& 361.8 & 57 			& 57 	& 0 	& 78 		& CR\\
$206.1^{\circ}$ & $42.4^{\circ}$ 	& 30.8 		& 1 			& 4 	&  2 	& 28 		& CR\\
$72.5^{\circ}$    & $63.2^{\circ}$ 		& 20.6 		& 100 		& 93 	& -7 	& 36  		&CR\\
$164.4^{\circ}$ & $40.0^{\circ}$ 	& 10.7 		& 58 			& 38 	& -21 & 41 	& CR\\
$172.5^{\circ}$ &   $44.3^{\circ}$   & 12.8   	&   60       	& 37  	& -23 & 23 		&CR \\
$10.1^{\circ}$    &  $54.9^{\circ}$      &   15.2   &   53     		& 38 	&-15 & 36 		& CR \\
$24.6^{\circ}$    & $48.8^{\circ}$       & 64.2 	& 64		 & 43 		& -20 & 56 	& CR\\
\hline
$85.0^{\circ}$   &   $20.2^{\circ}$      &   21.3   &   67     &   100     &   17 & 14 		& NC \\
$202.3^{\circ}$   &  $33.8^{\circ}$      &   24.5   &   53     &       1    &   -60 & 15 	& NC \\
\hline
$84.0^{\circ}$    &   $20.6^{\circ}$      &   10.1   &   32     &  100 	 &    58  & 10 	& RFI \\
$204.6^{\circ}$   &   $11.5^{\circ}$      &   19.2   &   24     &    0     &   -20  	& 5 	& RFI \\
$201.3^{\circ}$   &   $5.9^{\circ}$       &   20.2   &   16     &     2       &   2  	& 4 	& RFI \\
$327.7^{\circ}$ & $80.2^{\circ}$ 		& 11.9 		& 65 	& 75 		& -12 	& 9 	& RFI\\
%$206.3^{\circ}$   &   $42.2^{\circ}$      &   31.7   &   09     &          3              &    -6  & RFI \\    
\hline
\end{tabular}
\caption{Cosmic ray candidates, grouped by classification (CR: Cosmic Ray, NC: No Call, RFI: Radio Frequency Interference), subsequently sorted in order
of arrival time. Power is measured in multiples of the galactic background
power (estimated as the pulse profile estimate far before event arrival:median
across all events in the dataset, in spite of the fact that galactic background power changes as a function of local sidereal time). Fractional power received each
polarization estimated as the amplitude of a fit spatial Gaussian in a Gaussian+DC model. A rudimentary beam model was used for the expected
power for each polarization, but that beam model has only been weakly
validated and is subject to several instrumental effects. Polarization measurements have uncertainties < 20\%.}
\label{tab:Cosmic-ray-candidates,}
\end{table}

%%%%%%%%%%%%%%%%

\section{Sensitivity Analysis}
In this section we provide an analysis of the sensitivity to cosmic rays with the OVRO-LWA and check that the observations are consistent with expectations. As an additional check, we show that the cosmic ray candidates identified in this analysis are inconsistent impulsive transients associated with airplanes, which is the main source of potential false positives.

\subsection{Number of Events}

The number of events $N$ expected from an energy dependent flux $F(E)$ for an observing run with duration $T$ is given by
\begin{equation}
N=\int dE \ F(E)   \int_0^T dt \int_A \int_\Omega \hat{v}\cdot  d\vec{A} \ d\Omega \ P_\mathrm{det}(\vec{x}_\mathrm{core}, \hat{v}, E, t)
\end{equation}
where $\vec{x}_\mathrm{core}$ is the shower core ground intersection point integrated over an area $A$, $\hat{v}$ is the cosmic ray direction sampled over a range of solid angles $\Omega=2\pi$ corresponding to the visible part of the sky, and $P_\mathrm{det}$ is the probability of detection (including trigger and analysis efficiency), which is a function of shower core position, extensive air shower direction, energy, and time of observation. 
%\bigskip
%\bigskip

\subsection{Cosmic Ray Air Shower Simulations}
We use a US Standard atmosphere~\cite{united1976u} model of the altitude-dependent air density $\rho(h)$ for the OVRO-LWA site located in Owens Valley, California at $1.2$~km altitude. Given the cosmic ray direction $\hat{v}$, the depth $X$ is given by the integral over the slant column density
\begin{equation}
X(s)=\int^\infty_s ds' \ \rho(s'\hat{v}).
\end{equation}
Given the maximum shower depth of interaction $X_\mathrm{max}$, the coordinates of shower maximum are obtained from finding $s_\mathrm{max}$ such that $X_\mathrm{max}=X(s_\mathrm{max})$ and the assumed shower core position on the ground $\vec{x}_\mathrm{core}$. For the energy-dependent $X_\mathrm{max}$ we assume proton air showers approximated by 
\begin{equation}
X_\mathrm{max}(E) = \left[55 \log_{10}(E/\mathrm{eV})-240\right] \ \mathrm{g/cm^2}
\end{equation}
based on the trends from various simulations (see, for example,~\cite{2003NuPhS.122..364H}). For these simulations we assume a fixed value at each energy which is a valid approximation since, to first order, energy is much more important in determining the trigger threshold than the uncertainties and fluctuations in $X_\mathrm{max}$.

\begin{figure}[htbp!]
  \centering
   \includegraphics[width=0.8\linewidth]{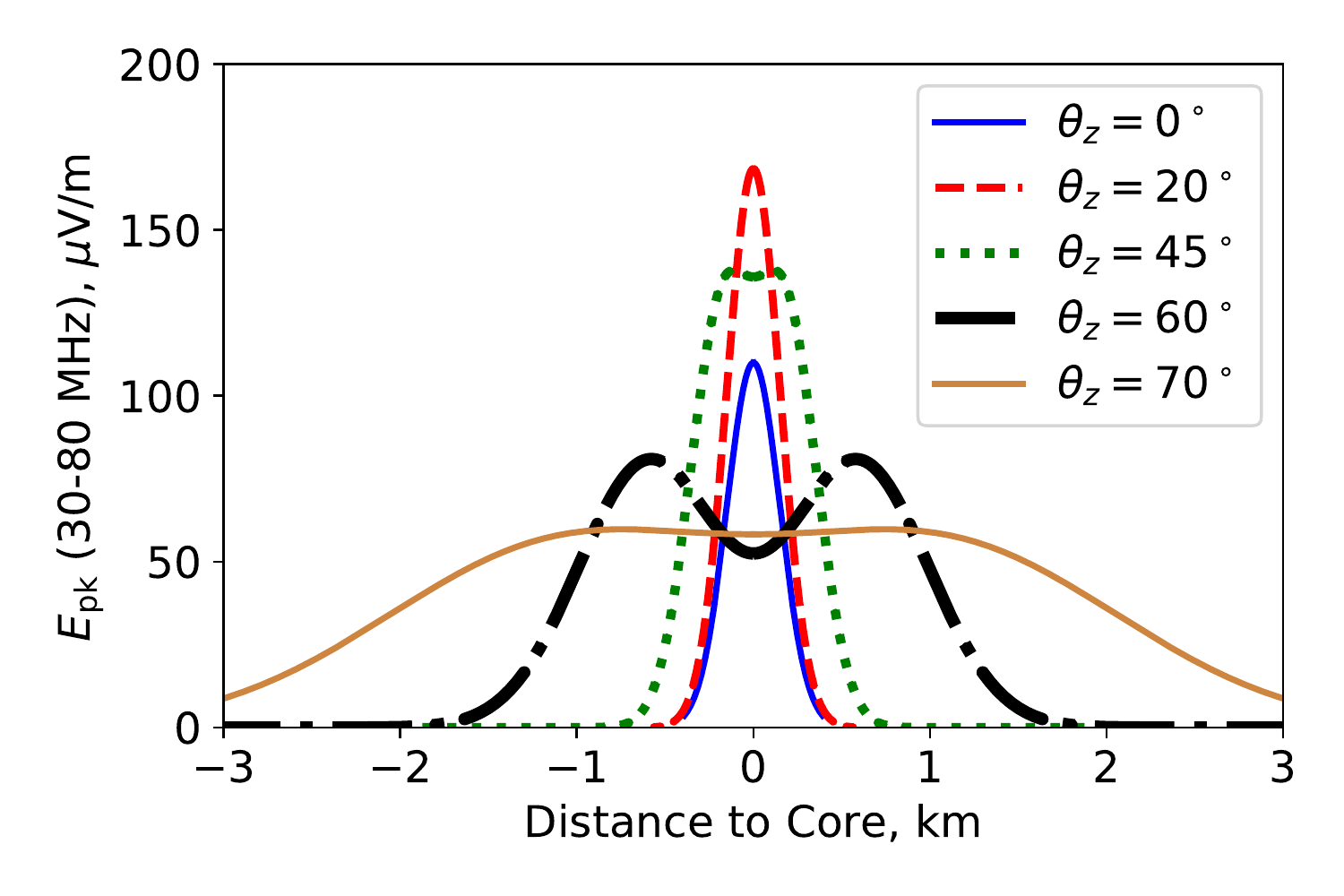} %
   \caption{Examples of cosmic ray air shower peak electric fields in the 30 - 80 MHz band as a function of distance from core. These beam patterns are for a $10^{17}$~eV proton air shower with  $X_\mathrm{max}$ corresponding to that energy and zenith angle $\theta_z$, observed at the OVRO-LWA altitude of 1.2~km.}
   \label{fig:radio_emission_profiles}
\end{figure}

For the air shower radio emission profiles, we used the results of~\cite{2014APh....59...29A}, and parameterized them as a double Gaussian shape as a function of shower inclination and view angle. The parameterization, projected onto ground distance to the core, is shown in Figure~\ref{fig:radio_emission_profiles}.

%-----------------------------------------------------------------
%-----------------------------------------------------------------

The electric field strength from the air shower is directly proportional to energy and inversely proportional to the distance between the antenna and $X_\mathrm{max}$. The polarization of the air shower is given by 
\begin{equation}
\hat{p} = \frac{\hat{v}\times \vec{B}}{|\hat{v}\times \vec{B}|},
\end{equation}
based on the Lorentz force law. The geomagnetic field at the Owens Valley site is $\vec{B}=(22.506, \ 5.012, \ 42.7) \ \mathrm{\mu T}$ in a (+N$|$-S), (+E$|$-W), (+D$|$-U) coordinate system obtained from the NOAA online geomagnetic field calculator\footnote{\url{https://www.ngdc.noaa.gov/geomag/calculators/magcalc.shtml#igrfwmm}}, based on the International Geomagnetic Reference Field~\cite{2015EP&S...67...79T}. In these simulations, the azimuthal directions are sampled uniformly with the polarization taken into account as discussed above. 

\subsection{Detector Model}
\begin{figure}[t!]
  \centering
   \includegraphics[width=0.8\linewidth]{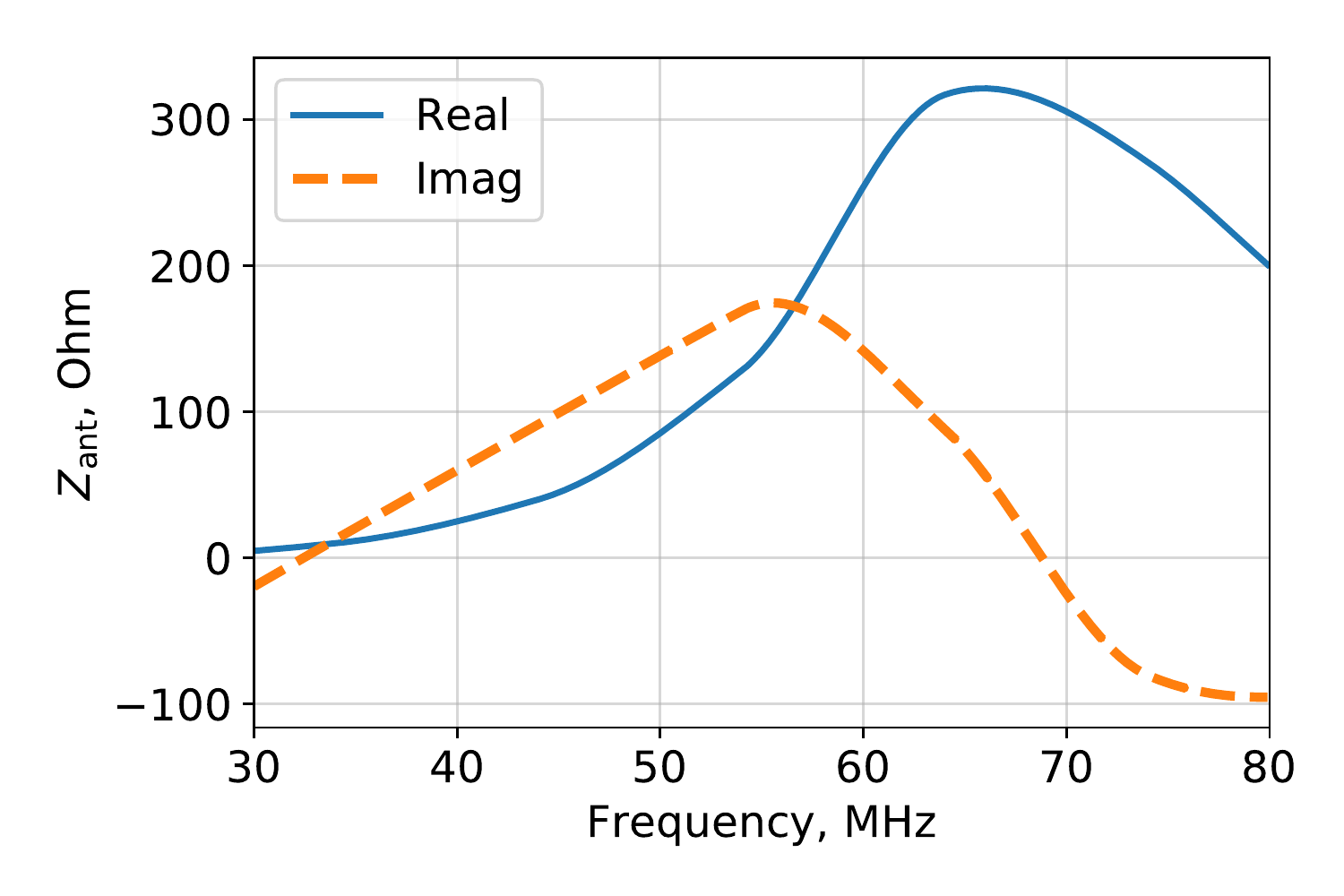} %
   \caption{Real and imaginary components of the LWA antenna impedance.}
   \label{fig:LWA_impedance}
\end{figure}

\begin{figure}[t!]
  \centering
   \includegraphics[width=0.8\linewidth]{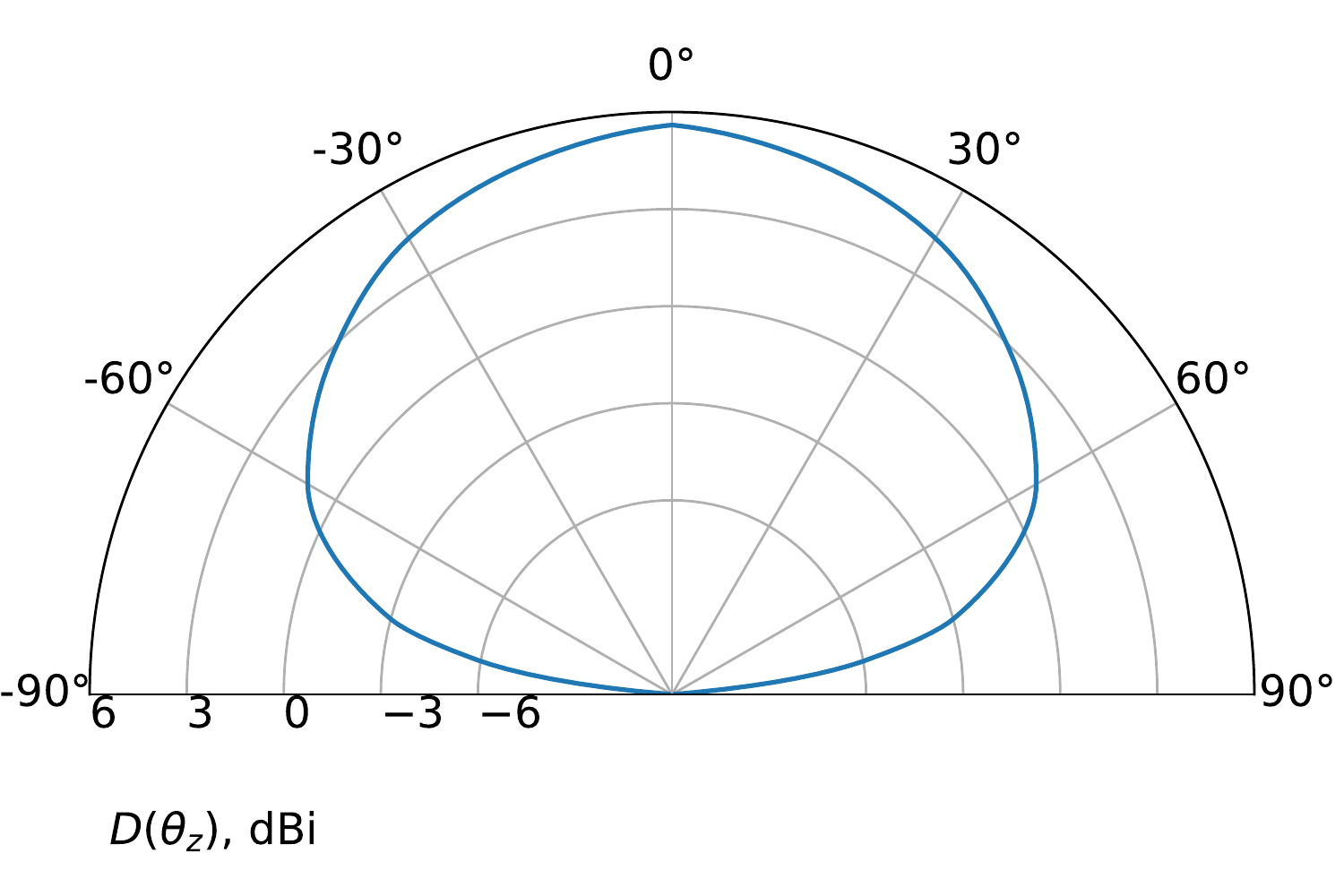} %
   \caption{LWA antenna beam pattern as a function of zenith angle $\theta_z$.}
   \label{fig:antenna_beam_pattern}
\end{figure}

With the air shower signal generation model above, the electric field of each antenna in the array is estimated based on it position, the shower core location $\vec{x}_\mathrm{core}$ and the shower direction $\hat{v}$. We convert the peak electric field to peak voltage at the antenna front-end, according to $V_{pk}=E_{pk}\langle h_\mathrm{eff} \rangle$, where $\langle h_\mathrm{eff} \rangle \simeq 0.47 \ \mathrm{m}$ is the average effective height of the LWA receiver in the 30-80 MHz band. As a simplifying approximation, we do not take into account the frequency-dependence of the electric field pulse spectrum. In this 30-80 MHz band, the electric field pulse spectrum is flat for many geometries of interest, although it can vary depending on the distance to the core position and zenith angle~\cite{2012APh....35..325A, 2012PhRvD..86l3007A}. The wavelength-dependent effective height is given by 
\begin{equation}
h^2_\mathrm{eff}(\lambda, \theta_z) = \frac{4R_A}{Z_0}\frac{\lambda^2}{4\pi}\frac{|Z_\mathrm{in}|^2}{|Z_A+Z_\mathrm{in}|^2} \mathcal{F}(\lambda) D(\theta_z).
\end{equation}
The LWA antenna impedance $Z_A$ (with real component $R_A$) is shown in Figure~\ref{fig:LWA_impedance}. The input impedance for the front=end electronics is $Z_\mathrm{in} = 100 \ \Omega$. The impedance of free space is $Z_0$ and $\lambda$ is the wavelength. Additional frequency dependent behaviors in the system, including filters for rejecting low frequency radiation, are captured in $\mathcal{F}(\lambda)$, which is obtained from a fit to the measured antenna noise power spectrum. The zenith-angle dependent antenna directivity $D(\theta_z)$, estimated with a NEC2 simulation at the central frequency of 55~MHz, is shown in Figure~\ref{fig:antenna_beam_pattern}. 

The noise voltage spectrum at the front end is given by 
\begin{equation}
|V|_\mathrm{noise}^2 = k_B T_\mathrm{sky} R_A \frac{|Z_\mathrm{in}|^2}{|Z_A+Z_\mathrm{in}|^2} \mathcal{F}(\lambda) + k_B T_\mathrm{sys} Z_\mathrm{in},
\end{equation}
where $k_B$ is the Boltzmann constant, the galactic noise $T_\mathrm{sky}$ is given by~\cite{1979MNRAS.189..465C}, and $T_\mathrm{sys}=250 \ \mathrm{K}$. The root-mean-square noise is $V_\mathrm{rms}=\sqrt{\int df|V|_\mathrm{noise}^2}$ over the frequency $f$ interval 30 - 80 MHz. The estimated value is $V_\mathrm{rms}\simeq 11.5 \ \mathrm{\mu Volts}$. 

%-----------------------------------------------------------------
%-----------------------------------------------------------------

%-----------------------------------------------------------------
%-----------------------------------------------------------------

\subsection{Trigger Model}
The trigger described in Section~\ref{sec:Firmware} used a running average of the receiver voltage power over 4 points (20 ns). We approximate the power signal-to-noise ratio of that procedure as $\mathrm{SNR}_p \simeq (V^2_\mathrm{pk}+4V^2_\mathrm{rms})/(4V^2_\mathrm{rms})$ to account for the fact that the peak voltage is typically in one sample while the rest are noise. We can just as easily simulate the trigger in terms of the amplitude signal-to-noise ratio $\mathrm{SNR}_a = V_\mathrm{pk}/V_\mathrm{rms}$, with the equivalent $\mathrm{SNR}_p = 1+\mathrm{SNR}_a^2/4$. The trigger is approximated by requiring that $\geq12$ antennas for the EW polarized channels or $\geq12$ antennas in NS polarized channel exceed a threshold SNR. To estimate the threshold SNR, relevant to cosmic-ray signals, we inspect the set of 16 events that were promoted to manual inspection. For each simulated event we identify the 12$^\mathrm{th}$ from the maximum SNR in the strongest polarization as a proxy. We find that the weakest event of the set has a threshold $\mathrm{SNR}_a=5.7$.  
 
\subsection{Expected Event Rates and Acceptance}
In addition to the trigger, the analysis imposes additional quality cuts (QC) to discriminate against RFI that affects the cosmic ray air shower detection efficiency. The QC include directional filtering, plane detection, and temporal clustering cuts and has an efficiency of $\sim 75\%$. This efficiency is dominated by directional filtering since the airplanes are transient and cover a relatively small amount of solid angle in the sky. At the manual inspection stage, the discrimination between cosmic ray air showers and RFI a Guassian power cut (GC) is applied that requires that the fraction of power in the two-dimensional Gaussian fitted to the distribution of antenna signals is $>20\%$. The GC discriminates between events that have the expected Gaussian shape (see Figure~\ref{fig:radio_emission_profiles}) from plane-wave RFI. Based on the simulations described in the previous section, this cut has an estimated efficiency of $\sim 60\%$. The total analysis efficiency is estimated to be $\sim 45\%$.

%-----------------------------------------------------------------
\begin{figure}[t!]
  \centering
   \includegraphics[width=0.8\linewidth]{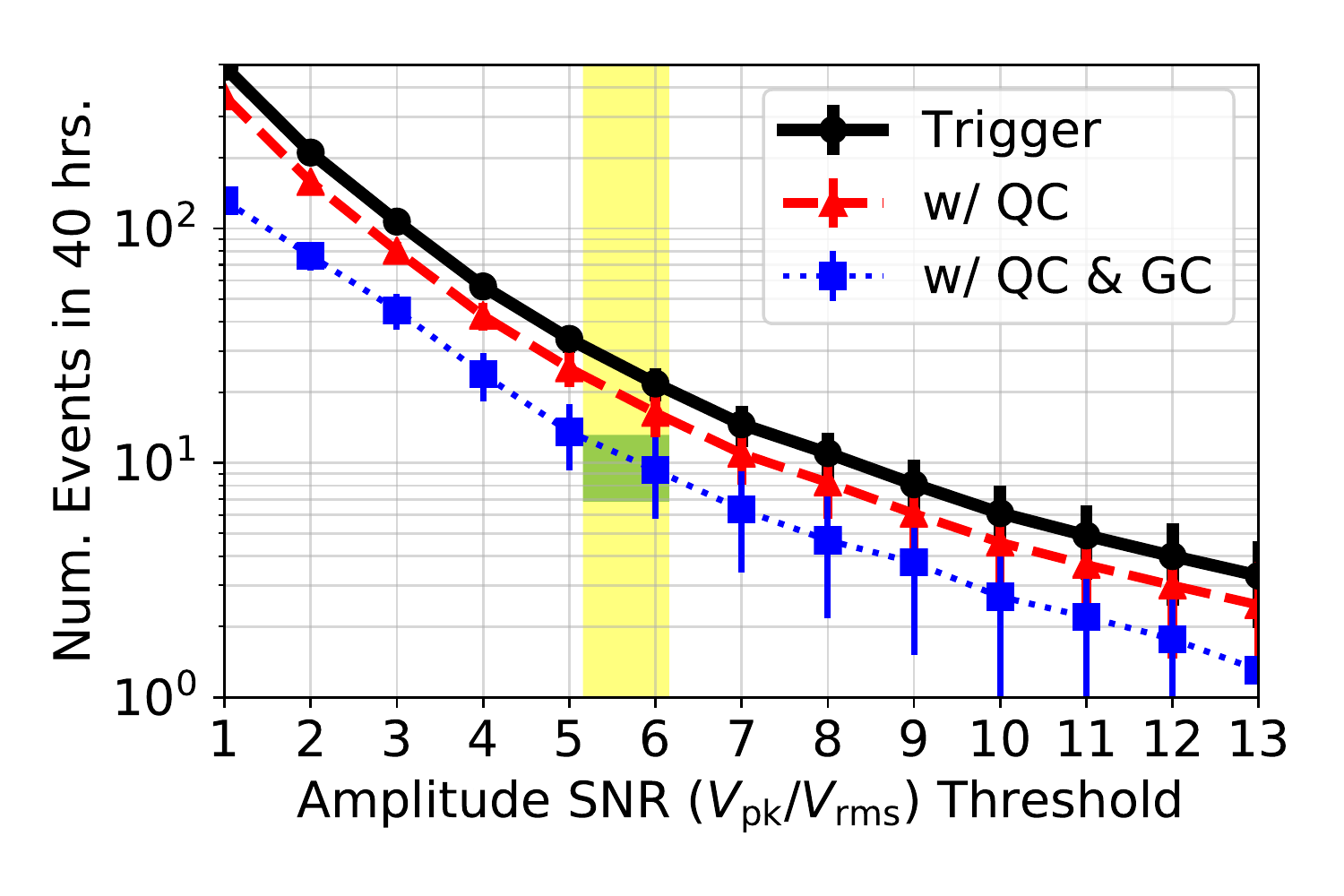} %
   \caption{Number of expected events in a 40 hour run as a function of peak amplitude trigger threshold and analysis cuts. The black line shows the number expected to trigger the array. The uncertainties shown are statistical only. The red line shows the number of triggered events that pass the quality cuts (labelled QC) and the red line includes both the QC and power in Gaussian cut (labelled GC). See text for details on the cuts. The light yellow region is the range of trigger thresholds based on the amplitude SNR threshold of 5.7 with an uncertainty corresponding to the noise voltage fluctuations $V_\mathrm{rms}$. The green region indicates the 10 events observed with a Poisson uncertainty range.}
   \label{fig:num_events}
\end{figure}
%-----------------------------------------------------------------

\begin{figure}[t!]
  \centering
   \includegraphics[width=0.8\linewidth]{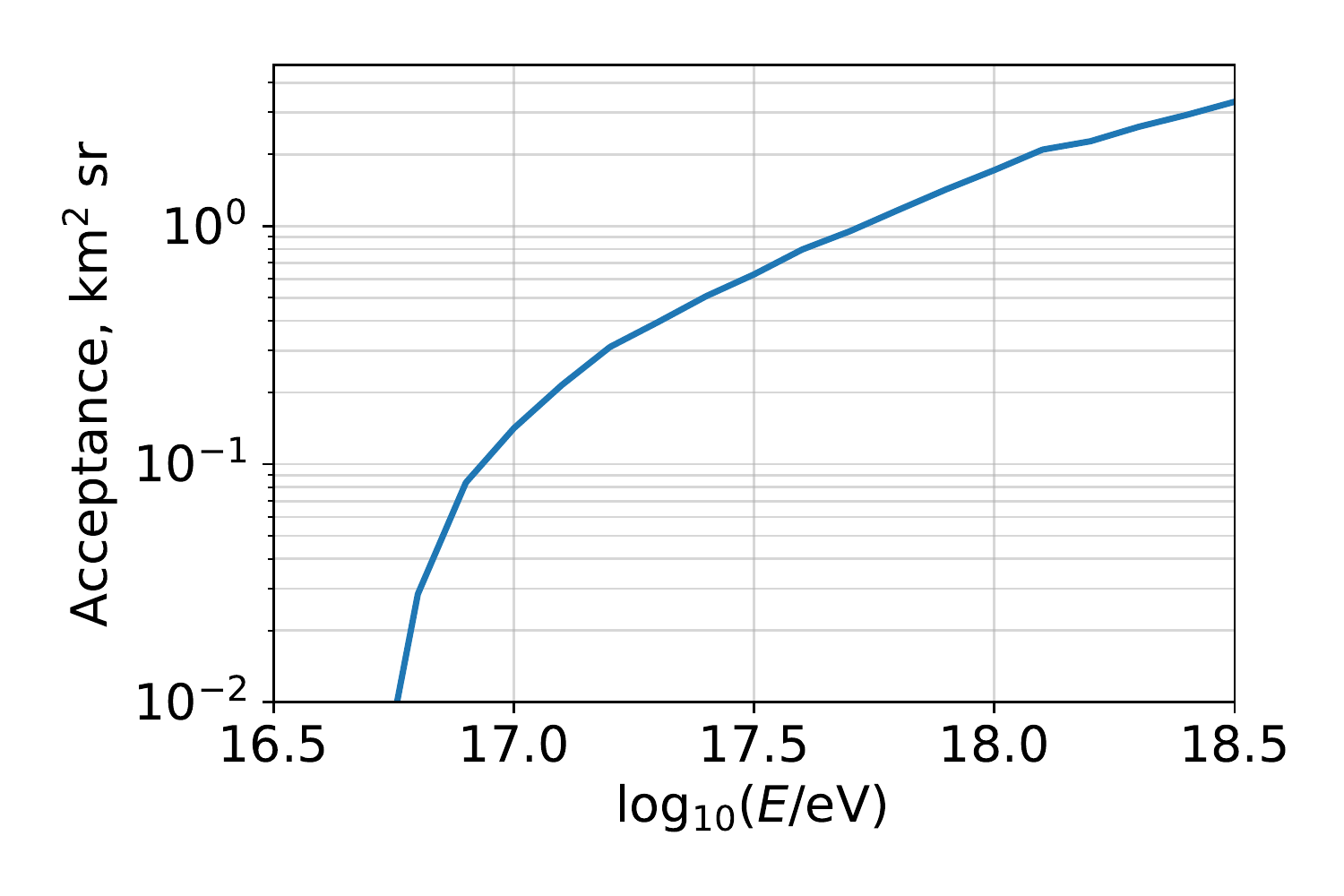} %
   \caption{Acceptance as a function of energy for cosmic ray events that trigger the OVRO-LWA detector ($\mathrm{SNR_a}>5.7$) including the analysis efficiency.}
   \label{fig:acceptance}
\end{figure}

\begin{figure}[t!]
  \centering
   \includegraphics[width=0.8\linewidth]{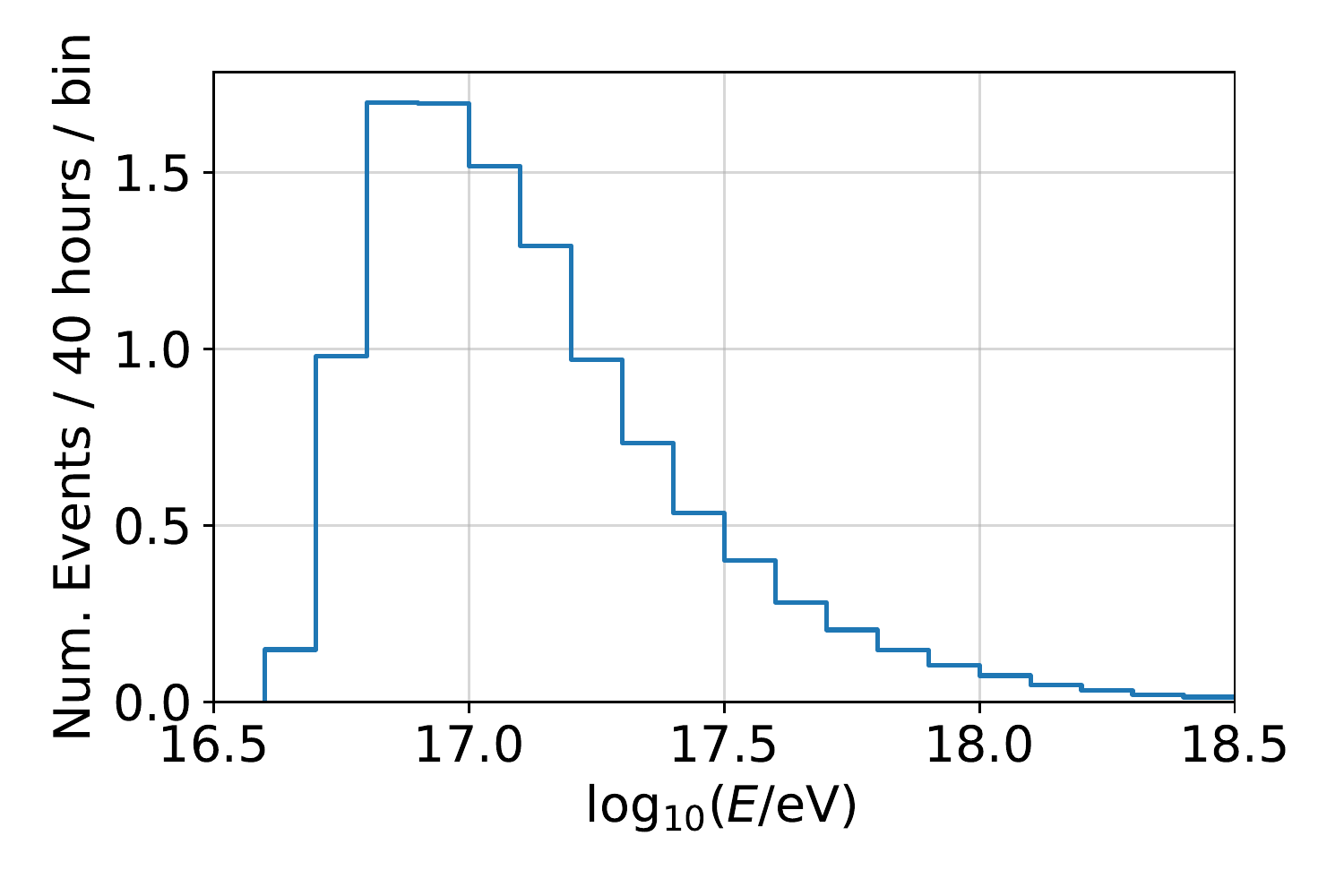} %
      \caption{Number of events function of energy for cosmic ray events that trigger the OVRO-LWA detector ($\mathrm{SNR_a}>5.7$) including the analysis efficiency.}
   \label{fig:energy_distrib}
\end{figure}

In Figure~\ref{fig:num_events}, we show the expected event rates based on the trigger model, the quality cuts (labelled QC) and the power in Gaussian fit cut (labelled GC). For the cosmic-ray flux, we use the Auger parameterization~\cite{2017ICRC...35..486F}. The 10 events detected and surviving selection cuts are consistent with the range of thresholds corresponding to the trigger rate and analysis efficiency. The acceptance and distribution of energies corresponding to the trigger threshold $\mathrm{SNR_a}=5.7$ are shown in Figures~\ref{fig:acceptance}~and~\ref{fig:energy_distrib}, respectively. The most likely energy of the events is expected to be $\sim 7\times10^{16}$~eV. 

%-----------------------------------------------------------------
%-----------------------------------------------------------------

%-----------------------------------------------------------------
%-----------------------------------------------------------------

The OVRO-LWA has several antennas extending across $\sim 2$~km separation from the core of the array which were not used in this demonstration. In future runs, it is expected that using the antennas extending over wider separation from the core will provide a significantly improved analysis efficiency. The longer baselines will improve the directional sensitivity to discriminate against stationary sources of RFI. The higher spatial coverage of the antennas will also improve the sensitivity to the Gaussian shape of the radio beam pattern of air showers.  

\subsection{Zenith Angle Distribution}
\begin{figure}[t!]
  \centering
   \includegraphics[width=0.8\linewidth]{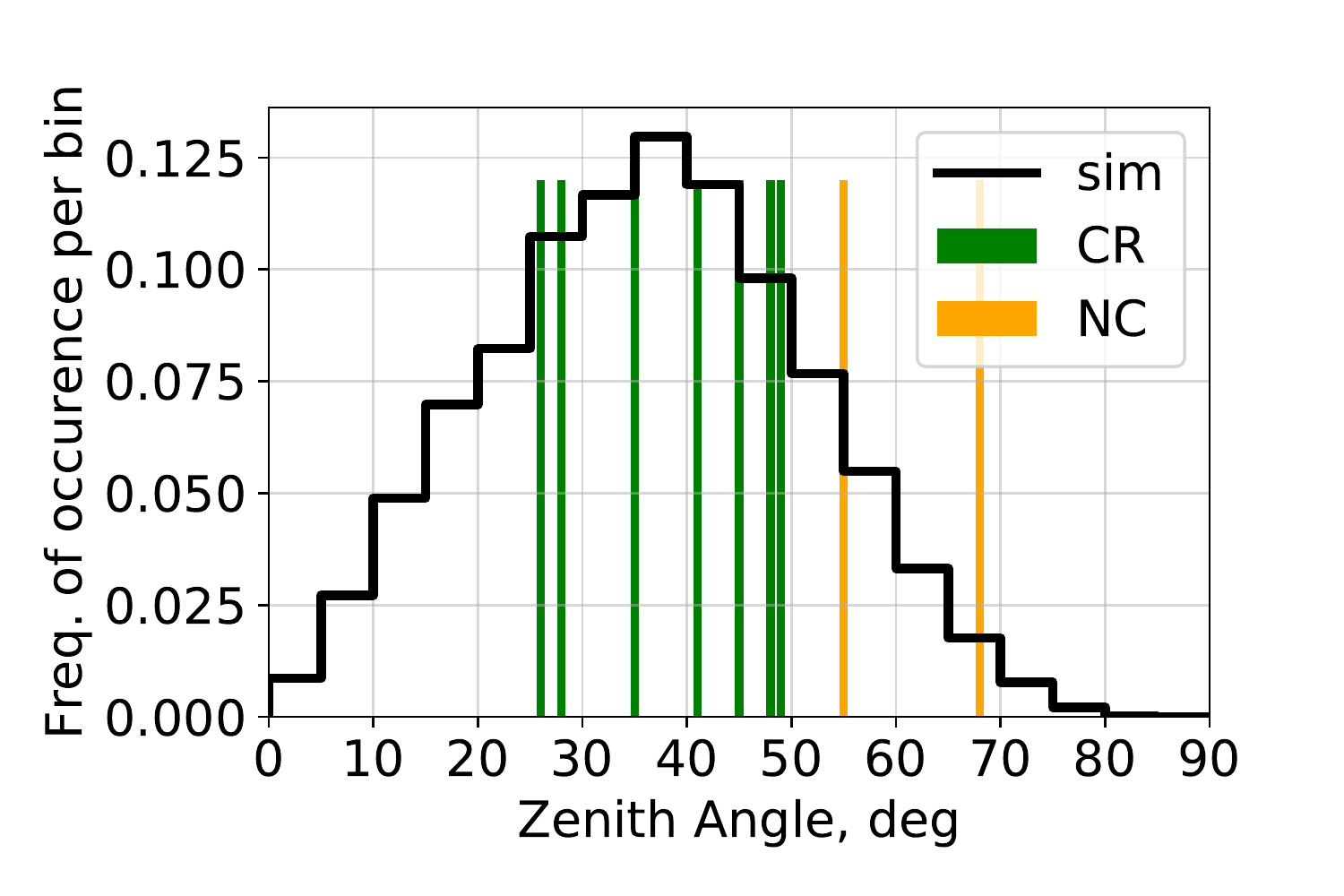} %
      \caption{Frequency of occurrence of cosmic rays that trigger the OVRO-LWA detector ($\mathrm{SNR_a}>5.7$) and analysis cuts. The zenith angles of cosmic ray candidates (CR) are shown in green bars with the ``no call" (NC) events shown in orange bars.}
   \label{fig:zenith_distrib}
\end{figure}

As an additional consistency check, we plot the simulated triggered distribution of zenith arrival directions, including the analysis cuts, in Figure~\ref{fig:zenith_distrib}. The 20\% power in Gaussian fit cut reduces the fraction of high zenith angle ($\theta_z>50^\circ$) events since these will, in general, appear to have a flat power distribution across the $\sim 0.2$~km diameter array (see Figure~\ref{fig:radio_emission_profiles}). Figure~\ref{fig:zenith_distrib} marks the best-fit zenith angle values for cosmic ray candidates (CR) and the no call (NC) events. The zenith angle distribution of cosmic ray candidate events are clearly clustered around the most likely range of values providing additional evidence that the CR events are indeed produced by cosmic ray air showers while disfavoring the cosmic-ray origin of the NC events.

%-----------------------------------------------------------------
%-----------------------------------------------------------------

\subsection{Airplane Background Characterization}
The main source of background for producing false positives are impulsive transients, the majority of which are associated with airplane events. We estimate the probability that such background events could be responsible for the cosmic-ray candidate events. 

\begin{figure}[t!]
  \centering
   \includegraphics[width=1.0\linewidth]{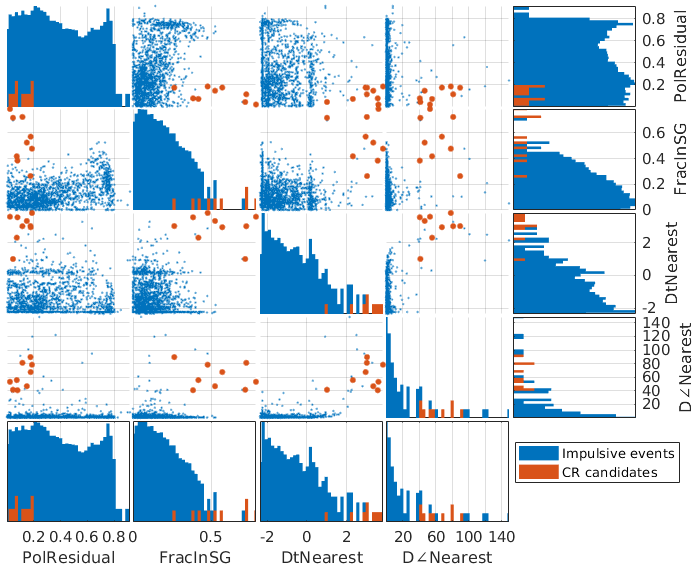} %
      \caption{Histograms and scatterplots of the variables used for characterizing background and cosmic ray candidates (see text for details).}
   \label{fig:bkg_pdf}
\end{figure}

We characterize this background in terms of four variables. The first two are the observables used to identify cosmic-ray candidates: the error between the observed and expected N/S polarization for an air shower (PolResidual) and the fraction of power in Gaussian (FracInSG). The third is the shortest time to between two events flagged (DtNearest), which is expected to be short for background sources and random for an air shower. The fourth is the angular distance between the event and its nearest neighbor in time (D$\angle$Nearest), which is also expected to be small for a background event and random for an air shower. In Figure~\ref{fig:bkg_pdf} we show the distribution of these variables for transient events (in blue) and the cosmic ray candidates in orange.

\begin{figure}[t!]
  \centering
   \includegraphics[width=1.0\linewidth]{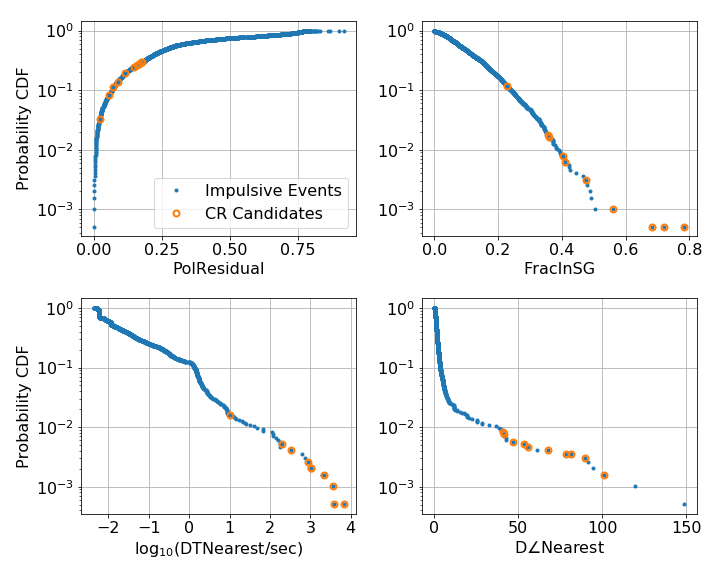} %
      \caption{Cumulative distribution function for the impulsive events and cosmic ray candidates in Figure~\ref{fig:bkg_pdf}}
   \label{fig:bkg_cdf}
\end{figure}

\begin{figure}[t!]
  \centering
   \includegraphics[width=1.0\linewidth]{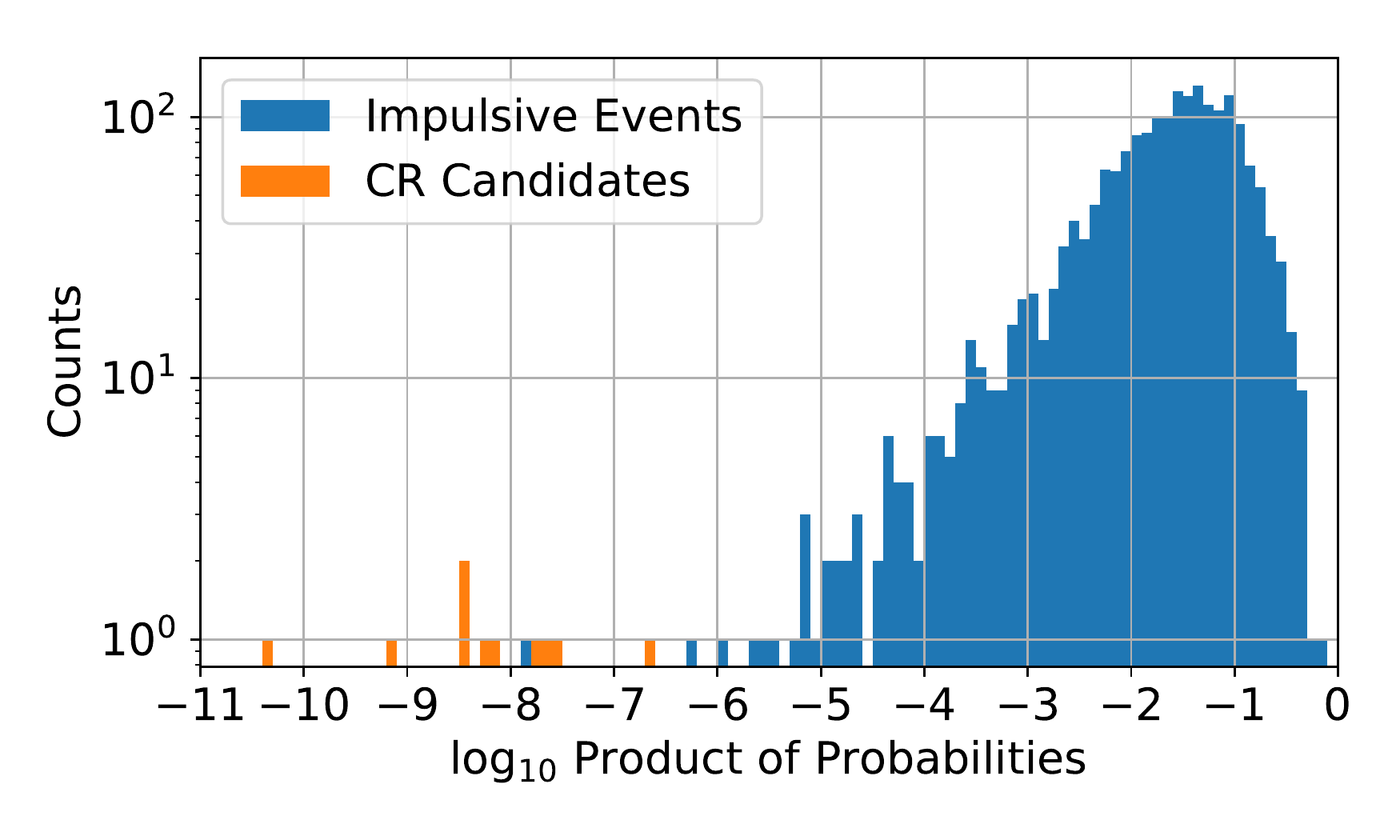} %
      \caption{Histogram of the combined probabilities shown in Figure~\ref{fig:bkg_cdf}. The non-CR candidate event was rejected due to clustering with a known flight path event though it was temporally isolated from other impulsive transients.}
   \label{fig:bkg_hist}
\end{figure}

Cosmic ray candidates are generally separated from the impulsive backgrounds but no single variable or combination of variables shows a strong separation. However, the combination of these variables does show a clear separation. To show this, we take the cumulative distribution function of the events (Figure~\ref{fig:bkg_cdf}) to estimate the probability of each variable. We then multiply the probabilities of each variable. The resulting distribution (Figure~\ref{fig:bkg_hist}) shows that the cosmic ray candidates are distinctly separated from the distribution of impulsive transients. 

The results in Figure~\ref{fig:bkg_hist} show one impulsive transient event flagged as a background that is consistent with a cosmic ray candidate. This event was rejected as a cosmic ray candidate because it clustered with the fitted curve of an airplane trajectory despite being well separated in time and distance from it's nearest event. This shows that our analysis has been conservative in flagging cosmic ray events. 

We also inspected the events flagged as backgrounds in the tails of the distribution in Figure~\ref{fig:bkg_hist} (x-axis values between -7 and -6). These events are rejected due to high values of PolResidual ($>$60\%). This  population of isolated impulsive transients that are not consistent with cosmic rays is clearly distinguishable from the cosmic-ray candidate events.

%-----------------------------------------------------------------
%-----------------------------------------------------------------

%-----------------------------------------------------------------
%-----------------------------------------------------------------

%-----------------------------------------------------------------
%-----------------------------------------------------------------

\section{Conclusion}

Purely RF detection and identification of cosmic rays has been demonstrated with the
OVRO-LWA. The use of custom FPGA firmware, centralized signal processing, dense antenna sampling and novel software post-processing made this demonstration possible. With its successful demonstration, however, replicating this result should not be challenging for many low-frequency arrays, including LOFAR, HERA, CHIME, and the SKA\citep{2015ICRC...34..309H}. Implementation would be especially easy for other LWA stations, which share much of the back-end hardware.

There were essentially two problems faced by cosmic ray self-trigger applications in civilization.  First, candidate impulsive events must be detected on-FPGA at a rate which does not saturate network bandwidth. Second, cosmic ray events must be discriminated from RFI -- especially in the case of airplanes, which have more complicated time-frequency structure.  Performance for the FPGA stage may be enhanced with the on-chip planar fit routines recommended in Section~\ref{sec:Future-experimental-design}, whereas it is systems which contain a dense component will get a substantial sensitivity and trigger rate benefit from hierarchical beamforming, described in the same section. The extraordinary properties of the air shower events demonstrated make successful detection of $\sim$10 events unquestionable, with another $\sim$2 being less clear. Simple changes can improve the missed detection rate from $\sim$40\% down to nearly 0\% for sufficiently energetic events above an elevation angle of $20^\circ{}$. Many steps which would make this improvement practical and economical were discussed in Section~\ref{sec:Future-experimental-design}.

This demonstration, although it can improve, lends credibility to future
arrays of RF-only air shower detectors for cosmic ray and neutrino
science. The techniques described here can also be used to measure the beam pattern of individual antennas in the array, a challenging goal for many low frequency radio arrays.

\section{Acknowledgements}

This material is based in part upon work supported by the National
Science Foundation under Grant AST-1654815 and AST-1212226, as well
as the NASA Solar System Exploration Virtual Institute cooperative
agreement 80ARC017M0006 and the DFG grant NE2031/1-1. Gregg Hallinan
acknowledges the support of the Alfred P. Sloan Foundation and the
Research Corporation for Science Advancement. The OVRO-LWA project
was initiated through the kind donation of Deborah Castleman and Harold
Rosen. 
Part of this work was carried out at the Jet Propulsion Laboratory, California Institute of Technology, under a contract with the National Aeronautics and Space Administration. Andres Romero-Wolf and Gregg Hallinan thank the Caltech/JPL President's and Director's Research and Development Fund. Copyright 2019. All rights reserved.

\clearpage

\appendix
\section{All events promoted to final cuts}

In order to better visualize the spatial power distribution of the events, these plots represent the power received as a function of antenna position in both the color \textit{and} area of each scatter-point, a departure from previous plots which contain a similar aesthetic. These figures follow the order in Table~\ref{tab:Cosmic-ray-candidates,}
\subsection{Events likely to originate from cosmic rays}

\begin{figure}[!htb]\centering
\includegraphics[width=0.8\columnwidth]{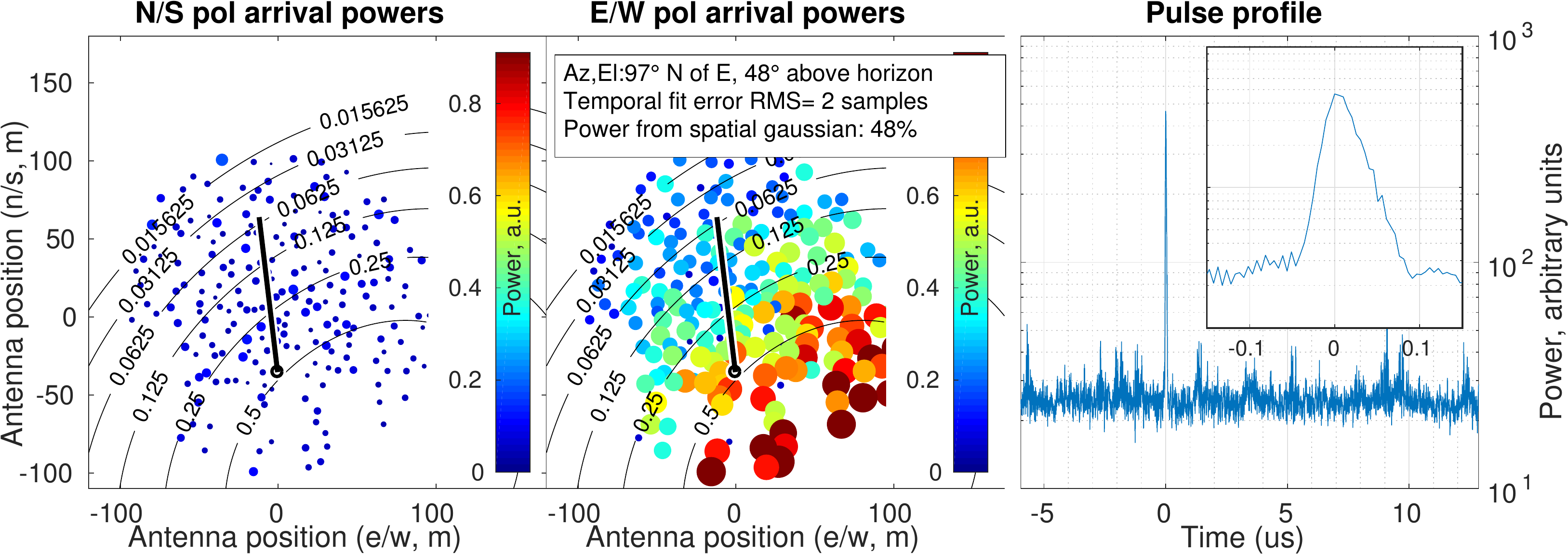}
\end{figure}

\begin{figure}[!htb]\centering
\includegraphics[width=0.8\columnwidth]{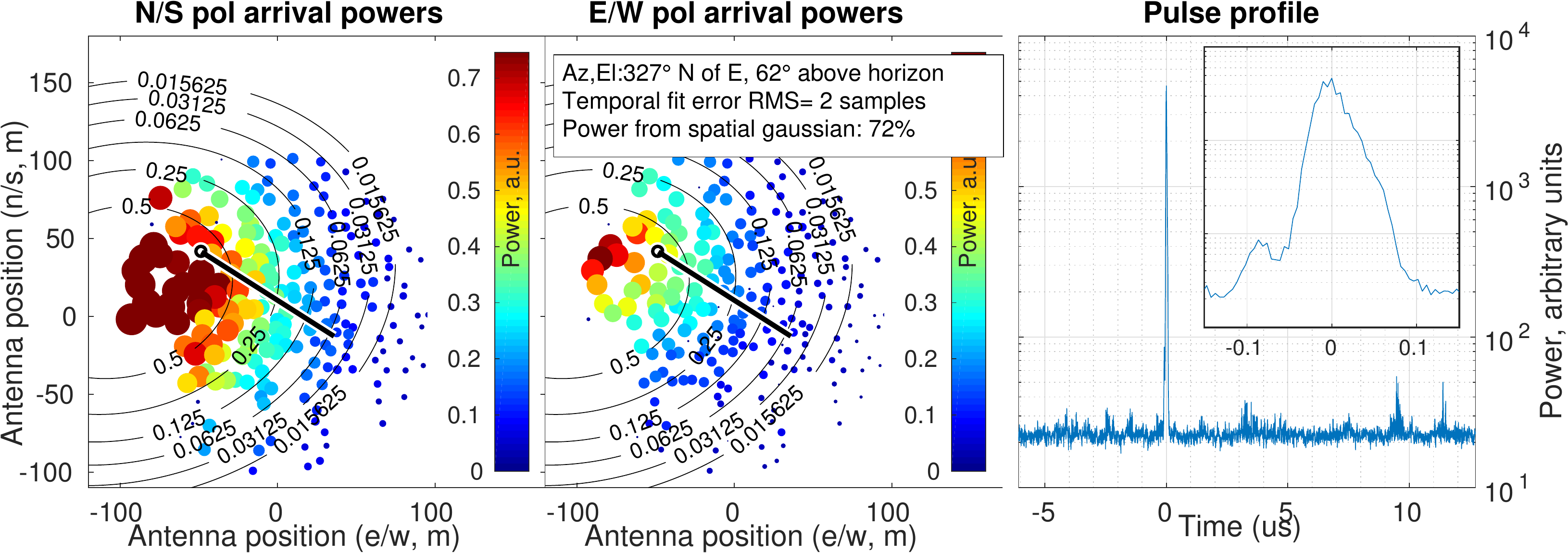}
\end{figure}

\begin{figure}[!htb]\centering
\includegraphics[width=0.8\columnwidth]{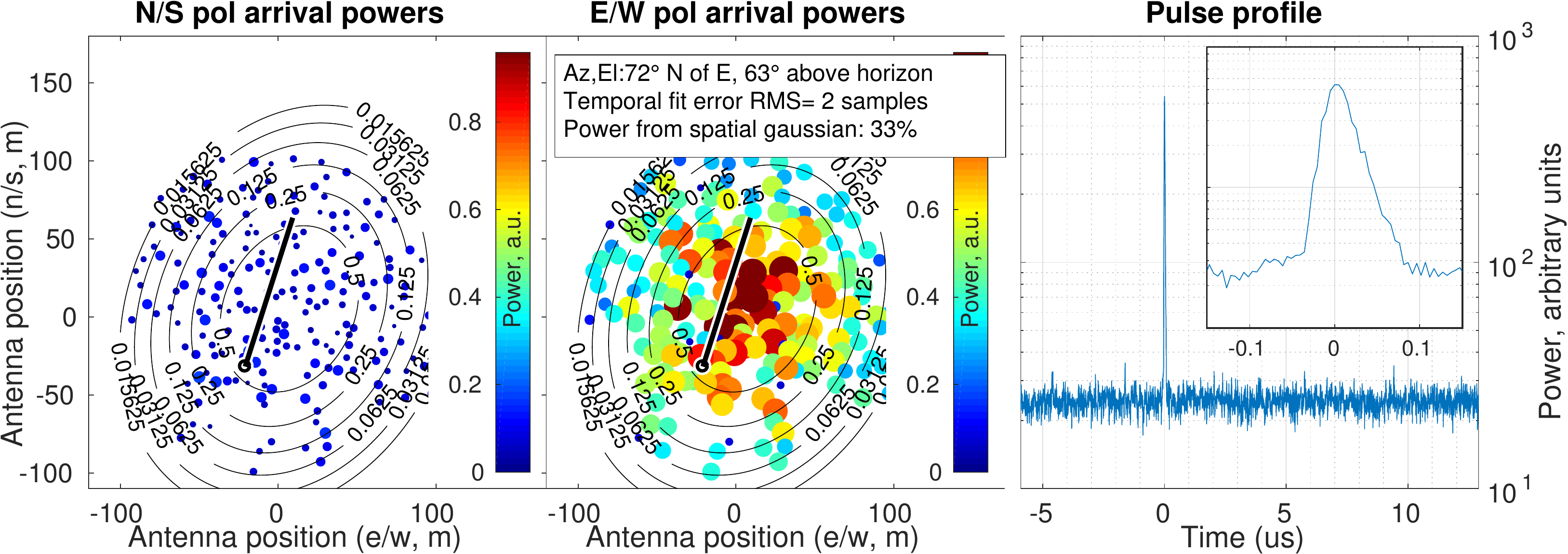}
\end{figure}

\begin{figure}[!htb]\centering
\includegraphics[width=0.8\columnwidth]{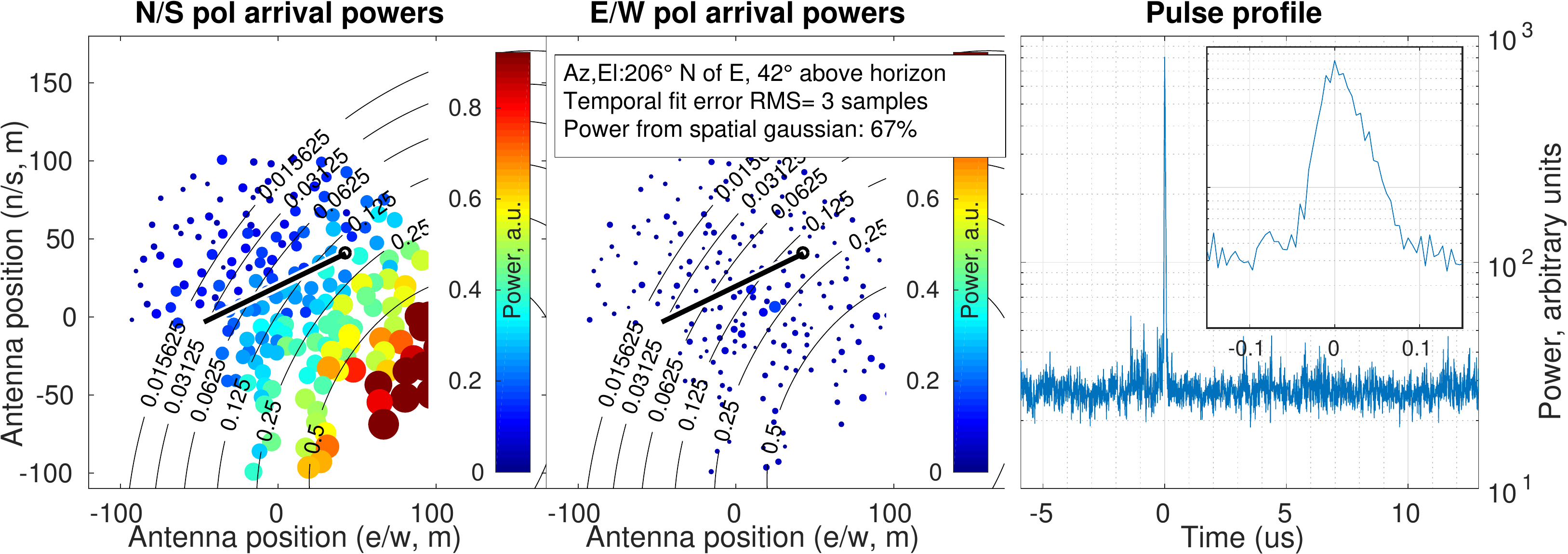}
\end{figure}

\begin{figure}[!htb]\centering
\includegraphics[width=0.8\columnwidth]{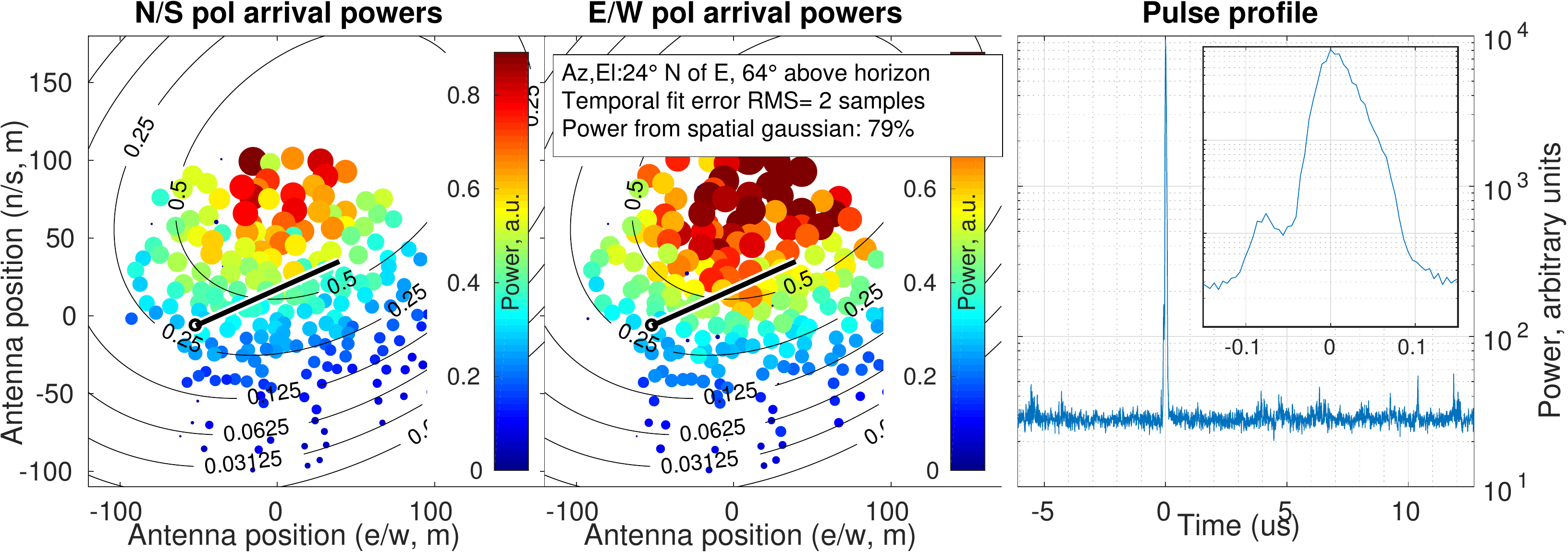}
\end{figure}

\begin{figure}[!htb]\centering
\includegraphics[width=0.8\columnwidth]{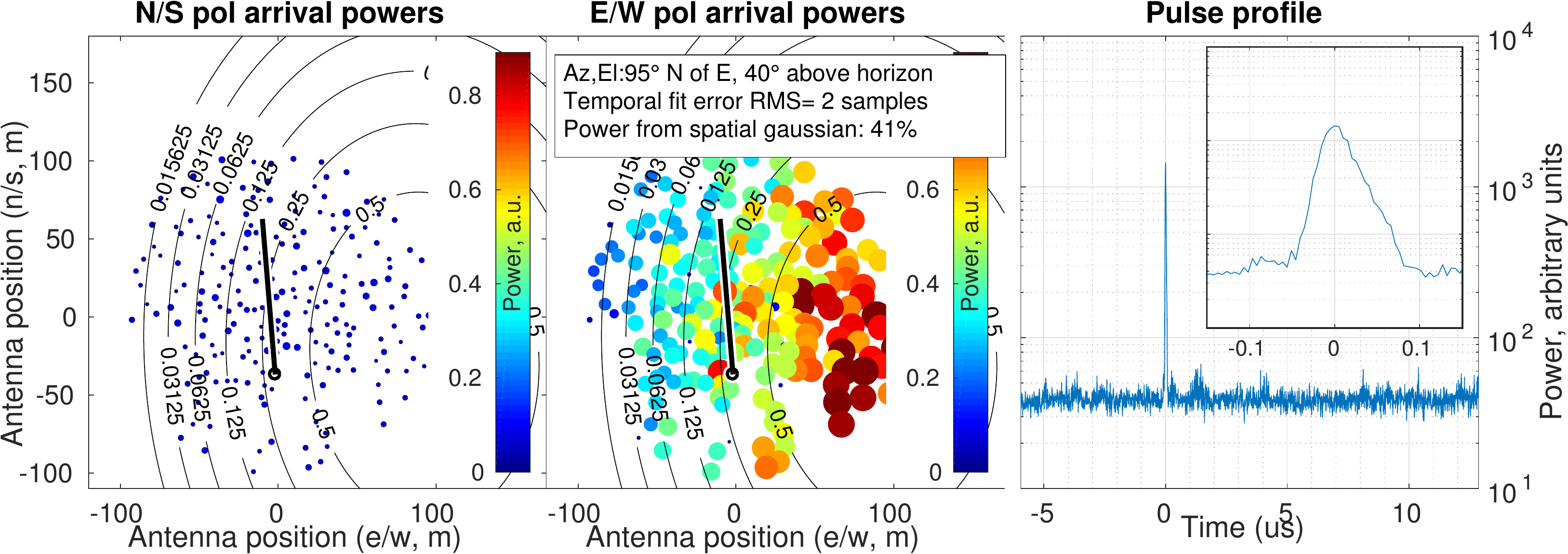}
\end{figure}

\begin{figure}[!htb]\centering
\includegraphics[width=0.8\columnwidth]{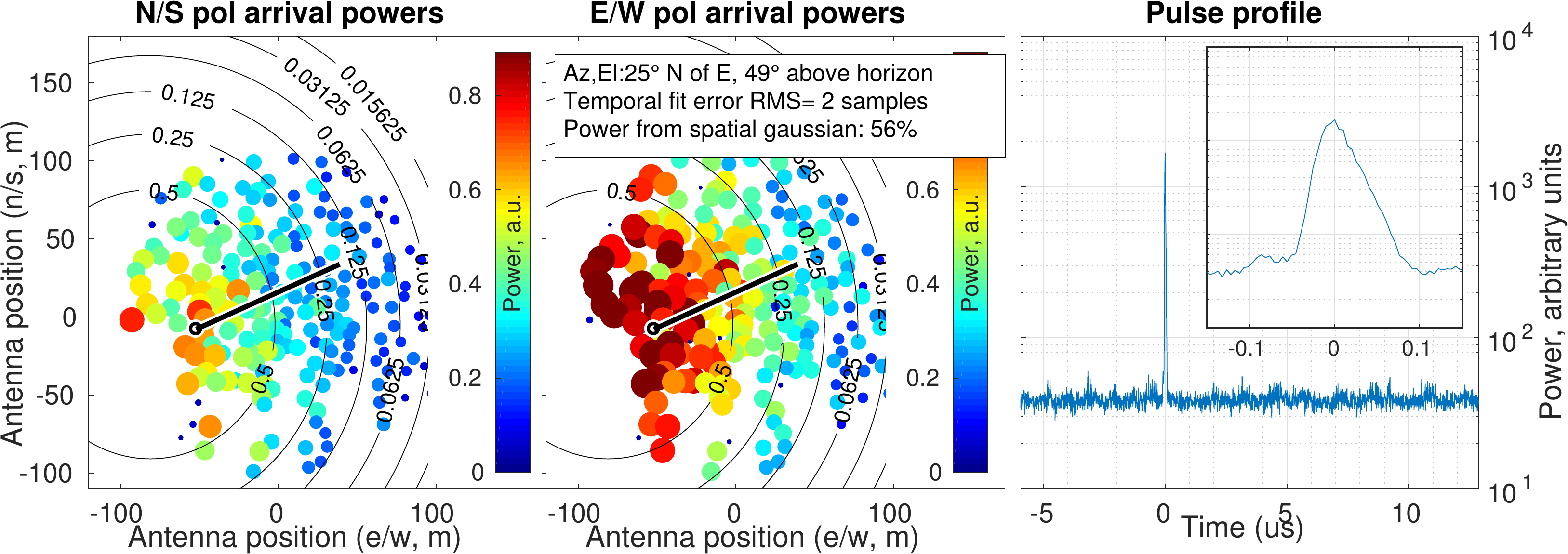}
\end{figure}

\begin{figure}[!htb]\centering
\includegraphics[width=0.8\columnwidth]{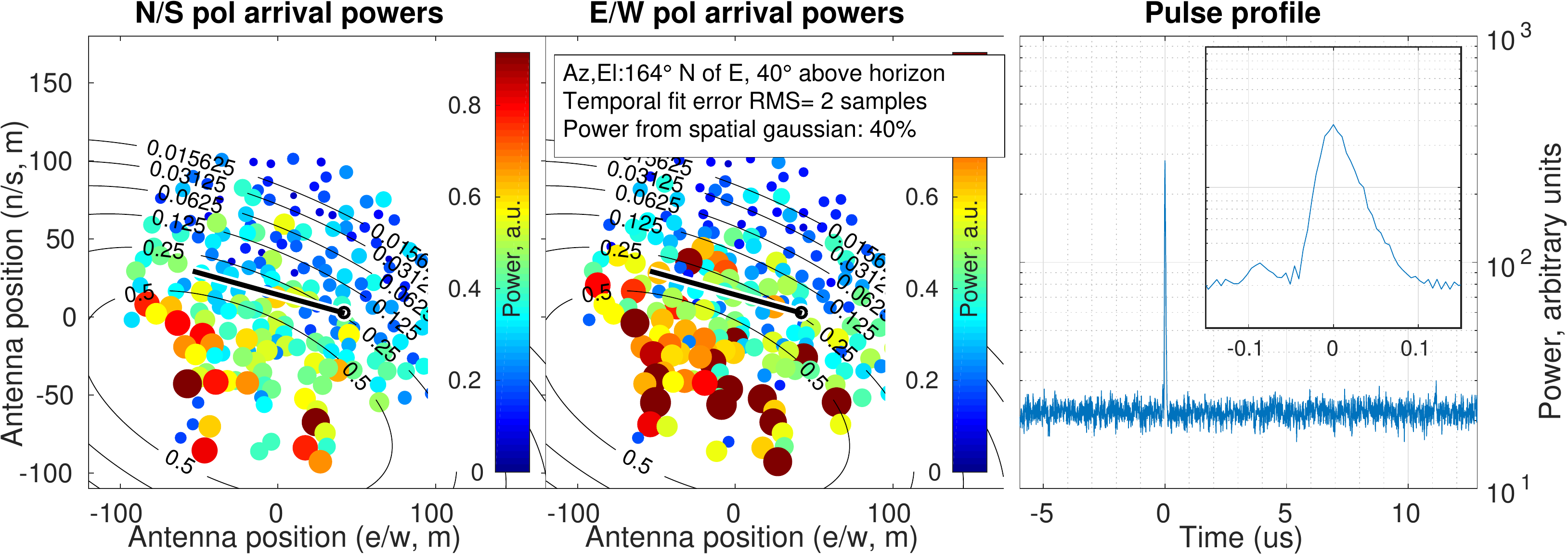}
\end{figure}

\begin{figure}[!htb]\centering
\includegraphics[width=0.8\columnwidth]{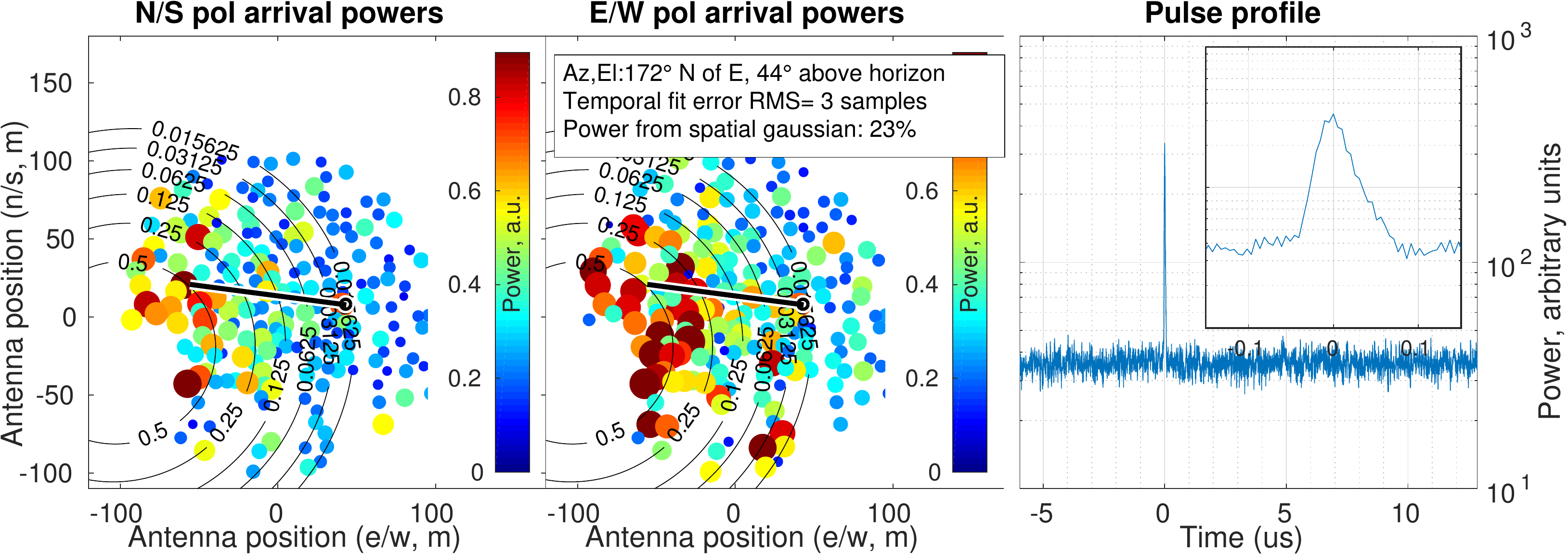}
\end{figure}

\begin{figure}[!htb]\centering
\includegraphics[width=0.8\columnwidth]{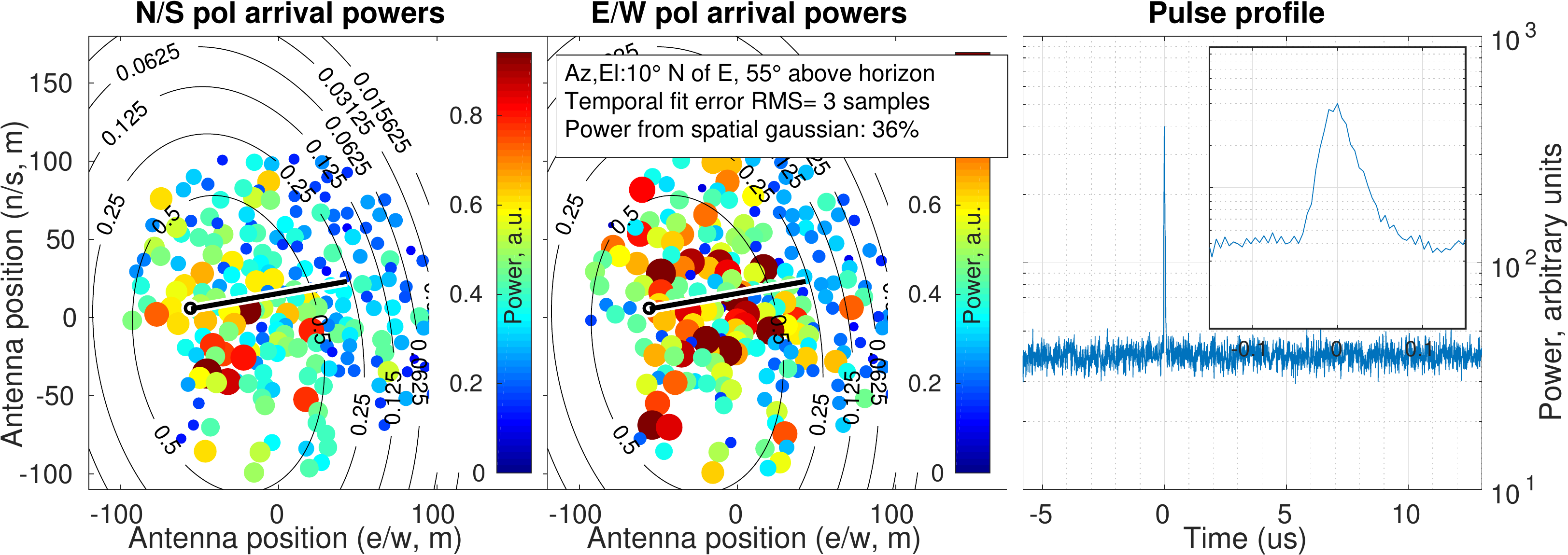}
\end{figure}

\clearpage

\subsection{Events which are inconclusive}

\begin{figure}[!htb]\centering
\includegraphics[width=0.8\columnwidth]{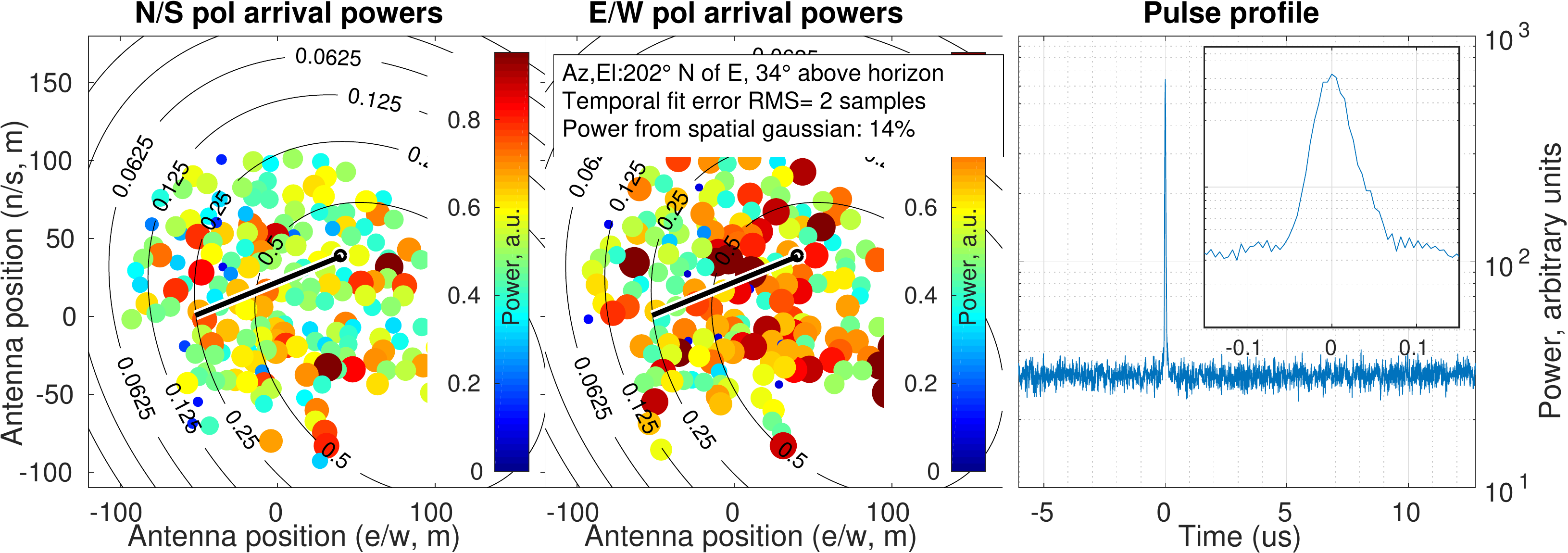}
\end{figure}

\begin{figure}[!htb]\centering
\includegraphics[width=0.8\columnwidth]{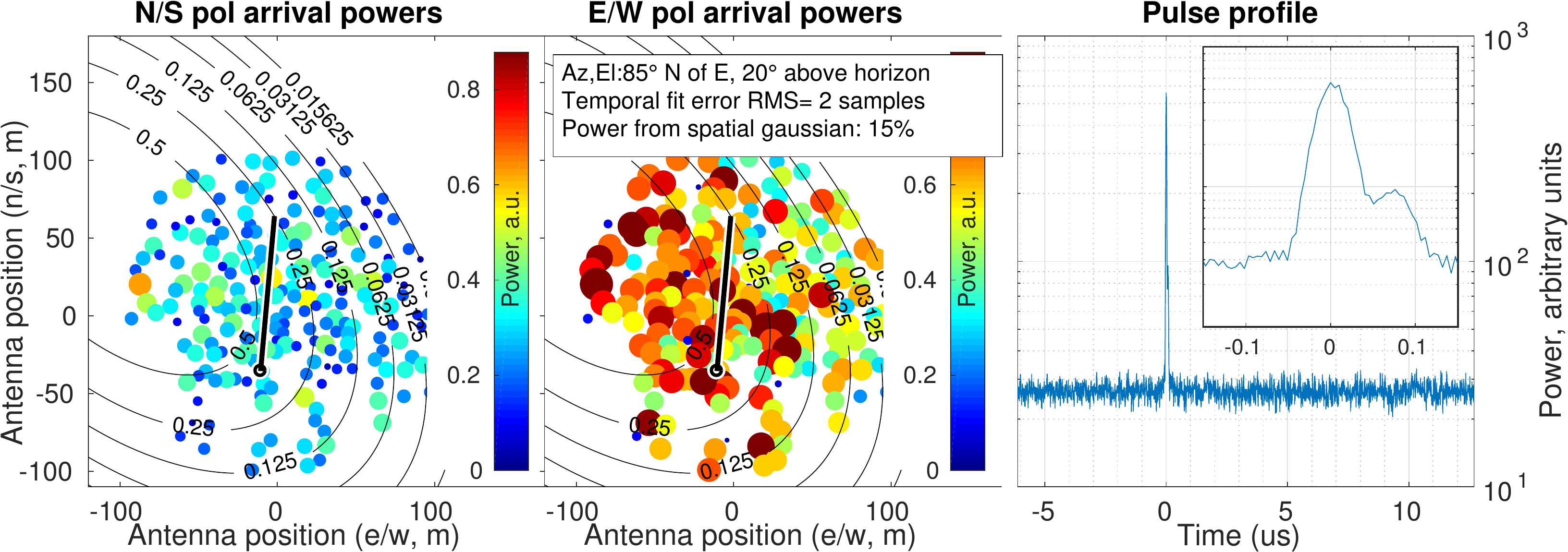}
\end{figure}

\clearpage

\subsection{Events likely to originate from man-made sources}

\begin{figure}[!htb]\centering
\includegraphics[width=0.8\columnwidth]{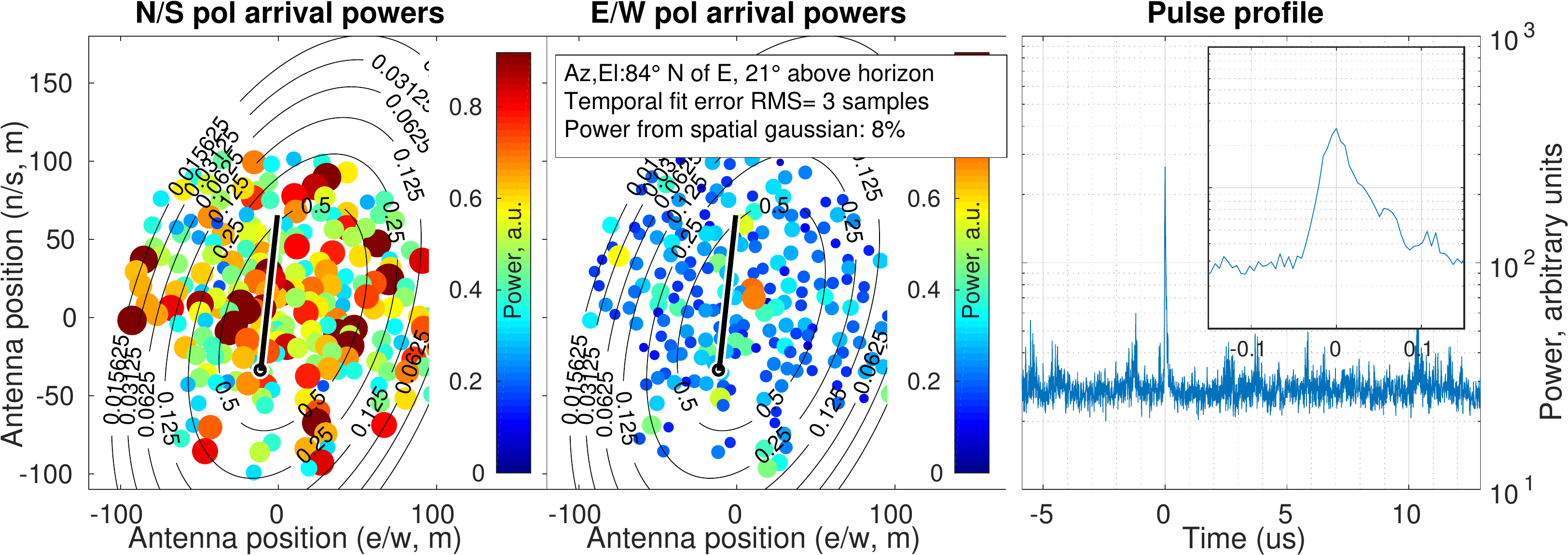}
\end{figure}

\begin{figure}[!htb]\centering
\includegraphics[width=0.8\columnwidth]{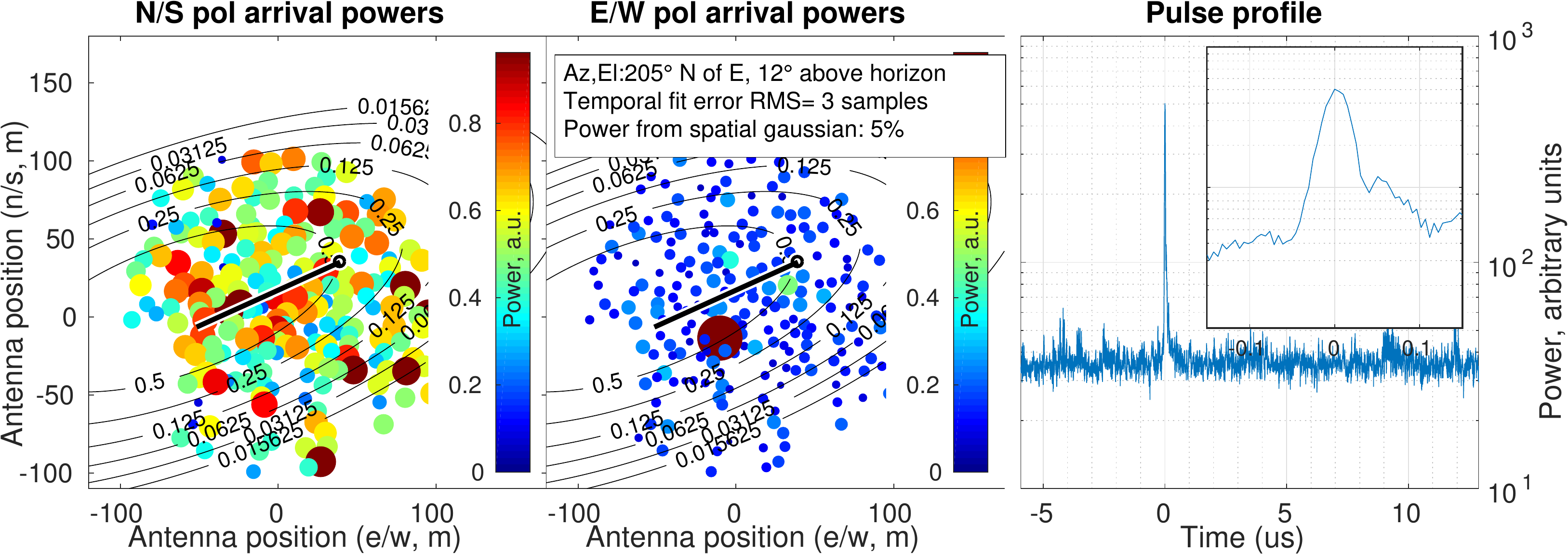}
\end{figure}

\begin{figure}[!htb]\centering
\includegraphics[width=0.8\columnwidth]{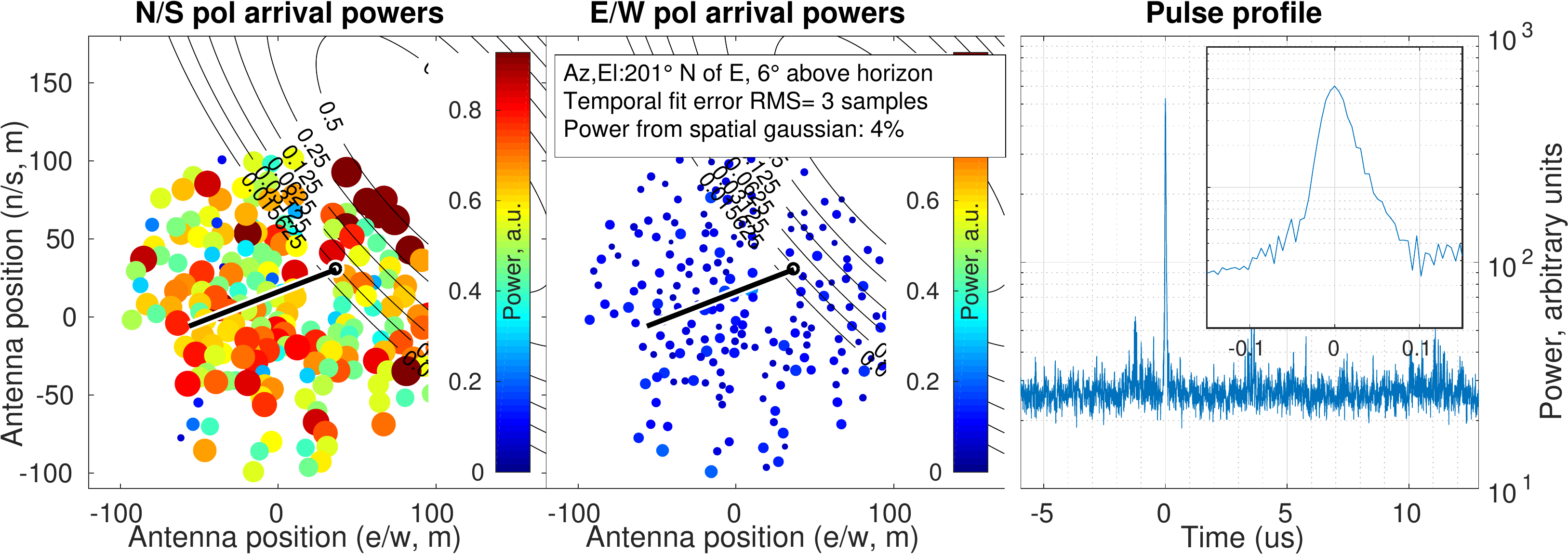}
\end{figure}

\begin{figure}[!htb]\centering
\includegraphics[width=0.8\columnwidth]{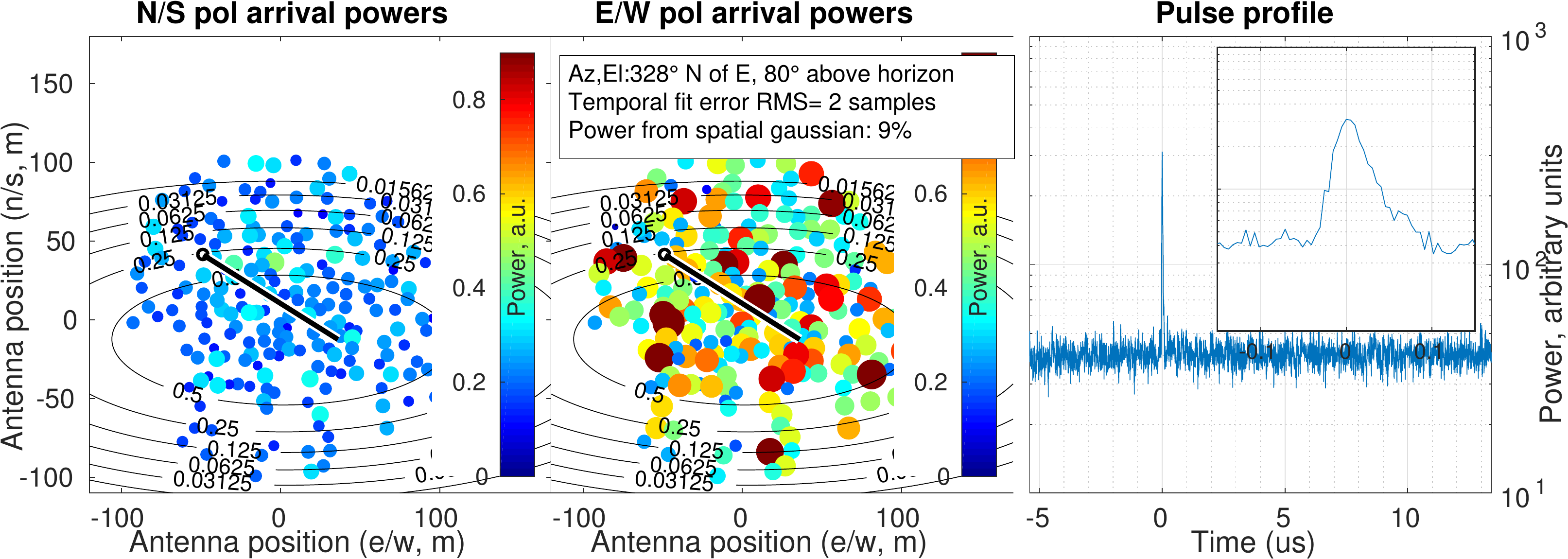}
\end{figure} 

\clearpage

\section{\label{sec:Mistakes-that-were}Discussion}

Since this is one of the first successful RF self-triggers of cosmic
rays with a radio telescope, a frank commentary on  ``lessons
learned'' merits discussion for future efforts.
\begin{itemize}
\item \textbf{Cross-correlation time-of-arrival method }The initial time-of-arrival
algorithm was far more complicated than the one presented in this
work, including estimation of the received signal profile with almost
no priors and then using the peak cross-correlation between that estimated
signal and all inputs as the time of arrival estimate~\citep{2005Natur.435..313F}. It was found that almost
all RFI events detected by this system had a strong impulsive component
(not surprising, considering that was what the FPGA detector was designed
to operate on). The technique described in Section \ref{enu:Software}
is faster and can give useful results at lower SNR, but exhibits larger
fitting residuals than the more sophisticated technique. That said,
the Hilbert envelope technique described in \citep{2013A&A...560A..98S}
would have likely have performed very well as well. It is likely that
a cross-correlation between input voltages would have worked best
of all (provided that a full cross-correlation of all inputs was performed).
This was prototyped and worked well, but the best way to extract the
relative time-of-arrival for each input was not obvious in the context
of not all inputs receiving an impulse, resulting in the use of the
techniques described above. If a voltage cross-correlation technique
was used however, precise extraction of individual antenna power estimates
could be performed by formulating the problem as a noisy Rank-1 matrix
completion problem, where each element in the matrix is the peak cross-correlation
power between two inputs, and the diagonal (to be solved for) is the
power seen by each input. Such a technique would largely remove the
additive impact of receiver noise on power estimates, leaving only
the astronomical background (this would likely give performance similar to imaging under the assumption that the image consists of only thermal noise and a point source). If cross-correlating voltages, remember
to correlate the analytic version of all signals (via a Hilbert transform,
for example). 
\item \textbf{Distribution of error sources }The RMS time-of-arrival residual
was about 4~samples with the arg-max time-of-arrival estimator, and
about 3~samples with the more cross-correlation technique
(available in the included code, but not described here for brevity).
In each case, about 0.5~samples of that residual was due to using
the arg-max of the signal power (or alternatively, cross-correlation
value), which does not consider the value of nearby samples. A better
technique would use some form of super-resolution, such as quadratic
interpolation of the peak location, to generate sub-sample resolution
time-of-arrival estimates. An additional $\sim1.8$~samples of the
RMS residual was due to bias that was systematic to each antenna \textendash{}
likely inaccuracies in either cable delay estimation or antenna position
estimation. Accurate estimation of these parameters would also have
improved system performance. Responding to this issue, cable-length
delays for each input were estimated using triggered RFI events. Great
care was assigned to avoiding biasing the delay estimates towards
common events. If this work were repeated, the same process would
be performed jointly on antenna positions as well. It is important
to note that with many FPGA boards and ADCs, the on-FPGA PLLs will
lock to a different clock cycle of the ADC data line on each FPGA
configuration. For this reason, it is possible that the effective
cable lengths will vary across FPGA configurations, requiring a new
cable length calibration each time.
\item \textbf{You want the original samples }Being able to process the raw
ADC voltages directly (as opposed to F-engine products, which lack
the needed temporal resolution, or beam-formed products which lack
the needed spatial coverage) was crucial for the success
of this project. Additionally, being able to program the FPGAs with
custom firmware greatly simplified the effort.
\item \textbf{Power line RFI is 60}\,\textbf{Hz phase dependent. }Although
it was not demonstrated in this work, many sources of terrestrial
RFI are correlated with the phase of the 50/60\,Hz mains voltage:
filtering event triggers on this signal may produce good RFI mitigation
results for some sources. Because $\lambda$ at 60\,Hz is $\sim$50,000\,km,
phase acquisition could be done at a central location and then transmitted
to a distributed processing system, provided adequate time synchronization.
\item \textbf{Matched filtering would have done a bit better. }Matched filters
are the optimal detector of a known signal in the presence of white
Gaussian noise \citep{1057571}. This would likely have performed
better than a moving average. Depending on the bandwidth of the instrument
and RFI, it is possible that the impulse response of the system can
be used as this matched filter (for OVRO-LWA, this was the case).
\item \textbf{Grouping of antennas. }All detected cosmic rays, as well as almost
all RFI sources, were considerably polarized. If grouping antennas
on FPGAs for detection purposes, it might have been helpful to address
polarization directly by taking the time-of-arrival for the highest
SNR input on each antenna or alternatively simply summing in power
across polarizations for the purpose of detection and localization.
Likewise, re-arranging antennas on FPGAs such that each FPGA only
contained inputs corresponding to one polarization could be an alternative
method to capitalize on this property. Doing this would give the direction-of-arrival
algorithm more unique locations with which to fit and presumably greater
spatial extent. Since most of the power was typically in one polarization
or the other for this work, sorting by polarization would not have
terrible consequences on sensitivity for most RFI events, most of
the time. None of this was explored, but it is worth thinking about the trade-offs of various antenna grouping schemes for future experiments. 
\item \textbf{Elevation filtering was not as important as expected}. Initial
filters rejected all events with an elevation angle less than $29^\circ$,
among many other filters which were tested at one point or another
in this work. As an experiment to see if the system as designed would
be effective at detecting $\tau$-neutrinos, this filter was relaxed
to permit all events above an elevation angle of $2^\circ$.
This resulted in more false positives (by a factor of $2\sim3$) being promoted to final cuts,
but no additional missed detections (which was a small risk due to
the time domain clustering algorithm). Adding more filter metrics
(reject events which lack data from over 1/4th of inputs due to instrumental
errors; reject events whose pulse profiles do not appear to represent band-limited
impulses) brought the RFI rejection performance to that which is shown
in the main body of this work. Because the algorithm works well down
to $\sim2^{\circ}$ elevation angle, it is likely suitable for discriminating
between RFI and $\tau$-neutrinos. 
\item \textbf{The 3D source fitting algorithm performed much better than
plane fitting but was computationally expensive.} The 3D algorithm
was written in MATLAB and likely very computationally inefficient.
Likewise, the faster but less accurate robust plane-fitting algorithm
was simply called from MATLAB without any special tuning. It is possible
that engineering could bridge the gap between these algorithms. Likewise, implementing the entire routine in a compiled language (as opposed to MATLAB) could render the entire issue irrelevant.
\item \textbf{These automatic RFI filtering metrics completely ignore the
distribution of antenna power as a function of position.} We held
the opinion that it would be too challenging to algorithmically discriminate
RFI from astronomical events without having examples of cosmic rays.
\item \textbf{There was an unexpected surplus of RFI at $15^\circ$ elevation angle}.  We put a fair amount of effort into solving this, and could not.  The AREA team has seen similar behavior \citep{arena:2018,Fliescher:82811}, also visible (but not addressed) in Figure 2 of \citep{2013NIMPA.725..133K}. Through inspection of the frequency of narrow-band RFI and comparison with imaging data and spectra, we are confident that the sampling frequency is correct.  Several sources known to be produced on the ground near the array are reconstructed to positions near the horizon. Geometrically, there are no geographical features near OVRO-LWA which are perceived as a $15^\circ$ elevation angle from the location of the array. Furthermore, hand-calculation of the elevation angle of these strange events results in solutions very similar to those produced in the automated fashion, confirming that result. Reflections against some subterranean surface were considered but rejected based on both signal strength arguments and the detection of these events by AREA. This author disagrees with the conclusions of \cite{Fliescher:82811}, which attribute the spurious elevation angle events with an ill-conditioned fit due to a lack of antenna coverage. Even after application of an appropriate cut, the number of spurious events is not substantially reduced, and the artifact is seen in this dataset at the OVRO-LWA which has excellent coverage. Figure~\ref{fig:elHist} shows the histogram of observed elevation angles in detail, while the upper-left section of Figure~\ref{fig:polarPlots} illustrates the behavior in a different perspective.   The issue remains a mystery, and resolving or characterizing it will likely be a prerequisite for many proposed neutrino experiments which are predicated on radio triggering.
\end{itemize}

\begin{figure}[hbt!]
\centering
\includegraphics[width=1\columnwidth]{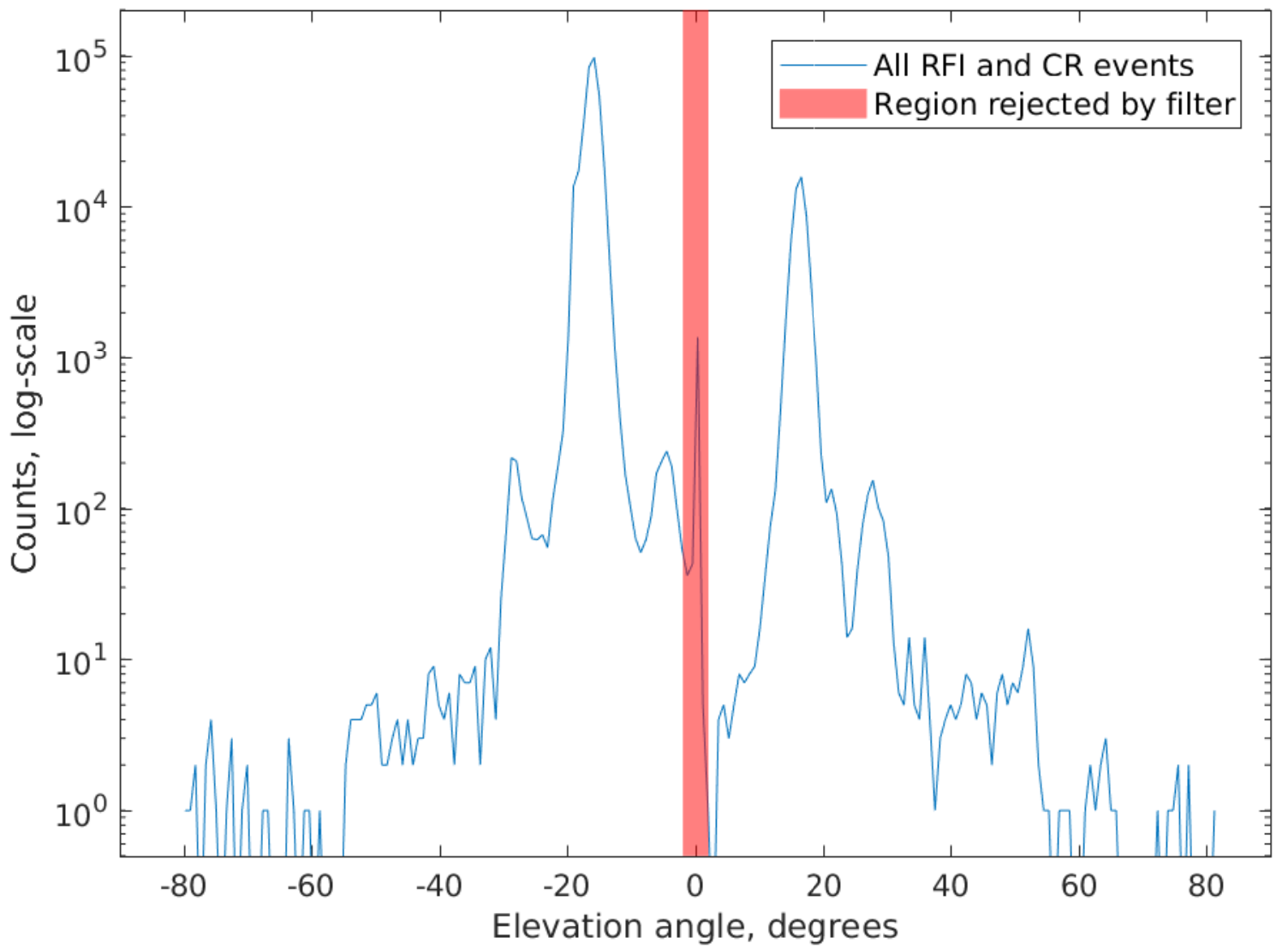}

\caption{Histogram of event elevation angles.  The array is nearly planar, producing a degeneracy between events above the array and below the array.  For all other elevation angle displays in this work, the absolute value of the elevation angle was taken.}
\label{fig:elHist}
\end{figure}

\section{\label{sec:Future-experimental-design}Future experimental design}

\begin{figure}[hbt!]
\centering
\includegraphics[width=0.8\columnwidth]{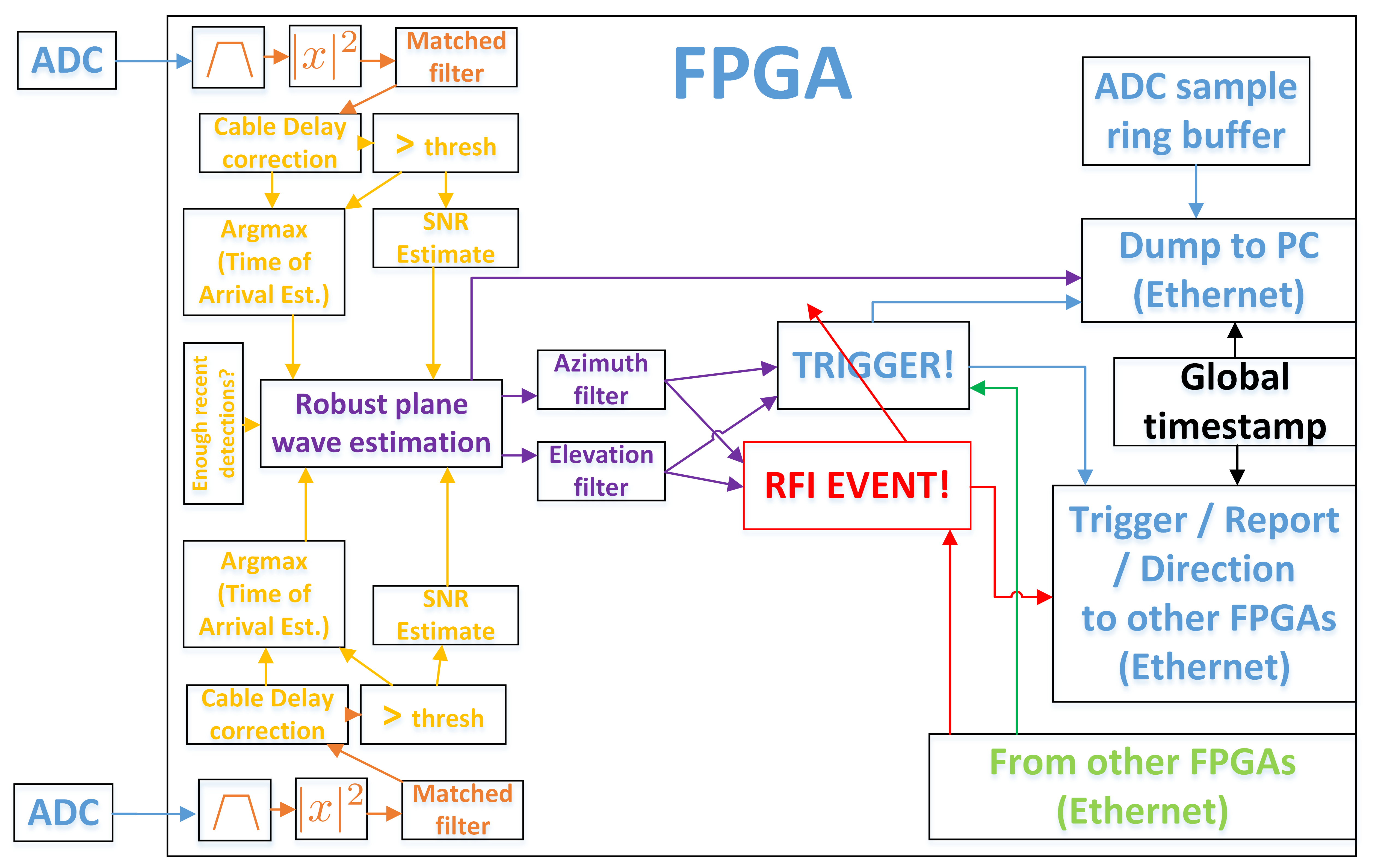}

\caption{Block diagram of a proposed, lower
bandwidth FPGA firmware. In prototypes, robust planar fitting vastly
outperforms the RFI filtering used in Section~\ref{sec:Firmware}.}
\label{fig:future_block_diagram}
\end{figure}

The lessons learned from this effort (and prior work in the field)
inform future experiments. Broadly speaking the method of using single
FPGAs for batch-processing baseband sampled antennas as described
in Section \ref{sec:Summary-of-technique} performed adequately for this experiment. Future systems (especially ones which are more compute or communications-limited)
should perform techniques similar to those described in Section \ref{sec:Firmware},
as well as \eqref{eq:noiseEst}, \eqref{eq:powEst}, and \eqref{eq:toaEstOne}
on-FPGA. Afterward, the subsequent steps of source direction fitting
with a plane-wave assumption should be performed on-board \citep{Josh2016}
(FPGA soft processors could also be used for this task), followed
by azimuth/elevation/fit-quality RFI blocking, which is sent to other
devices as a time-stamped message to other FPGAs, not unlike the trigger
packets used for regular detection above (Figure~\ref{fig:future_block_diagram}).
This would filter all except $\sim$1/10,000~events with minimal consequence
to cosmic ray detections, suitable for sites with relatively little
air traffic. At these ``low air traffic'' sites, the data pipeline
will simply saturate while the airplane is overhead, likely preventing
cosmic ray detections. For sites with heavier air traffic or severe
bandwidth limitations (such as those communicating event detections
over cellular), these ``Stage 2'' detections can instead be buffered
in on-board DRAM. After several minutes have passed, either time-domain
clustering or robust curve fitting on \{azimuth, elevation, time-stamp\}
as described in Section \ref{sec:Mistakes-that-were} can be used
to flag out airborne RFI sources, reducing event rate to $\sim$1/hr if
applied at the OVRO-LWA, which is slow enough for most arrays.

\begin{figure}[hbt!]
\centering
\includegraphics[width=0.8\columnwidth]{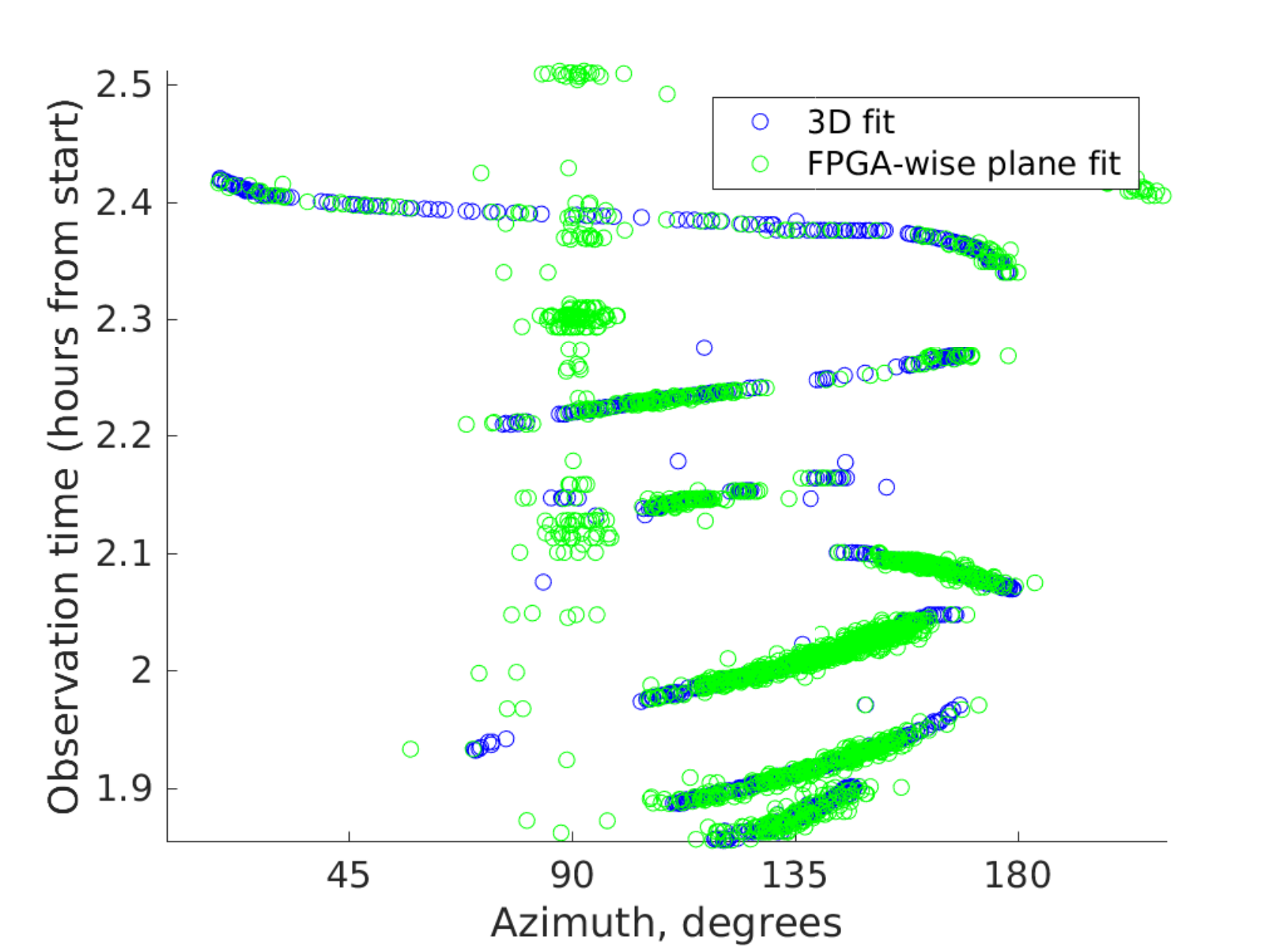}
\caption{Similar analysis to that performed
in Section~\ref{enu:Software}, overlaid on top of the results of local planar
fits from individual FPGAs as described in Section~\ref{sec:Future-experimental-design}.
Software issues resulted in the spurious events around Azimuth $90^{\circ}$,
which would not be present in a production system.}
\label{fig:scatterplot-zoom-overlay}
\end{figure}

In order to simulate the impact of an algorithm which uses on-chip
plane-fitting algorithms such as those described above, previously
collected data was re-processed. Antennas were re-clustered into groups
of 16 which were more optimal for RFI filtering, and source localization was performed using robust plane fitting,
with the omission of the better-performing 3D localization routines. A separate
plane fit was performed with each of the 16 groups of antennas, with
the lowest-residual fit being accepted as the ``true'' source direction.
The algorithm performed reasonably well \textendash{} worse than a
full fit, but would be a much more effective RFI filter than the methods
described in Section \ref{sec:Firmware}, especially considering that
PC post-processing can easily flag the remaining false positives once
all data is available. A quick demonstration of the feasibility of this technique is provided in Figure~\ref{fig:scatterplot-zoom-overlay}, which shows similar behavior and quality when performing a simplified direction of arrival routine on-chip.

A successful RF self-trigger system relies on having (on a processing
unit scale) both a sufficient density of antennas (to have coincident
detections on cosmic rays) and sufficient extent of antennas (for
the direction of arrival estimation). For the system and data analyzed
in this work, a $\sim$100\,m maximum antenna separation would have
been suitable for source localization on these high elevation angle events. For arrays which collect all
signals at a central location, this suggests arranging inputs on FPGAs
such that they have at least roughly this extent in both dimensions.
The narrow E/W extent of the OVRO-LWA antenna-FPGA groupings appears
to have made it less suitable for this style of filtering. 

\begin{figure}[hbt!]
\centering
\includegraphics[width=0.8\columnwidth]{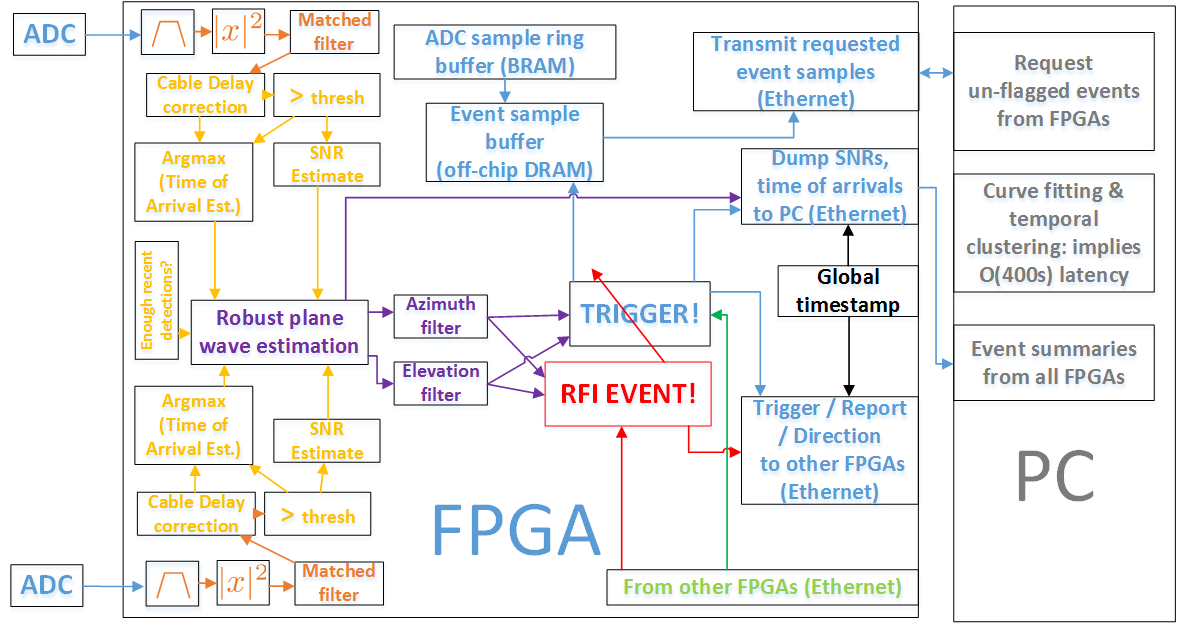}
\caption{Block diagram of a proposed, lower
bandwidth FPGA firmware, with the use of an additional PC co-processor to further reduce
bandwidth by avoiding the transmission of airplane-driven or temporally clustered events.}
\label{fig:future_block_diagram-1}
\end{figure}

For arrays which only collect post-trigger data, those same processors
could be placed in the field, handling small groups of antennas. The
same processing for centralized arrays could then be performed by
sending the \{azimuth, elevation, time-stamp\} data over cellular
to a central server which, after rejecting all airplanes, would trigger
a dump from the local DRAM buffers to the central server (Figure~\ref{fig:future_block_diagram-1}).

Alternatively, beamforming the array in various directions would make near-optimal use of the collecting area of the array. Doing this right is tricky because you cannot simply beamform the entire array in each direction: the event may strike only a subset of the array antennas. Therefore, an intelligent trade-off will have to be struck between spatial resolution and detection SNR, optimized for the expected CR footprint on the array. After an initial detection is made in some direction, beamforming groups of antennas in the array in the direction implied by wavefront fitting will maximize SNR for CR parameter extraction. Using hierarchical beamforming \citep{5613258,2000prat.conf..265H} will make the computation of beams economical with minimal SNR cost.  

Plans at OVRO include near-term expansion to 352~antennas, as well
as expansion to 2048~dipoles on a timescale of $\sim$5~years. Both
arrays will span a 2.6\,km diameter area and will include extensive
air-shower detection as a continuous commensal of operation. Using
the lessons learned from this work, the techniques described in this
section, as well as Section \ref{sec:Mistakes-that-were}, will be
applied directly in the design of that cosmic ray engine.

\subsection{Beam-mapping of radio-astronomy arrays using cosmic ray events}

Understanding the individual antenna response of low frequency radio-astronomy arrays can be very challenging.  For higher frequency telescopes, steering the dish across a bright point source allows for the response to be mapped.  However, low frequency antennas are not large compared to the wavelength of radiation that they receive, and so individual antennas/receivers typically see a substantial fraction of the sky at any given time.  Additionally, these antennas are fixed and cannot be steered beyond what the motion of the earth facilitates. Techniques to reduce the sky to a single point source include using man-made drones as artificial point sources or taking advantage of the variability of radio pulsars to make a time-domain filter which rejects the rest of the sky.  The drone technique suffers from challenges mapping the antenna pattern of the drone, and current techniques only map the amplitude of each antennas beam.  On a similar note, pulsar holography \citep{holography} suffers from a limited selection of pulsars which have adequate SNR (limiting the region of the beam which can be mapped).
Cosmic ray events provide an alternate technique by which the antenna beams can be mapped.  Each event listed in this work has an individual antenna SNR of greater than five for a substantial portion of the array, and several events were ten times that bright. First, a cosmic ray event would be detected and its parameters (such as the direction of arrival, energy and $X_{max}$ ) extracted.  A model of the complex-valued gain seen by each antenna (to a scalar power constant and a scalar time of arrival, likely as a function of frequency) is extracted.  This is compared to the actual complex-valued gain seen by each antenna.  This allows the relative gains of antennas to be mapped.  The technique described here does not provide an absolute gain in each direction, but does greatly simplify the process of mapping individual dipole beams, provided a sufficient density of cosmic ray events striking the array. Making this work in practice will be challenging\textendash{}the power received (and therefore resultant SNR) will vary as a function of event and antenna.  Cosmic ray parameter estimation errors (including the direction of arrival) will contribute to errors in beam modeling, especially for cosmic rays with low received power. There will be a frequency-resolution/SNR trade-off.
One good approach might be to grid events spatially.  On an event-by-event basis (again, as a function of frequency, per an FFT), remove delays implied by geometric effects and a hyperbolic or conical beaming model at a sub-sample resolution.  Integrate the expected and actual signal received after removing a model of thermal noise.  After integrating a sufficient number of events, take the complex-valued ratio between the expected and actual received signal, as a function of antenna, frequency, and direction.  That value will be an estimate of the complex-valued relative gains of each antenna in the given direction.
On a related note, this same technique could be performed on the $\sim$ 5000 airplane events which were detected in this dataset (with the added challenge that the source signal in those cases is unknown).  It is possible that these airplane-driven events are not isotropic, but AERA (the only other group known to investigate this in any capacity\textendash{}it appears that they were trying to synchronize their antennas with better accuracy) only saw one useful airplane per week\citep{2016JInst..11P1018T}, whereas OVRO-LWA appears to see dozens per day.  This warrants further investigation.

\bibliography{ownpubs.bib}
\bibliographystyle{model1-num-names}






% \end{thebibliography}

\end{document}